\documentclass[preprint, floatfix, showkeys]{revtex4-2}

\usepackage[utf8]{inputenc}
\usepackage{bm}
\usepackage[version=4]{mhchem}
\usepackage{rotating}
\usepackage{multirow}
\usepackage{amsmath}
\usepackage{graphicx}
\usepackage{hyperref}
\usepackage{booktabs}
\usepackage{adjustbox}
\usepackage{subfigure}
\usepackage{array}
\usepackage{cleveref}
\usepackage[english]{babel}
\crefname{equation}{Eq.}{Eqs.}
\Crefname{equation}{Equation}{Equations}

\DeclareUnicodeCharacter{2212}{-}
\DeclareUnicodeCharacter{3B1}{\ensuremath{\alpha}}
 \DeclareUnicodeCharacter{394}{$\Delta$}
 \DeclareUnicodeCharacter{F8FF}{\textbullet}
\DeclareUnicodeCharacter{202A}{}
\DeclareUnicodeCharacter{202C}{}

\usepackage{xcolor}
\usepackage{chemnum}

\usepackage{chemformula}
\usepackage{ulem}

\begin{document}

\title{\textbf{Understanding oxide surface stability: Theoretical insights from silver chromate} 
}%

\author{A. Facundes}
\author{T. T. Dorini}%
\author{T. W. von Zuben}
\author{M. A. San-Miguel}
\email{smiguel@unicamp.br}
\affiliation{Department of Physical Chemistry, Institute of Chemistry, Universidade Estadual de Campinas, Campinas, São Paulo, Brazil}

\date{\today}

\begin{abstract}
Understanding the thermodynamic stability of surfaces under realistic environmental conditions is a central challenge in predicting materials behavior from first principles. Here we employ a first-principles atomistic thermodynamics framework based on density functional theory (DFT) to systematically assess the relative stability of several surface orientations and terminations of silver chromate (\ce{Ag2CrO4}). The surface Gibbs free energies were computed as functions of oxygen and silver chemical potentials, enabling the construction of a comprehensive surface phase diagram. This methodology captures the interplay between atomic-scale structural motifs and the external thermodynamic environment, providing a robust route to predict surface stability beyond vacuum conditions. Silver chromate serves as a model system to illustrate the method, revealing that the coordination of surface chromium–oxygen clusters governs the relative stability of competing terminations. The approach further allows the prediction of morphology evolution via Wulff constructions, establishing a consistent theoretical framework for connecting first-principles energetics to equilibrium crystal shapes.
\end{abstract}



\keywords{Semiconductors, Silver chromate \ce{Ag2CrO4}; Surface termination, DFT thermodynamic calculations.}
\maketitle

	\newpage
	\section{\label{sec:introduction}Introduction}

In recent years, the family of ternary silver-based oxide semiconductors, particularly those with the general formula \ce{Ag2MO4} (where M represents transition metals such as Cr, Mo, W, and others), has been extensively studied due to their technological potential. 
These materials have found applications in various fields, including photoluminescent materials \cite{de_santana_silver_2014, teodoro2022connecting}, sensors \cite{silva_novel_2014, guo2022sensitive}, photocatalysts \cite{soofivand_silver_2013, arslan_hamat_experimental_2020, xu_photocatalytic_2015, assis_surface-dependent_2021, gouveia2022ag2wo4, gouveia2022surfactant, alvarez2020selective, lipsky2020unraveling}, and antimicrobial agents \cite{foggi_synthesis_2017, assis_surface-dependent_2021, ong_role_2017, alvarez2020selective, lipsky2020unraveling}.
Particularly, the photocatalytic degradation of pollutants in aqueous systems using silver chromate (\ce{Ag2CrO4}) as a photocatalyst has been a significant focus of research \cite{souza2024boosted, du2024visible, ren2021hollow}. This interest is due to its high photon absorption efficiency \cite{assis_surface-dependent_2021}, which is associated with its crystalline structure and favorable electronic band structure that promotes the formation of a band gap of approximately 1.75 eV \cite{assis_surface-dependent_2021}.

Unlike \ce{TiO2}, which primarily absorbs UV light ($\lambda <$ 400 nm), \ce{Ag2CrO4} extends its photoresponse into the visible region due to its narrower band gap, enabling more efficient solar spectrum utilization \cite{deng_facile_2016, xu_enhanced_2015, liu_microwave_2012}.

In general, the degradation of organic compounds by silver oxide semiconductors is attributed to the formation of reactive oxygen species (ROS), which include hydroxyl ($\cdot$OH) and superoxide radicals ($\cdot$O$_2^{-}$), hydrogen peroxide (\ce{H2O2}), perhydroxyl radical ($\cdot$O$_2$H), and singlet oxygen ($^1$\ce{O2}) on the catalyst surface. This process occurs through the interaction of water molecules and oxygen in the air with photogenerated electrons and holes generated in the system \cite{nosaka_generation_2017}.
More specifically, ROS formation occurs through the interaction of water molecules with photogenerated holes ($\text{h}^+$) in the valence band (VB) and molecular oxygen with photogenerated electrons ($\text{e}^-$) in the conduction band (CB):

\begin{equation}
\begin{aligned}
\ce{O2 + e^- &-> $\cdot$ O2^{-}} \\
\ce{$\cdot$ O2^- + H^+ &-> $\cdot$ O2H} \\
\ce{$\cdot$ O2H + H^+ &-> H2O2} \\
\ce{H2O2 + e^- &-> $\cdot$ OH + OH^{-}} \\
\ce{H2O + h^+ &-> $\cdot$ OH + H^+} \\
\ce{OH^- + h^+ &-> $\cdot$ OH}
\end{aligned}
\end{equation}

Once formed, these ROS play a crucial role in environmental remediation, as they react directly with organic pollutants through oxidation reactions, effectively neutralizing contaminants and eliminating pathogenic microorganisms and bacteria \cite{nosaka_generation_2017, feizpoor_novel_2017, luo_rational_2016, zhang2021reactive, bayati2024comprehensive, dai2024hydroxyl, shoudho2024influence, osazee2024urgent}. Recent computational and experimental studies have focused on elucidating the atomic-scale mechanisms of ROS formation and their correlation with surface structure \cite{lipsky_tale_2023, wang2024insights, cao2025mechanistic, gonzalez2025density, ta2025buffering, jiang2024highlight}. These investigations include density functional theory (DFT) \cite{hohenberg_inhomogeneous_1964, kohn_self-consistent_1965} calculations of surface reactivity, in-situ spectroscopic studies of radical formation, and structure-activity relationship analyses. However, fundamental understanding of how specific surface terminations and atomic arrangements influence ROS generation pathways and photocatalytic efficiency remains limited in the literature.

In a previous study \cite{assis_surface-dependent_2021}, the photocatalytic activity of \ce{Ag2CrO4} was proved to be highly surface dependent, since different samples of the same material have shown different photocatalytic efficiency in the degradation of both rhodamine B (as a model of organic polution of water streams) and \textit{Candida glabrata} cells (as a model of pathogenic agent). 
The observed differences stem from the intrinsic structural complexity of the \ce{Ag2CrO4} crystal, which features a variety of network-forming and network-modifying clusters (such as [\ce{CrO4}], [\ce{AgO4}] and [\ce{AgO6}] ) at its surfaces. This diversity of structural motifs gives rise to multiple possible terminations with distinct coordination environments and stability, which critically govern the surface reactivity and, consequently, the ability of the material to generate ROS.

The previous investigation \cite{assis_surface-dependent_2021} examined only one surface termination for each crystallographic orientation, leaving unexplored the potential influence of different termination compositions within the same orientation on system stability and photocatalytic performance. This limitation represents a significant knowledge gap, as the relative thermodynamic stability and surface reactivity of various termination configurations remain uncharacterized \cite{gouveia2024, gouveia2021modulating, gouveia2025morphology}. Understanding these structure-property relationships is essential for establishing the fundamental principles that govern the surface physics and chemistry of \ce{Ag2CrO4}, thereby enabling the rational design of optimized photocatalytic materials with enhanced ROS generation capabilities. 

First-principles studies have provided significant insights into the structural, electronic, and energetic properties of silver chromate \cite{kushwaha_investigation_2017, lacerda_diagnosis_2022, meena_performance_2019} through systematic investigation of bulk crystalline phases \cite{santamaria-perez_phase_2013}, surface energies \cite{silva_theoretical_2016}, electronic band structures \cite{zhang_illustration_2015}, and defect formation mechanisms \cite{fabbro_understanding_2016}. These investigations, predominantly based on DFT, have elucidated fundamental aspects such as the relationship between crystal structure and photocatalytic activity, the nature of surface reconstructions, and the mechanisms underlying charge carrier dynamics. However, standard DFT calculations are performed within the framework of static lattice approximations at 0 K, neglecting thermal effects and finite temperature dynamics that are crucial for realistic surface behavior. In practical photocatalytic applications, \ce{Ag2CrO4} surfaces operate under ambient or elevated temperatures where thermal fluctuations significantly influence surface reconstruction, atomic mobility, and thermodynamic stability. Incorporating these finite-temperature and pressure-dependent effects into theoretical frameworks presents considerable computational challenges, as it requires going beyond conventional DFT approaches to include methods such as ab initio molecular dynamics, thermodynamic integration, or statistical mechanical treatments of surface phase diagrams. 

To overcome these limitations, first-principles atomistic thermodynamics, developed by Reuter and Scheffler \cite{reuter_composition_2001,rogal2007ab}, provides a robust framework for incorporating finite temperature and pressure effects into surface stability analysis. This methodology utilizes DFT total energies as input to construct surface phase diagrams by calculating the Gibbs free energy of different surface configurations as a function of external conditions. The approach achieves this by coupling the surface system with molecular reservoirs (such as O$_2$, H$_2$O, or other atmospheric species) through their respective chemical potentials, thereby enabling the prediction of thermodynamically stable surface structures under realistic environmental conditions. 
This enables a more realistic analysis of the stability of solid surfaces, while still relying on purely theoretical and quantum mechanical calculations \cite{stampfl_catalysis_2002, reuter_ab_nodate, lee2024rise}. 
This approach has demonstrated remarkable success and reliability across diverse pure metalic \cite{suleiman2025iron, muller2020shape} and metal oxide systems, such as \ce{BiFeO3} \cite{dai_thermodynamic_2017}, \ce{SrTiO3} \cite{heifets_electronic_2007, bottin_stability_2003}, \ce{BaZrO3} \cite{heifets_density_2007}, \ce{Ag2MoO4} \cite{dorini2025mapping}, Ni-rich Ni-Co-Mn oxide cathodes \cite{liu_ab_2020}, and \ce{BaFe12O19} \cite{pobervznik2023surface}, providing quantitatively accurate predictions of surface stability under operating conditions. 

To address these fundamental knowledge gaps and establish comprehensive structure-property relationships for \ce{Ag2CrO4}, the present investigation employs a systematic combination of DFT calculations and first-principles thermodynamics to examine the complete range of low-index crystallographic orientations ((100), (010), (001), (110), (101), (011), and (111)) and their different surface terminations. We placed greater emphasis on the mathematical and physical description of the model, also demonstrating that vibrational and entropic contributions of the solid phases are, in fact, negligible at our level of theory, an aspect that has not been explicitly accounted in recent literature \cite{suleiman2025iron, muller2020shape,dai_thermodynamic_2017,dorini2025mapping,liu_ab_2020,pobervznik2023surface, wu2021stability, wu2021thermodynamic, yang2023first, cai2016first}. We also worked with a much broader set of surface terminations than usually reported \cite{reuter_composition_2001,reuter_ab_nodate,suleiman2025iron,muller2020shape,dai_thermodynamic_2017,heifets_electronic_2007,bottin_stability_2003, heifets_density_2007,pobervznik2023surface,wu2021stability,wu2021thermodynamic, yang2023first,liu_ab_2020, cai2016first}, which enables, for the first time, a quantitative assessment of how environmental conditions (including temperature and oxygen partial pressure) influence the relative stability and surface composition of different \ce{Ag2CrO4} faces. Through the construction of temperature and pressure-dependent surface phase diagrams and Wulff morphologies, this work provides critical insights into the thermodynamic driving forces that govern surface reconstruction and cluster formation ([\ce{AgO6}], [\ce{AgO4}], and [\ce{CrO4}]) under realistic operating conditions. The resulting structure-stability maps will not only establish the theoretical foundation necessary for rational design of \ce{Ag2CrO4} photocatalysts with optimized surface configurations and enhanced ROS generation capabilities, but also enable general reasonings towards the comprehension of the physics behind the stability of ternary oxides.

\section{\label{sec:methodology}Methodology}
\subsection{\label{subsec:computational-details}Computational details}

	All DFT calculations were performed using the Vienna Ab initio Simulation Package (VASP) \cite{kresse_efficient_1996, kresse_ultrasoft_1999}. The Perdew-Burke-Ernzerhof functional with generalized gradient approximation (PBE-GGA) \cite{perdew_generalized_1996, perdew_atoms_1992} and the projector augmented wave (PAW) method \cite{blochl_projector_1994} were used to describe the core-valence electrons interactions. The valence electron configurations were: silver (Ag) 4d$^{10}$ 5s$^1$ (11 electrons), chromium (Cr) 3d$^5$ 4s$^1$ (6 electrons), and oxygen (O) 2s$^2$ 2p$^4$ (6 electrons). The cutoff energy for the plane wave expansion was determined through a convergence test of the system's total energy. 
    The Brillouin zone was sampled using the Monkhorst-Pack method scheme \cite{monkhorst}  centered at the $\Gamma$-point, with the $k$-point grid density for bulk calculations set to ensure a precision of at least 0.03$\times2\pi/\text{\AA}$ in reciprocal space. 
    The system was relaxed using the conjugate gradient algorithm, employing the Gaussian smearing method with a small sigma value (e.g., 0.05 eV) to determine the partial occupancy of the wave functions.

	The convergence of the bulk system was achieved when the Hellmann-Feynman forces on each atom were less than 0.001 eV/{\AA} and the total free energy change between successive ionic steps was below $10^{-5}$ eV. The resulting optimized bulk unit cell parameters were then employed to construct surface slabs with symmetric terminations along the (100), (010), (001), (110), (101), (011), and (111) crystallographic orientations. This cleavage process was performed using the Pymatgen library \cite{tran2016surface}. For these slab models, the relaxation was also made using the conjugate gradient algorithm with the same sigma value of the bulk system for the Gaussian smearing method, but a different set of convergence criteria was applied: atomic positions were relaxed until residual forces were less than 0.01 eV/{\AA} per atom, and the total energy change was below $10^{-6}$ eV.

	The selected slabs were chosen based on the variability of generated terminations to obtain a comprehensive sample of systems, limited to a maximum of 9 distinct slabs per crystallographic orientation to ensure computational feasibility. For the optimization of each slab, the same plane-wave cutoff energy established for the bulk system was utilized. However, the $k$-point grids for slab calculations were generated using VASPKIT \cite{wang_vaspkit_2021} to maintain a $k$-point density in the $a$ and $b$ planar directions comparable to the bulk calculations (precision of 0.03$\times2\pi/\text{\AA}$). In contrast, a single $k$-point was employed in the $c$-direction (perpendicular to the surface plane) due to the vacuum. 
 To prevent interactions between periodic images of the slab along the $z$-axis, a vacuum layer of at least 15 {\AA} was introduced. During the relaxation of the slab structures, the ions in the central layers were kept fixed in their bulk-truncated positions, while a sufficient number of layers near both the upper and lower surfaces were allowed to fully relax. All calculations and subsequent analysis were conducted using an initial state of PS-TEROS code \cite{psteros}, developed in our research group, serving as a validation of the program. This framework automates the entire workflow, from slab generation and VASP simulations to the thermodynamic analysis and plotting of surface stability, ensuring both consistency and reproducibility throughout the study.
	

		\subsection{\label{subsec:structural-model}Structural model}

\begin{figure}[h]
			\centering
			\includegraphics[scale=0.5]{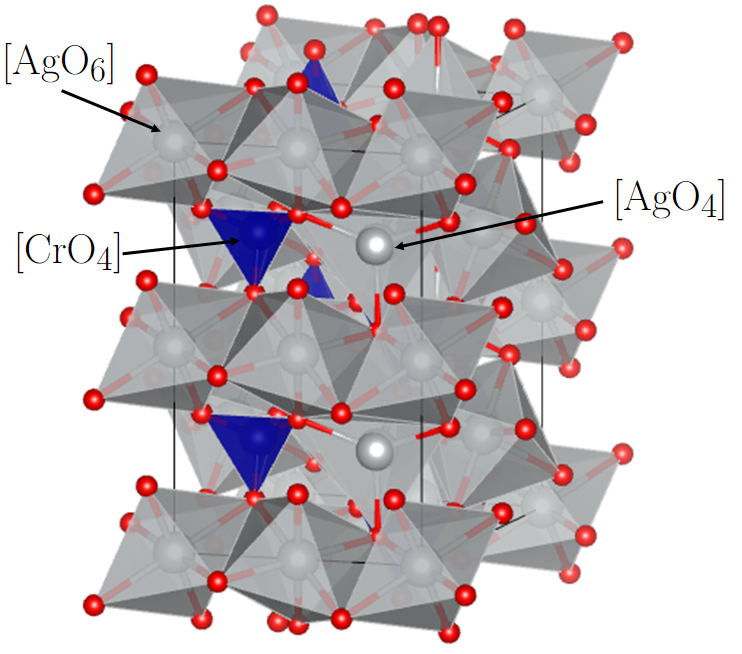}
			\caption{Representation of a \ce{Ag2CrO4} bulk and its clusters.}
			\label{fig:ag2cro4-bulk}
		\end{figure}

		The silver chromate crystal (\ce{Ag2CrO4}), with an orthorhombic geometry, is 
		composed of three-dimensional repetitions of elongated [\ce{AgO4}] clusters, 
		distorted octahedral [\ce{AgO6}] clusters, and tetrahedral [\ce{CrO4}] clusters 
		(\Cref{fig:ag2cro4-bulk}). Additionally, the [\ce{AgO4}] clusters have bond angles ranging from 
		86° to 117°, while the [\ce{AgO6}] and [\ce{CrO4}] clusters exhibit bond angles 
		around 88° and 110°, respectively \cite{assis_surface-dependent_2021}.

		The lattice parameters of the relaxed structure obtained in this work are very 
		close to those found in experimental and theoretical investigations reported in 
		the literature employing the PBE-GGA functional 
		(as indicated in \Cref{tab:struc-param}), with an average percentage error  
		of approximately ± 1$\%$. 

		\begin{table}[!h]
			\caption{Structural parameters optimization and comparison with data from the 
			literature.}
			\centering
			\begin{tabular}{c|c|c|c|c}
				\hline
				\textbf{Reference} & \textbf{a ({\AA})} & \textbf{b ({\AA})} & \textbf{c ({\AA})} & \textbf{V ({\AA}$^3$)} \\ 
				\hline 
				This work & 10.237 & 7.002 & 5.628 & 403.412 \\ 
        \cite{santamaria-perez_phase_2013} (Theoretical)  & 10.224 & 7.025 & 5.653 & 406.200 \\ 
        \cite{fabbro_understanding_2016} (Theoretical)  & 10.117 & 7.011 & 5.618 & 400.730 \\ 
        \cite{assis_surface-dependent_2021} (Theoretical) & 10.194 & 7.001 & 5.634 & 400.090 \\ 
        \cite{fabbro_understanding_2016} (Experimental) & 10.066 & 7.025 & 5.540 & 391.830 \\ 
        \cite{santamaria-perez_phase_2013} (Experimental)  & 10.065 & 7.013 & 5.538 & 390.900 \\ 
				\hline
			\end{tabular} 
			\label{tab:struc-param}
		\end{table}

\begin{figure}[h]
			\centering
			\includegraphics[scale=0.45]{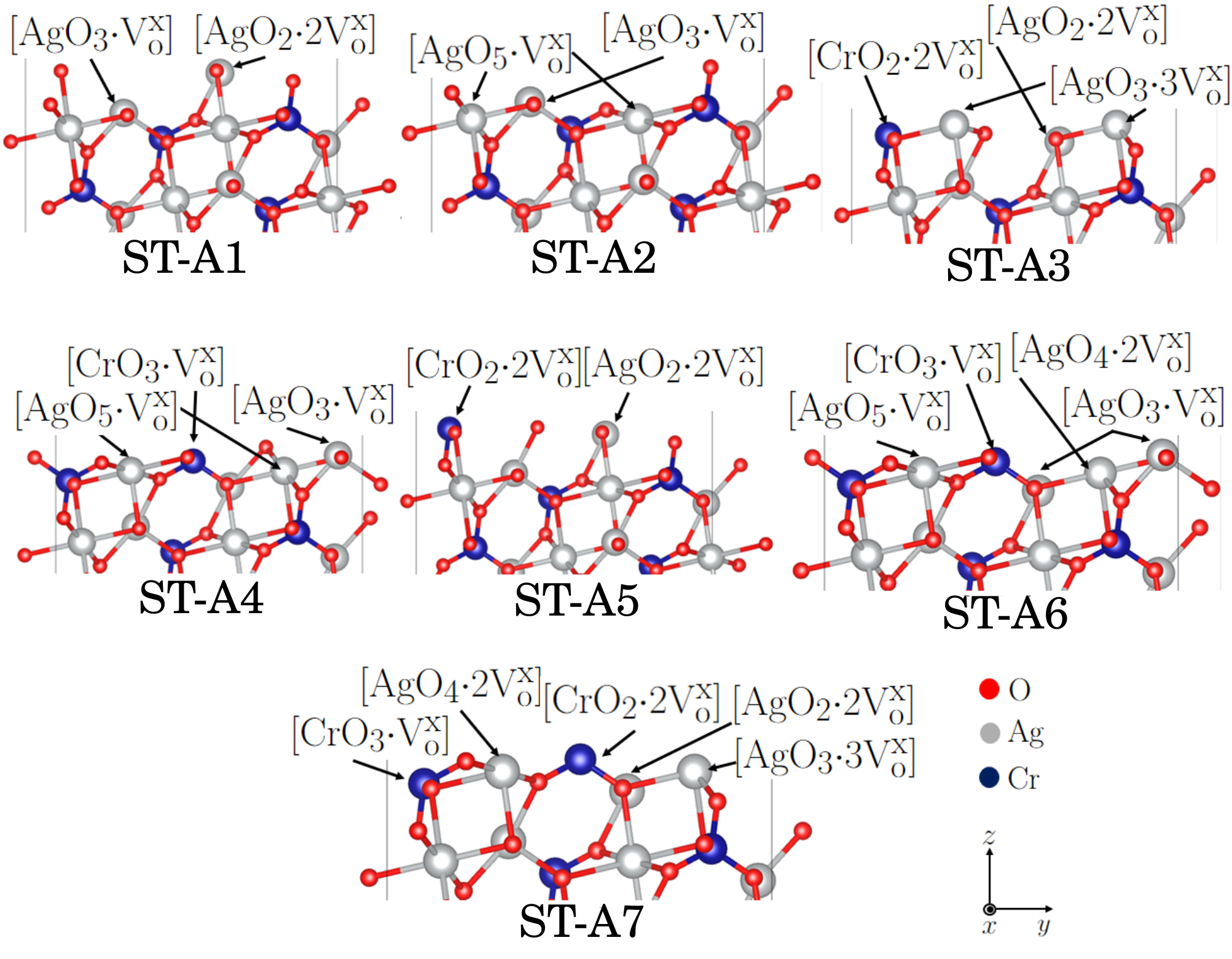}
			\caption{A representation of the unrelaxed slab models for the  
			(110) orientation. The different surface clusters are presented using the 
			Kröger-Vink notation.}
			\label{fig:110-slabs}
		\end{figure}

		From the optimized bulk structure, the surface models were cut out along the (110), 
		(111), (100), (010), (001), (011), and (101)  directions based on previously 
		published results in the literature regarding the photocatalytic activities of the 
		material \cite{assis_surface-dependent_2021}. 
		Symmetric slabs were chosen due to their ease of achieving electronic 
		convergence compared to asymmetric slabs, and because they do not require dipole 
		corrections. Thus, considering the different surface terminations, a total of 46 slab models were built. 
		\Cref{fig:110-slabs} displays the seven terminations for the (110) orientation, and the terminations for the other orientations can be visualized in Figures S1 - S10, Section I, of the Supplementary Materials (SM).
		
		The details about its structural general characterization in terms of 
		surface incomplete metallic clusters and the number of neutral oxygen vacancies 
		($\mathrm{V_o^x}$), atomic species, and of surface clusters are shown in 
		\Cref{tab:generalchar-table1} and \Cref{tab:generalchar-table2}, showing 
		how different the systems here studied can be from each other, not only 
		varying considerably  the number of neutral oxygen vacancies from zero to 
		eleven, but also the amount and coordination degree of surface clusters, 
		such as in more complex terminations like ST-A7 (from (110) orientation) 
		with clusters [\ce{AgO4}$\cdot2\mathrm{V^o_x}$], [\ce{AgO2}$\cdot2\mathrm{V^o_x}$], 
		[\ce{AgO3}$\cdot3\mathrm{V^o_x}$], [\ce{CrO3}$\cdot\mathrm{V^o_x}$], 
		[\ce{CrO2}$\cdot2\mathrm{V^o_x}$] on its surface, or more simple ones, such as 
		the ST-D2 termination (from (001) orientation), with only one 
		[\ce{AgO3}$\cdot\mathrm{V^o_x}$] cluster and fully coordinated [\ce{AgO4}], [\ce{AgO6}], and [\ce{CrO4}] clusters. 
		Additionally, the terminations ST-C6 (orientation (100)) and ST-D1 (orientation (001)) are only made 
		of clusters with complete coordination.

			\begin{table}[h]
				\caption{Structure general characterization of the terminations of 
				the (110), (111), and (100) orientations.}
				\centering
				\resizebox{\textwidth}{!}{%
				\begin{tabular}{cccccccccc}
					\hline
					\textbf{Orientation} & \textbf{Termination} & \textbf{Structures} & \textbf{$\mathrm{V_o^x}$} & \textbf{\ce{Ag}} & \textbf{\ce{Cr}} & \textbf{O} & \textbf{[\ce{AgO6}]} & \textbf{[\ce{AgO4}]} & \textbf{[\ce{CrO4}]} \\
					\hline
					\multirow{7}{4em}{(110)} & ST-A1 & [\ce{AgO3}$\cdot\mathrm{V^o_x}$], [\ce{AgO2}$\cdot2\mathrm{V^o_x}$], 2\ce{O} & 3 & 2 & - & 7 & - & 2 & - \\
					& ST-A2 & [\ce{AgO3}$\cdot\mathrm{V^o_x}$], 2[\ce{AgO5}$\cdot\mathrm{V^o_x}$], \ce{O} & 3 & 3 & - & 14 & 2 & 1 & - \\
					& ST-A3 & 2[\ce{AgO3}$\cdot3\mathrm{V^o_x}$], [\ce{AgO2}$\cdot2\mathrm{V^o_x}$], [\ce{CrO2}$\cdot2\mathrm{V^o_x}$] & 10 & 3 & 1 & 10 & 2 & 2 & 1 \\
					& ST-A4 & 2[\ce{AgO5}$\cdot\mathrm{V^o_x}$], [\ce{AgO3}$\cdot\mathrm{V^o_x}$], [\ce{CrO3}$\cdot\mathrm{V^o_x}$] & 4 & 3 & 1 & 16 & 2 & 1 & 1 \\
					& ST-A5 & [\ce{AgO2}$\cdot2\mathrm{V^o_x}$], [\ce{CrO2}$\cdot2\mathrm{V^o_x}$], 2\ce{O} & 4 & 1 & 1 & 6 & - & 1 & 1 \\
					& ST-A6 & [\ce{AgO5}$\cdot\mathrm{V^o_x}$], [\ce{AgO4}$\cdot2\mathrm{V^o_x}$], 2[\ce{AgO3}$\cdot\mathrm{V^o_x}$], [\ce{CrO3}$\cdot\mathrm{V^o_x}$] & 6 & 4 & 1 & 18 & 3 & 2 & 1 \\
					& ST-A7 & [\ce{AgO4}$\cdot2\mathrm{V^o_x}$], [\ce{AgO2}$\cdot2\mathrm{V^o_x}$], [\ce{AgO3}$\cdot3\mathrm{V^o_x}$], [\ce{CrO3}$\cdot\mathrm{V^o_x}$],  [\ce{CrO2}$\cdot2\mathrm{V^o_x}$] & 10 & 3 & 2 & 15 & 2 & 1 & 2 \\
					\hline
					\hline
					\multirow{7}{4em}{(111)} & ST-B1 & 3[\ce{AgO3}$\cdot\mathrm{V^o_x}$], [\ce{AgO5}$\cdot\mathrm{V^o_x}$], [\ce{AgO4}$\cdot2\mathrm{V^o_x}$] & 6 & 5 & - & 18 & 2 & 3 & - \\
					& ST-B2 & 2[\ce{AgO3}$\cdot\mathrm{V^o_x}$], [\ce{AgO4}$\cdot2\mathrm{V^o_x}$], 2[\ce{AgO2}$\cdot2\mathrm{V^o_x}$], [\ce{AgO3}$\cdot3\mathrm{V^o_x}$] & 11 & 6 & - & 17 & 2 & 4 & - \\
					& ST-B3 & [\ce{AgO3}$\cdot3\mathrm{V^o_x}$], [\ce{AgO5}$\cdot\mathrm{V^o_x}$], 2[\ce{AgO3}$\cdot\mathrm{V^o_x}$], [\ce{CrO3}$\cdot\mathrm{V^o_x}$] & 7 & 4 & 1 & 17 & 2 & 2 & 1 \\
					& ST-B4 & 2[\ce{AgO2}$\cdot2\mathrm{V^o_x}$], 2[\ce{AgO3}$\cdot3\mathrm{V^o_x}$], [\ce{AgO3}$\cdot\mathrm{V^o_x}$], 2\ce{O} & 11 & 5 & - & 15 & 2 & 3 & - \\
					& ST-B5 & [\ce{AgO5}$\cdot\mathrm{V^o_x}$], [\ce{AgO2}$\cdot2\mathrm{V^o_x}$], [\ce{AgO3}$\cdot\mathrm{V^o_x}$], 2[\ce{AgO3}$\cdot3\mathrm{V^o_x}$], 2\ce{O} & 7 & 5 & - & 18 & 3 & 2 & - \\
					& ST-B6 & 2[\ce{AgO2}$\cdot2\mathrm{V^o_x}$], 2[\ce{AgO5}$\cdot\mathrm{V^o_x}$], [\ce{CrO3}$\cdot\mathrm{V^o_x}$] & 7 & 4 & 1 & 17 & 2 & 2 & 1 \\
					& ST-B7 & [\ce{AgO2}$\cdot2\mathrm{V^o_x}$], 2[\ce{AgO5}$\cdot\mathrm{V^o_x}$], [\ce{CrO3}$\cdot\mathrm{V^o_x}$] & 5 & 3 & 1 & 15 & 2 & 1 & 1 \\
					\hline
					\hline
					\multirow{7}{4em}{(100)} & ST-C1 & [\ce{AgO3}$\cdot\mathrm{V^o_x}$], 3\ce{O} & 1 & 1 & - & 6 & - & 1 & - \\
					& ST-C2 & [\ce{AgO2}$\cdot2\mathrm{V^o_x}$], 2[\ce{AgO4}$\cdot2\mathrm{V^o_x}$] & 6 & 3 & - & 10 & 2 & 1 & - \\
					& ST-C3 & 2[\ce{AgO4}$\cdot2\mathrm{V^o_x}$] & 4 & 2 & - & 8 & 2 & - & - \\
					& ST-C4 & 2[\ce{AgO3}$\cdot3\mathrm{V^o_x}$] & 6 & 2 & - & 6 & 2 & - & - \\
					& ST-C5 & [\ce{AgO3}$\cdot\mathrm{V^o_x}$], [\ce{CrO3}$\cdot\mathrm{V^o_x}$] & 2 & 1 & 1 & 6 & - & 1 & 1 \\
					& ST-C6 & - & - & - & - & - & - & - & - \\
					& ST-C7 & [\ce{AgO2}$\cdot2\mathrm{V^o_x}$], [\ce{CrO2}$\cdot2\mathrm{V^o_x}$] & 4 & 1 & 1 & 4 & - & 1 & 1 \\
					\hline
				\end{tabular}}%
				\label{tab:generalchar-table1}
			\end{table}

			\newpage
		
			\begin{table}[h]
				\caption{Structure general characterization of the terminations of 
				the (001), (010), (011), and (101) orientations.}
				\centering
				\resizebox{\textwidth}{!}{%
				\begin{tabular}{cccccccccc}
					\hline
					\textbf{Orientation} & \textbf{Termination} & \textbf{Structures} & \textbf{$\mathrm{V_o^x}$} & \textbf{\ce{Ag}} & \textbf{\ce{Cr}} & \textbf{O} & \textbf{[\ce{AgO6}]} & \textbf{[\ce{AgO4}]} & \textbf{[\ce{CrO4}]} \\
					\hline
					\multirow{4}{4em}{(001)} & ST-D1 & - & - & - & - & - & - & - & - \\
					& ST-D2 & [\ce{AgO3}$\cdot\mathrm{V^o_x}$] & 2 & 1 & - & 3 & - & 1 & - \\
					& ST-D3 & [\ce{AgO3}$\cdot\mathrm{V^o_x}$], [\ce{AgO5}$\cdot\mathrm{V^o_x}$] & 2 & 2 & - & 8 & 1 & 1 & - \\
					& ST-D4 & 2[\ce{AgO3}$\cdot\mathrm{V^o_x}$], [\ce{AgO5}$\cdot\mathrm{V^o_x}$], 2[\ce{CrO3}$\cdot\mathrm{V^o_x}$] & 5 & 3 & 2 & 17 & 1 & 2 & 2 \\
					\hline
					\hline
					\multirow{4}{4em}{(010)} & ST-E1 & 2[\ce{AgO4}$\cdot2\mathrm{V^o_x}$] & 4 & 2 & - & 8 & 2 & - & - \\
					& ST-E2 & 2[\ce{AgO3}$\cdot3\mathrm{V^o_x}$] & 6 & 2 & - & 6 & 2 & - & - \\
					& ST-E3 & 2[\ce{AgO3}$\cdot\mathrm{V^o_x}$] & 2 & 2 & - & 6 & - & 2 & - \\ 
					& ST-E4 & 2[\ce{AgO3}$\cdot\mathrm{V^o_x}$], 2[\ce{CrO3}$\cdot\mathrm{V^o_x}$] & 4 & 2 & 2 & 12 & - & 2 & 2 \\
					\hline
					\hline
					\multirow{8}{4em}{(011)} & ST-F1 & 2[\ce{AgO2}$\cdot2\mathrm{V^o_x}$], 2[\ce{AgO4}$\cdot2\mathrm{V^o_x}$] & 8 & 4 & - & 12 & 2 & 2 & - \\
					& ST-F2 & 2[\ce{AgO4}$\cdot2\mathrm{V^o_x}$], 2\ce{O} & 4 & 2 & - & 10 & 2 & - & - \\
					& ST-F3 & 2[\ce{AgO3}$\cdot3\mathrm{V^o_x}$], [\ce{AgO3}$\cdot\mathrm{V^o_x}$], 2\ce{O} & 7 & 3 & - & 11 & 2 & 1 & - \\
					& ST-F4 & 2[\ce{AgO4}$\cdot2\mathrm{V^o_x}$], [\ce{AgO3}$\cdot\mathrm{V^o_x}$] & 5 & 3 & - & 11 & 2 & 1 & - \\
					& ST-F5 & 2[\ce{AgO3}$\cdot3\mathrm{V^o_x}$], [\ce{AgO3}$\cdot\mathrm{V^o_x}$], 2[\ce{CrO3}$\cdot\mathrm{V^o_x}$] & 9 & 3 & 2 & 15 & 2 & 1 & 2 \\
					& ST-F6 & 2[\ce{AgO3}$\cdot\mathrm{V^o_x}$], 2[\ce{CrO3}$\cdot\mathrm{V^o_x}$] & 4 & 2 & 2 & 6 & - & 2 & 2 \\ 
					& ST-F7 & 2[\ce{AgO2}$\cdot2\mathrm{V^o_x}$], 2[\ce{AgO5}$\cdot\mathrm{V^o_x}$], 2[\ce{CrO3}$\cdot\mathrm{V^o_x}$] & 8 & 4 & 2 & 16 & 2 & 2 & 2 \\
					& ST-F8 & 2[\ce{AgO2}$\cdot2\mathrm{V^o_x}$], 2[\ce{AgO5}$\cdot\mathrm{V^o_x}$], 2[\ce{CrO3}$\cdot\mathrm{V^o_x}$] & 8 & 4 & 2 & 16 & 2 & 2 & 2 \\
					\hline
					\hline
					\multirow{9}{4em}{(101)} & ST-G1 & 2[\ce{AgO5}$\cdot\mathrm{V^o_x}$] & 2 & 2 & - & 10 & 2 & - & - \\
					& ST-G2 & [\ce{AgO5}$\cdot\mathrm{V^o_x}$], [\ce{AgO4}$\cdot2\mathrm{V^o_x}$] & 3 & 2 & - & 9 & 2 & - & - \\
					& ST-G3 & [\ce{AgO5}$\cdot\mathrm{V^o_x}$], [\ce{AgO3}$\cdot3\mathrm{V^o_x}$] & 4 & 2 & - & 8 & 2 & - & - \\
					& ST-G4 & 2[\ce{AgO3}$\cdot\mathrm{V^o_x}$] & 2 & 2 & - & 6 & - & 2 & - \\
					& ST-G5 & [\ce{AgO3}$\cdot3\mathrm{V^o_x}$], [\ce{AgO4}$\cdot2\mathrm{V^o_x}$] & 5 & 2 & - & 7 & 2 & - & - \\
					& ST-G6 & 2[\ce{AgO3}$\cdot\mathrm{V^o_x}$], 2[\ce{CrO3}$\cdot\mathrm{V^o_x}$] & 4 & 2 & 2 & 12 & - & 2 & 2 \\
					& ST-G7 & 2[\ce{AgO5}$\cdot\mathrm{V^o_x}$], [\ce{AgO3}$\cdot\mathrm{V^o_x}$], [\ce{CrO3}$\cdot\mathrm{V^o_x}$] & 4 & 3 & 1 & 16 & 2 & 1 & 1 \\
					& ST-G8 & 3[\ce{AgO3}$\cdot\mathrm{V^o_x}$], 2[\ce{CrO3}$\cdot\mathrm{V^o_x}$] & 5 & 3 & 2 & 15 & - & 3 & 2 \\
					& ST-G9 & 2[\ce{AgO3}$\cdot\mathrm{V^o_x}$], [\ce{CrO3}$\cdot\mathrm{V^o_x}$] & 3 & 2 & 1 & 9 & - & 2 & 1 \\
					\hline
				\end{tabular}}%
				\label{tab:generalchar-table2}
			\end{table}
	
		\subsection{\label{subsec:thermodynamic-model}Thermodynamic model}
			The thermodynamic model here derived was based on previous works 
			\cite{reuter_composition_2001,rogal2007ab,bottin_stability_2003} and standard statistical mechanics formalism \cite{mcquarrie}. 
			The surface Gibbs free energy ($\gamma$) of a \ce{Ag2CrO4} symmetric slab 
			with surface area $A$ in contact with a pure atmosphere (for simplification) in oxygen molecules can be given as
			\begin{equation}
				\gamma(T, P) = \displaystyle\frac{1}{2A}\left[\Gamma(T, P) - 
				\Delta\mu_{\ce{Ag}}(T, P)N_{\ce{Ag,Cr}} - \Delta\mu_{\ce{O}}N_{\ce{O,Cr}}\right],
				\label{eq:gamma}
			\end{equation}
			where
			\begin{multline}
				\Gamma(T, P) = G^{\mathrm{slab}}(T, P) - N_{\ce{Cr}}g^{\mathrm{bulk}}_{\ce{Ag2CrO4}}(T, P) -
				g^{\mathrm{bulk}}_{\ce{Ag}}(T, P)N_{\ce{Ag,Cr}} - \frac{1}{2}g_{\ce{O2}}^{\text{gas}}(T, P)N_{\ce{O,Cr}},
				\label{eq:Gamma}
			\end{multline}
			which accounts for the Gibbs free energy of the slab ($G^{\mathrm{slab}}$) 
			in relation to pure phases, i.e., \ce{Ag2CrO4} bulk ($g^{\mathrm{bulk}}_{\ce{Ag2CrO4}}$), 
			Ag bulk ($g^{\mathrm{bulk}}_{\ce{Ag}}$) and free oxygen molecule ($g_{\ce{O2}}^{\text{gas}}$), 
			also dependent on the number of the atomic species $N_i$, with
			\begin{gather}
				N_{\ce{Ag,Cr}} = N_{\ce{Ag}} - 2N_{\ce{Cr}} 
				\label{eq:N_Ag_Cr}\\
				N_{\ce{O,Cr}} = N_{\ce{O}} - 4N_{\ce{Cr}}.
				\label{eq:N_O_Cr}
			\end{gather}

			The chemical potentials $\Delta\mu_i$ in \Cref{eq:gamma} are defined here as 
			\begin{gather}
				\Delta\mu_{\ce{Ag}}(T, P) = \mu_{\ce{Ag}}(T, P) - g_{\ce{Ag}}^{\text{bulk}}(T, P)
				\label{eq:delta_mu_ag}
				\\\Delta\mu_{\ce{O}}(T, P) = \mu_{\ce{O}}(T, P) - \frac{1}{2}g_{\ce{O}}^{\text{gas}}(T, P),
				\label{eq:delta_mu_o}
			\end{gather}
			that is, the difference of energies of the species in the silver chromate phase 
			in relation to its pure phase. In \Cref{eq:delta_mu_ag} and 
			\Cref{eq:delta_mu_o}, the approximation $\mu_i^{\text{pure}} 
			\approx g_i^{\text{pure}}$ is being made considering the contact with a 
			massive thermodynamic reservoir. These equations are very useful to find 
			well-defined theoretical limits for the chemical potentials $\Delta\mu_i$. 
			To ensure our results are physically meaningful, our analysis explicitly includes 
			the stability of competing oxides, like \ce{Ag2O} and \ce{Cr2O3}, to define the 
			conditions in which \ce{Ag2CrO4} is the stable phase. Thus, 
			observing \Cref{eq:delta_mu_ag} and \Cref{eq:delta_mu_o}, it is 
			clear that the silver and oxygen atoms would be favoured to stay in the \ce{Ag2CrO4} 
			lattice since its respective energies are lower than the energies of its pure phases. 
			Therefore,
			\begin{gather}
				\Delta\mu_{\ce{Ag}}(T, P) < 0
				\label{eq:delta_mu_ag_limit}
				\\\Delta\mu_{\ce{O}}(T, P) < 0.
				\label{eq:delta_mu_o_limit}
			\end{gather}

			Additionally, there is the constraint 
			\begin{equation}
				2\Delta\mu_{\ce{Ag}}(T, P) + 4\Delta\mu_{\ce{O}}(T, P) > E_{\ce{Ag2CrO4}}^f,
				\label{eq:deltamu_higherlimit}
			\end{equation}
			where $E_{\ce{Ag2CrO4}}^f$ is the formation energy of the silver chromate. 
			This inequality, together with \Cref{eq:delta_mu_ag_limit} and 
			\Cref{eq:delta_mu_o_limit}, guarantees that all the thermodynamic conditions 
			here studied imply the existence of a \ce{Ag2CrO4} crystal (its formation is 
			favoured), and they form well-defined theoretical limits for the 
			difference of chemical potentials.

			The $\Delta\mu_{\ce{O}}(T, P$) quantity was computed as described in 
			\cite{rogal2007ab}, given by the partition function of a free oxygen molecule 
			considered as an ideal gas using the Born-Oppenheimer approximation:
			\begin{multline}
				\displaystyle\Delta\mu_{\ce{O}}(T, P) = -\frac{1}{2}k_BT\bigg\{
					\ln\left[\left(\frac{2\pi m}{h^2}\right)^{3/2}\frac{(k_BT)^{5/2}}{P}\right] + 
					\ln\left(\frac{k_BT}{\sigma^sB_0}\right) - \\
					\ln\left[1 - \exp\left(\frac{\hbar\omega_0}{k_BT}\right)\right] +
					\ln\left(I^{\text{spin}}\right)\bigg\},
					\label{eq:delta_mu_o_sm}
			\end{multline}
			where $m$ is the mass of an oxygen molecule, $\sigma^s$ is a symmetry number that 
			indicates the number of distinguishable orientations that the molecule can have, 
			$B_0$ is the rotation constant, $\omega_0$ is its fundamental vibrational 
			mode, $I^{\text{spin}}$ is its spin degeneracy, and $k_B$ and $h$ are the 
			Boltzmann's and Planck's constants, respectively. The terms multiplied by the 
			$-\frac{1}{2}k_BT$ factor are the translational, rotational, vibrational, 
			and electronic contributions, respectively. 
			To match exactly the limits of $\Delta\mu_{\ce{O}}$ defined from the 
			\Cref{eq:delta_mu_o} and \Cref{eq:deltamu_higherlimit} using 
			\Cref{eq:delta_mu_o_sm}, we set $P = 10^{-6}$ atm. 

			Since an analytical expression derived from statistical mechanics for the 
			silver atoms is not as simple as for the oxygen atoms, mainly because of 
			phonon contributions from the silver cell, as a simplification, the 
			$\Delta\mu_{\ce{Ag}}(T, P)$ quantity was estimated using the limits from 
			\Cref{eq:delta_mu_ag_limit} and \Cref{eq:deltamu_higherlimit}, 
			as will be explained later.

			At last, all the Gibbs free energy terms from the solid (or pure) phases were 
			computed as
			\begin{multline}
				G(T, P, N_i) = E^T(V, N_i) + F^{\text{vib}}(T, V, N_i) + 
				F^{\text{conf}}(T, V, N_i) + PV(T, P, N_i),
			\end{multline}
			where $E^T$ is the internal energy, $F^{\text{vib}}$ is the Helmholtz 
			vibrational energy, $F^{\text{conf}}$ is the Helmholtz conformational energy, 
			and $PV$ is the pressure-volume dependent term. In the case of \ce{Ag2CrO4} bulk,  
			the internal energy contribution is much larger than the other ones 
			(see Figures S11 and S12 and discussions in the Section II.C of the SM), which are 
			about (or less than) 10 meV, the same order of magnitude as 
			the numerical errors from the DFT calculations. Additionally, since 
			the results are based on energy differences, it is expected that numerical error cancellations occur. 
			That way, we assumed that all slabs present the same order of magnitude 
			of $F^{\text{vib}}$, $F^{\text{conf}}$ and $PV$ of the bulk system and we 
			neglected all energy terms, except the internal energy one, which 
			can be interpreted as the total energy of the system computed from a DFT calculation. 
			Thus, \Cref{eq:Gamma}, \Cref{eq:delta_mu_ag} and \Cref{eq:delta_mu_o} 
			are approximated to, respectively,
			\begin{gather}
				\Gamma \approx E^{\mathrm{slab}} - E^{\mathrm{bulk}}_{\ce{Ag2CrO4}}N_{\ce{Cr}} -
				E^{\mathrm{bulk}}_{\ce{Ag}}N_{\ce{Ag,Cr}} - \frac{1}{2}E_{\ce{O2}}^TN_{\ce{O,Cr}}
				\label{eq:Gamma_E}
				\\\Delta\mu_{\ce{Ag}}(T, P) \approx \mu_{\ce{Ag}}(T, P) - E_{\ce{Ag}}^T
				\label{eq:delta_mu_ag_E}
				\\\Delta\mu_{\ce{O}}(T, P) \approx \mu_{\ce{O}}(T, P) - \frac{1}{2}E_{\ce{O}}^T.
				\label{eq:delta_mu_o_E}
			\end{gather}

			Similarly to \Cref{eq:deltamu_higherlimit}, other binary 
			oxides may be formed on the surface of the silver chromate 
			crystal depending on the thermodynamic condition, like 
			\ce{Ag2O} and \ce{Cr2O3}. Hence,
			\begin{gather}
				2\Delta\mu_{\ce{Ag}}(T, P) + \Delta\mu_{\ce{O}}(T, P) < E_{\ce{Ag2O}}^f,
				\label{eq:ag2o}
			\end{gather}
			or, alternatively for a \ce{Cr2O3} crystal,
			\begin{equation}
				4\Delta\mu_{\ce{Ag}}(T, P) + 5\Delta\mu_{\ce{O}}(T, P) >  2E_{\ce{Ag2CrO4}}^{f} - E_{\ce{Cr2O3}}^f.
				\label{eq:cr2o3}
			\end{equation}
			
			A detailed derivation of this model can be found in Section II of the SM.

	\section{\label{sub:results-discussions}Results and Discussions}
		\subsection{\label{subsec:surface-stability}Surface stability}
\begin{figure}[h]
				\centering
				\includegraphics[scale=0.95]{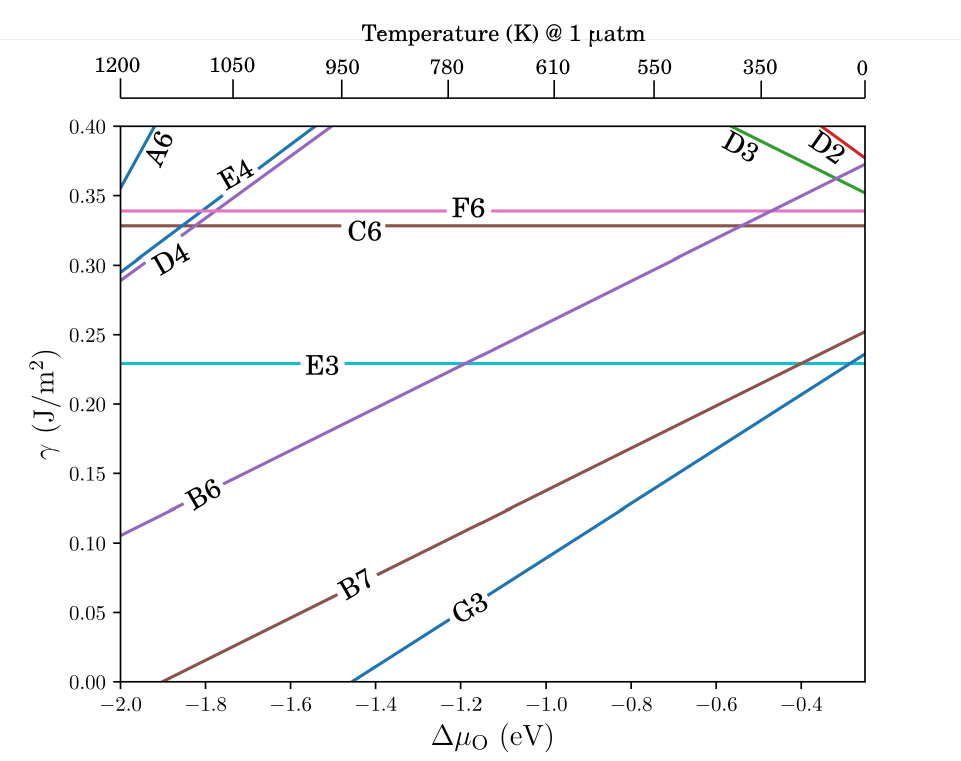}
				\caption{Surface Gibbs free energies as a function of oxygen 
				chemical potential for the 11 most stable symmetric terminations. 
				$\Delta\mu_{\ce{Ag}}$ is set to 0 eV.}
				\label{fig:gamma-general}
			\end{figure}

			First, we calculated the $\gamma$ values for all the 46 terminations as 
			a function only of $\Delta\mu_{\ce{O}}$, assuming no variation in the 
			chemical potential of silver with the aim of analysing the influence 
			of only the temperature (from \Cref{eq:delta_mu_o_sm}) on the 
			surface energy. Setting $\Delta\mu_{\ce{Ag}} = 0$ eV means that 
			the silver atoms in the silver chromate phase are in thermodynamic 
			equilibrium with silver atoms in a pure metallic phase 
			(from \Cref{eq:delta_mu_ag}). Accordingly, $\Delta\mu_{\ce{O}} = 0$ eV 
			implies a thermodynamic equilibrium between oxygen atoms in the lattice and free 
			oxygen molecules. The limit $\Delta\mu_i \rightarrow 0$ can be understood as 
			the addition of silver or oxygen atoms to the surface from the external 
			environment. The removal of such species implies a more negative $\Delta\mu_i$. We 
			will call these scenarios ``rich" and ``poor conditions", respectively. 
			The results for the 11 most stable slabs are depicted in \Cref{fig:gamma-general}. 
			We limited here the $\gamma$ values to 0.4 J/m$^2$ to help with 
			visualization. The temperature-chemical potential correspondence 
			values were calculated using \Cref{eq:delta_mu_o_sm}.


			Under this thermodynamic constraint, it is shown that the termination ST-G3 
			((101) orientation) is the most stable one, followed by ST-B7 ((111) orientation), 
			until a temperature very close to 0 K, when the ST-E3 ((010) orientation) has the 
			lowest surface Gibbs free energy. This shows that the ST-G3 structure is the least 
			sensitive to variations of the surrounding atmosphere, since it is the most 
			favoured one for both rich and poor oxygen conditions (i.e., for any temperature).
			
			In this study, $\Gamma$ (\Cref{eq:Gamma} or \Cref{eq:Gamma_E}) quantifies 
			the energetic deviation of the slab from the bulk system, which is related to the 
			system's enthalpy. \Cref{tab:general_info1} and 
			\Cref{tab:general_info2} show these $\Gamma$ values for each studied 
			termination, and indicating that the greater $\Gamma$, the higher the energetic cost 
			it is to generate the slab from the bulk, suggesting a less stable 
      		termination. This quantity dictates the stability of the 
			system when the \ce{Cr}:\ce{O} ratio in the system is 1:4, preserving the 
			bulk stoichiometry. This ratio is satisfied 
			if $N_{\ce{O,Cr}} = 0 \implies \gamma(T, P) = \gamma = \Gamma$, which is 
			the case of the terminations ST-C6, ST-E3, and ST-F6 in \Cref{fig:gamma-general}. 
			This is an interesting consequence of the thermodynamic model, which 
			states that the surface Gibbs free energy only varies with the chemical potential 
			of oxygen and silver if the system has structural defects 
			of chromium atoms in relation to silver and oxygen, respectively. If the material is 
			perfectly stoichiometric, the surface Gibbs free energy will depend only on 
			the $\Gamma$ quantity (or, indirectly, on the enthalpy), which is simplified here 
			as a constant, and, therefore, $\gamma$ does not change with the external 
			conditions.
		
			\begin{table}[htbp]
				\centering
				\caption{Excess of \ce{Ag} and \ce{O} atoms with respect to \ce{Cr} 
				atoms, and $\Gamma$ values for all terminations of the orientations 
				(110), (111), (100), and (001).}
				\begin{tabular}{ccccc}
					\hline
					\textbf{Orientation} & \textbf{Termination} & $\bm{N_{\ce{Ag,Cr}}}$ & 
					$\bm{N_{\ce{O,Cr}}}$ & $\bm{\Gamma}$ (\textbf{eV}) \\
					\hline
					\multirow{7}{4em}{(110)} & ST-A1 & 2 & 6 & 6.20 \\
					& ST-A2 & 0 & 2 & 3.14 \\
					& ST-A3 & 2 & 0 & 8.75 \\
					& ST-A4 & 0 & 0 & 7.26 \\
					& ST-A5 & $-$2 & 0 & 9.95 \\
					& ST-A6 & 0 & $-2$ & 6.57 \\
					& ST-A7 & $-$2 & $-6$ & 16.36 \\
					\hline
					\hline
					\multirow{7}{4em}{(111)} & ST-B1 & 6 & 10 & 5.69 \\
					& ST-B2 & 6 & 4 & 6.84 \\
					& ST-B3 & 2 & 6 & 10.02 \\
					& ST-B4 & 4 & 2 & 4.11 \\
					& ST-B5 & 2 & 2 & 5.97 \\
					& ST-B6 & $-$2 & $-2$ & 5.38 \\
					& ST-B7 & $-$4 & $-2$ & 3.80 \\
					\hline
					\hline
					\multirow{7}{4em}{(100)} & ST-C1 & 4 & 8 & 6.52 \\
					& ST-C2 & 4 & 4 & 3.60 \\
					& ST-C3 & 2 & 4 & 2.82 \\
					& ST-C4 & 2 & 0 & 4.17 \\
					& ST-C5 & 0 & 0 & 7.16 \\
					& ST-C6 & 6 & 0 & 1.61 \\
					& ST-C7 & $-2$ & $-4$ & 14.98 \\
					\hline
					\hline
					\multirow{4}{4em}{(001)} & ST-D1 & 0 & 8 & 8.10 \\
					& ST-D2 & 0 & 6 & 3.82 \\
					& ST-D3 & 0 & 2 & 2.87 \\
					& ST-D4 & 0 & $-2$ & 6.59 \\
					\hline
				\end{tabular}
				\label{tab:general_info1}
			\end{table}

			\begin{table}[htbp]
				\centering
				\caption{Excess of \ce{Ag} and \ce{O} atoms with respect to \ce{Cr} 
				atoms, and $\Gamma$ values for all terminations of the orientations 
				(010), (011), and (111).}	
				\begin{tabular}{ccccc}
					\hline
					\textbf{Orientation} & \textbf{Termination} & $\bm{N_{\ce{Ag,Cr}}}$ & 
					$\bm{N_{\ce{O,Cr}}}$ & $\bm{\Gamma}$ (\textbf{eV}) \\
					\hline
					\multirow{4}{4em}{(010)} & ST-E1 & 4 & 8 & 7.05 \\
					& ST-E2 & 2 & 6 & 3.67 \\
					& ST-E3 & 0 & 2 & 1.65 \\
					& ST-E4 & $-4$ & $-2$ & 10.56 \\
					\hline
					\hline
					\multirow{8}{4em}{(011)} & ST-F1 & 6 & 8 & 8.80 \\
					& ST-F2 & 2 & 8 & 4.69 \\
					& ST-F3 & 2 & 4 & 7.35 \\
					& ST-F4 & 2 & 4 & 5.02 \\
					& ST-F5 & 2 & 0 & 5.53 \\
					& ST-F6 & $-2$ & 0 & 3.89 \\
					& ST-F7 & $-2$ & $-4$ & 13.72 \\
					& ST-F8 & $-2$ & $-4$ & 13.34 \\
					\hline
					\hline
					\multirow{9}{4em}{(101)} & ST-G1 & 4 & 10 & 4.79 \\
					& ST-G2 & 4 & 8 & 6.10 \\
					& ST-G3 & 4 & $-4$ & 5.29 \\
					& ST-G4 & 0 & 6 & 4.50 \\
					& ST-G5 & 4 & 2 & 4.19 \\
					& ST-G6 & 0 & 4 & 9.19 \\
					& ST-G7 & 2 & 2 & 3.82 \\
					& ST-G8 & $-2$ & $-2$ & 11.31 \\
					& ST-G9 & $-4$ & $-2$ & 2.91 \\
					\hline
				\end{tabular}
				\label{tab:general_info2}
			\end{table}
  
			Setting $\Delta\mu_{\ce{Ag}} = 0$ eV implies that 
			$\gamma(T, P) = \Gamma - \Delta\mu_{\ce{O}}N_{\ce{O,Cr}}$, 
			a first order equation with $N_{\ce{O,Cr}}$ as the angular coefficient. 
			Considering that, from the theoretical limits defined above, the values of 
			$\Delta\mu_{\ce{O}}$ are always negative, a directly proportional dependence 
			of $\gamma$ on $\Delta\mu_{\ce{O}}$ requires $N_{\ce{O,Cr}} > 0$, that is, 
			from \Cref{eq:N_Ag_Cr}, there is an excess of oxygen atoms in 
			the system with respect to the number of chromium atoms, illustrated 
			by the terminations ST-A6, ST-B6, ST-B7, ST-D4, ST-E4, and ST-G3. 
			An inversely proportional dependence implies a deficiency of oxygen atoms, 
			which is the case of ST-D2 and ST-D3. 
      
      The physical insight of this behaviour can be given by deriving $\gamma(T, P)$ with respect 
      to the temperature under constant pressure, resulting in
      \begin{equation}
        \displaystyle \left(\frac{\partial\gamma}{\partial T}\right)_P = \frac{1}{2A}S_{\ce{O}}N_{\ce{O,Cr}},
        \label{eq:dgamma/dT}
      \end{equation}
      where we used the thermodynamic relation $\displaystyle S_{\ce{O}} = -\frac{\partial\mu_{\ce{O}}}{\partial T}$, 
      with $S_{\ce{O}}$ being the entropy of an oxygen atom, which is always positive.
      The oxygen atoms in the gas phase will always have higher entropy than the 
      ones in the solid phase, since they have more degrees of freedom and, 
      consequentely, more microstates to distribute its energy. Therefore, a system 
      with more oxygen atoms with respect to chromium ones (e.g., $N_{\ce{O,Cr}} > 0$) 
      is destabilized as temperature increases, because the O atoms in the lattice 
      would tend to go to the gas phase, or the entropically more favoured state. 
      Consequently, maintaining the oxygen-rich termination is energetically more demanding 
      (i.e., $\partial\gamma/\partial T > 0$ and $\gamma$ increases).

      The rationale is analogous for the slabs with less oxygen with respect to 
      chromium atoms ($N_{\ce{O,Cr}} < 0$). At higher temperatures, the 
      oxygen-poor termination does not require any extra energy to be maintained, as it is 
      already aligned with the entropic tendency of oxygen to remain in 
      the gas phase, i.e., $\gamma$ decreases with $T$. 
      The $\gamma$ profiles are shown individually for each orientation 
      in Section IV of the SM.		

\begin{figure}[h]
				\centering
				\includegraphics[scale=0.5]{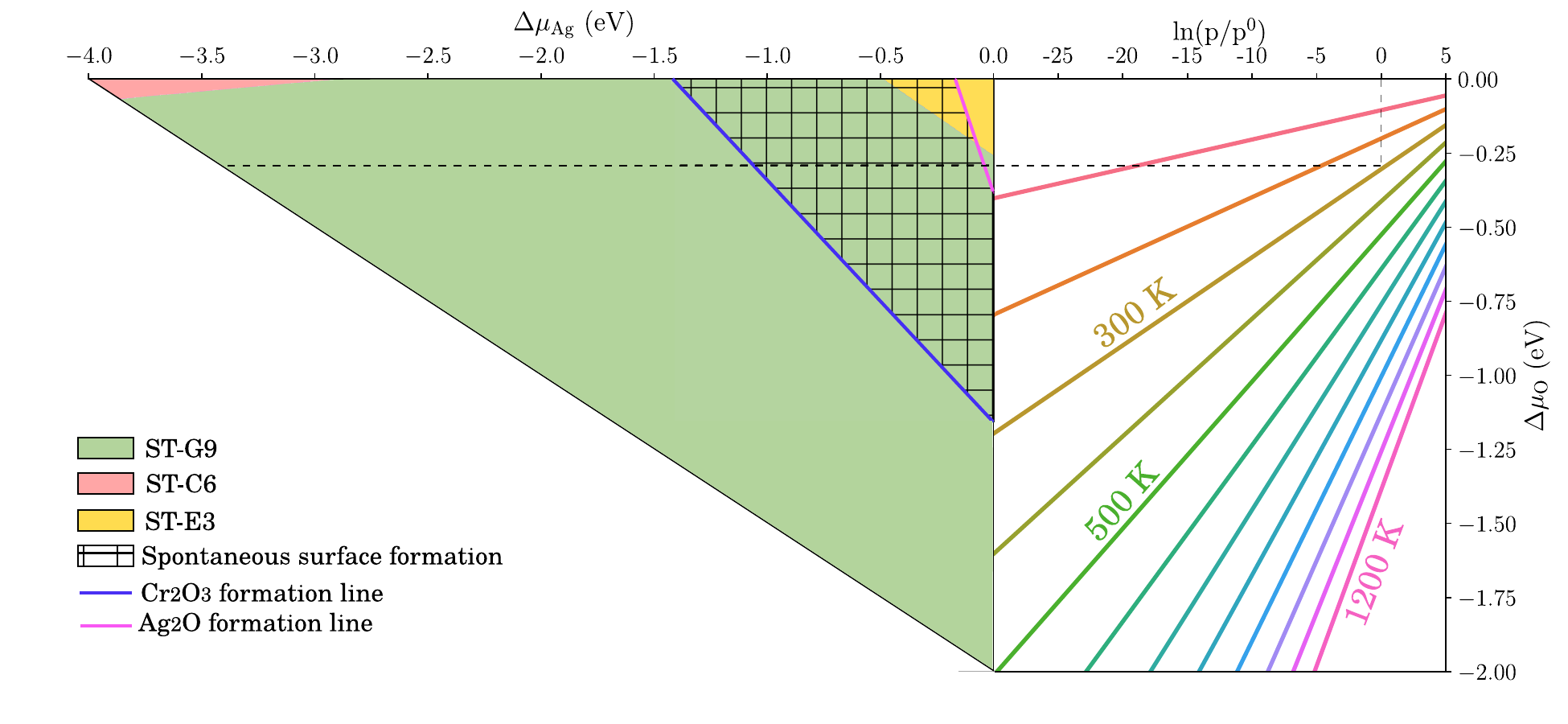}
				\caption{Left side: surface diagram showing the regions of most stable terminations. The considered slabs form spontaneously only in the hatched region. Above the pink line, there is \ce{Ag2O} formation on the slab's surface. Below the blue line, there is \ce{Cr2O3} formation on the slab's surface. Right side: relation between temperature and pressure. The black dotted line corresponds to standard temperature and pressure conditions.}
				\label{fig:general-diagram}
			\end{figure}

			Adding the dependence of $\Delta\mu_{\ce{Ag}}$ to the analysis, 
			\Cref{fig:general-diagram} shows the stability diagram for all 46 
      terminations compiled (the individual ones can be found in Section 
      IV of the SM).
			As can be seen, from the 46 systems, only three compete for the 
			highest stability. Under silver and oxygen highly rich 
			conditions, the ST-E3 termination ((010) orientation) is 
			the most stable, with a small stability area. Decreasing 
			the silver and oxygen content, there is a rapid change of 
			favourabiliy to the ST-G9 termination, which is the most 
			stable one at almost every thermodynamic condition. Only 
			at oxygen-rich and silver-poor conditions, 
			another shift of stability occurs to the ST-C6 termination, 
			which also has a very short stability area. 


			The hatched region in the diagram represents the spontaneous formation of 
			\ce{Ag2CrO4} crystal. Above this region (delimited by the pink line), from 
			\Cref{eq:ag2o}, \ce{Ag2O} is formed on the silver chromate slab, while, from 
			\Cref{eq:cr2o3}, below the hatched region there is the formation of \ce{Cr2O3} 
			on the slab (delimited by the blue line). Hence, it is shown that only the 
			ST-G9 and ST-E3 terminations are expected to form spontaneously. 
			Meanwhile, the ST-C6 termination would be modified by the presence of 
			\ce{Cr2O3}, as well as the ST-E3 termination in a silver and 
			oxygen-rich condition, by the presence of \ce{Ag2O}.
			
			Additionally, from the right-hand side of Figure \ref{fig:general-diagram} we conclude that under near-ambient conditions (1 atm and temperatures close to 300 K), which are particularly relevant for environmental and biotechnological applications, the ST-G9 termination is expected to be the most thermodynamically favorable, as indicated by the dashed line. This result suggests that ST-G9 is the surface termination most likely to be exposed under realistic operating conditions, providing a physically relevant reference for future experimental and applied studies.

			Terminations containing a higher concentration of neutral oxygen vacancies tend to be more electron-rich, since the removal of highly electronegative atoms (such as oxygen) leaves excess electronic charge at the surface. These electron-rich regions act as reduction sites, thereby favoring electron-driven processes and the photogeneration of reducing species. Conversely, terminations with a lower concentration of neutral oxygen vacancies tend to exhibit electron-depleted regions around the central metallic cations, giving rise to oxidative sites that favor hole-driven reactions. Therefore, different thermodynamic conditions, such as oxygen-poor or oxygen-rich environments and variations in silver chemical potential, are expected to directly influence the photocatalytic behavior of the material.
        
        	Molecular adsorption at the surface is highly sensitive to the uncompensated charge distribution of each termination, implying that the potential for ROS generation may differ significantly among distinct surface terminations, even for the same crystallographic orientation. Although no explicit electronic-structure calculations were performed in the present study, a qualitative discussion of these effects can nevertheless be proposed.
        
        	Under highly silver-rich conditions ($\Delta\mu_{\ce{Ag}} = 0$ eV), variations in temperature combined with fluctuations in the surface Gibbs free energy of up to 0.4 J/m$^2$ lead to the stabilization of multiple surface terminations across different crystal orientations, as shown in Figure \ref{fig:gamma-general}. Each of these terminations exhibits a distinct concentration of neutral oxygen vacancies, as well as different arrangements of silver and chromium surface clusters, as illustrated in Figure S1–S9 (Section 1 of the Supplementary Material). These results provide theoretical evidence that the propensity for ROS formation on \ce{Ag2CrO4} surfaces is intrinsically linked to surface thermodynamics. For example, the ST-G9 termination ((101) orientation), which is the most stable under these conditions, contains four neutral oxygen vacancies distributed over both octahedral and tetrahedral silver clusters. The ST-B7 termination ((111) orientation), the second most stable, contains seven $\mathrm{V_o^x}$, whereas the closely related ST-B6 structure contains two additional $\mathrm{V_o^x}$, making it the most undercoordinated termination within this set. A similar trend is observed for the (001) orientation, where the ST-D2, ST-D3, and ST-D4 terminations contain one, three, and six $\mathrm{V_o^x}$, respectively. In contrast, the fully coordinated ST-C6 termination may also form under these conditions.

        	Consequently, each termination is expected to exhibit a distinct efficiency for ROS generation. In particular, reducing ROS species, such as superoxide, peroxide, and trapped electrons, are more likely to form via the adsorption of oxygen and water molecules on highly undercoordinated terminations, such as ST-B6, ST-B7, and ST-D4. Conversely, oxidative ROS species, including hydroxyl radicals, singlet oxygen, and photogenerated holes, are expected to be favored on more coordinated terminations, such as ST-C6, ST-D2, and ST-D3.
	
        	When the silver chemical potential is varied (Figure \ref{fig:general-diagram}), under spontaneous surface formation conditions, i.e. for crystals exposing either the ST-E3 ((010) orientation) or ST-G9 terminations, the same trend is observed, suggesting a preferential promotion of hole-driven reactions. Under other conditions, however, it becomes difficult to draw definitive conclusions, as the predicted precipitation of \ce{Ag2O} and \ce{Cr2O3} on the stable terminations may significantly alter the surface structure. Such precipitates could strongly modify the local coordination environment and the distribution of neutral oxygen vacancies, effects beyond the scope of the present work.
        
        	Despite the qualitative nature of this discussion, it is important to emphasize that structural distortions arising from incomplete surface clusters break local symmetry and disrupt the periodic potential experienced by electrons. As a result, the electronic states become confined to the distorted clusters, resembling molecular-like orbitals rather than extended Bloch states. This spatial confinement leads to discrete electronic energy levels, which manifest as localized states within the forbidden band gap \cite{kittel2018}. Such effects may substantially alter the electronic character of the surface, such that a single crystal orientation could, in principle, exhibit terminations with metallic, semiconducting, or (even if less likely) insulating behavior. This variability is expected to directly affect photocatalytic ROS formation. A detailed characterization of the electronic structure of each surface termination, however, lies beyond the scope of the present study and represents an important direction for future work aimed at establishing a quantitative link between surface termination and photocatalytic activity in \ce{Ag2CrO4}.
		
			Although the present study is strictly theoretical, the predicted stable surface terminations can, in principle, be examined experimentally using well-established surface-sensitive techniques. X-ray photoelectron spectroscopy (XPS) can provide direct information on the surface composition and oxidation states of Ag, Cr, and O, enabling the distinction between oxygen-rich and oxygen-poor terminations and the identification of oxygen vacancy–related features \cite{henrich1994surface, briggs2003surface}. Ultraviolet photoelectron spectroscopy (UPS) and scanning tunneling spectroscopy (STS) may further probe surface electronic states, including defect-induced levels within the forbidden band gap \cite{hufner2013photoelectron, diebold2003surface}.
        
        	In addition, scanning tunneling microscopy (STM) and low-energy electron diffraction (LEED) can offer insights into surface reconstructions and symmetry changes associated with different terminations \cite{van2012surface, freund2008oxide}, while electron paramagnetic resonance (EPR) can be used to identify paramagnetic centers related to oxygen vacancies and trapped electrons \cite{weil2007electron, pacchioni2003oxygen}. Importantly, in situ or ambient-pressure XPS measurements performed under controlled temperature and oxygen partial pressure conditions provide a natural experimental framework to connect the predicted thermodynamic stability diagrams with realistic operating environments \cite{salmeron2008ambient}.

			\subsection{\label{subsec:general-view}Integrated discussion on surface stability}
				To identify a general rationale for the observed stability trends, we analyzed the structural differences between the unrelaxed and relaxed slab models for all terminations. All figures illustrating these differences are provided in the SM (Figures S1–S10, Section I), along with a brief description of our main observations for each orientation (Section V of the SM), which led us to the following conclusions.
Considering all 46 terminations collectively, we observed that the most unstable systems are characterized by undercoordinated clusters, particularly of the types [\ce{AgO}$\cdot5\mathrm{V_o^x}$], [\ce{AgO2}$\cdot4\mathrm{V_o^x}$], and [\ce{CrO2}$\cdot2\mathrm{V_o^x}$], or by the exclusive presence of silver atoms on the surface. Additionally, structural distortion plays a clear role, with some terminations being poorly bonded to the underlying layers.
In contrast, most systems with intermediate stability expose chromium clusters that are completely or nearly completely coordinated. Many unrelaxed systems already feature these clusters or develop them during the relaxation process. In these cases, we observed transformations such as [\ce{CrO2}$\cdot2\mathrm{V_o^x}$] $\rightarrow$ [\ce{CrO3}$\cdot\mathrm{V_o^x}$] or [\ce{CrO3}$\cdot\mathrm{V_o^x}$] $\rightarrow$ [\ce{CrO4}], totaling 10 systems. An additional seven systems retained clusters with a single oxygen vacancy, which appears to be the threshold for stabilizing the slab.
Based on these observations, we propose four criteria that can explain, individually or collectively, the observed stability trends:
($i$) terminations exposing highly coordinated chromium clusters ([\ce{CrO4}] and [\ce{CrO3}$\cdot\mathrm{V_o^x}$]), with few broken \ce{Cr}-\ce{O} bonds, tend to be more stable;
($ii$) terminations exposing chromium clusters with a coordination of 2 tend to be unstable;
($iii$) terminations exposing low-coordination silver octahedral clusters ([\ce{AgO2}$\cdot4\mathrm{V_o^x}$] and [\ce{AgO}$\cdot5\mathrm{V_o^x}$]) tend to be unstable;
($iv$) terminations exhibiting a lower degree of surface distortion relative to the unrelaxed structure tend to be more stable.

In a recent study \cite{dorini2025mapping}, the stability of \ce{Ag2MoO4} surfaces was also investigated using a similar approach based on first-principles atomistic thermodynamics. That work concluded that terminations retaining intact \ce{Mo}-\ce{O} bonds during relaxation were the most stable, which aligns with criterion ($i$). Among the 46 systems analyzed here, no cases were found in which the coordination number of surface chromium clusters decreased, unlike silver tetrahedral and octahedral clusters, for which such reductions were frequent.

As in silver molybdate, chromium in silver chromate acts as the lattice former, serving as the element responsible for connecting other species to form the repeating units of the crystal structure and, consequently, the three-dimensional lattice \cite{kittel2018}. This role is favoured when [\ce{CrO4}] clusters remain intact (or nearly so). The less coordinated the surface chromium becomes, the higher the energetic cost of reconstructing the oxide, since chromium atoms must undergo greater distortions to recover coordination from neighboring oxygen atoms.
In contrast, silver clusters exhibit greater flexibility in coordination. We frequently observed tetrahedral clusters such as [\ce{AgO2}$\cdot2\mathrm{V_o^x}$], [\ce{AgO3}$\cdot\mathrm{V_o^x}$], and [\ce{AgO4}], as well as octahedral clusters like [\ce{AgO3}$\cdot3\mathrm{V_o^x}$], [\ce{AgO4}$\cdot2\mathrm{V_o^x}$], [\ce{AgO5}$\cdot\mathrm{V_o^x}$], and [\ce{AgO6}] in the relaxed systems, including the most stable ones. Silver acts as a lattice modifier, adapting its coordination in response to structural changes required by chromium to stabilize the structure. This flexibility is consistent with the higher tolerance of silver clusters to defects. Additionally, in silver chromate, the chromium atoms have partially filled 3$d$ orbitals, which strongly hybridize with oxygen 2$p$ orbitals \cite{kushwaha_investigation_2017, fabbro_understanding_2016}, producing more localized bonding states with significant energy stabilization, but highly anisotropic coupling. Reducing coordination perturbs the orbital overlap and breaks the local electronic symmetry, leading to a pronounced splitting of the $d$-levels and an increase in the total electronic energy. By contrast, the silver atoms feature a diffuse, nearly free 5$s$ valence electron \cite{kushwaha_investigation_2017, fabbro_understanding_2016}, resulting in weaker, more isotropic coupling with oxygen and, hence, the electronic density distribution remains largely uniform, minimally perturbing the energy of low-coordination \ce{Ag} clusters and producing a smaller energetic penalty for geometric modifications.


Taken together with the findings reported in \cite{dorini2025mapping}, these observations suggest that lattice-generating elements in a crystalline system largely determine stability through the coordination state of exposed clusters. This serves as an initial step toward formulating a general theoretical model for predicting crystal stability. Based on the simpler slab models of \ce{Ag2MoO4} compared to those of \ce{Ag2CrO4}, we conclude that such influences strongly depend on the complexity of the surface crystal structure of the material.

		\subsection{\label{subsec:wulff-reconstruction}Effect of thermodynamic conditions on crystal morphologies}
		The variation in thermodynamic conditions leads to the stabilization of different surface terminations, as previously discussed. Consequently, it is expected that the morphology of an ideal crystal will also vary under these conditions. In general, in the literature, such morphologies are commonly generated by artificially modifying the surface energy of each face \cite{assis_surface-dependent_2021, lipsky2023tale, gouveia2023hinge, gouveia2023back, gouveia2022ag2wo4, gouveia2025morphology, gouveia2021modulating, gouveia2022surfactant, lacerda2021dft, lacerda2021linbo3, teodoro2022connecting} to obtain alternative Wulff shapes. In this work, we demonstrate that multiple ideal morphologies naturally arise under varying thermodynamic conditions, as defined here, and we provide the atomic structure (termination) associated with each exposed facet. This is possible because the thermodynamic analysis previously discussed enables us to predict the most stable termination for each orientation (\Cref{fig:wulff-contcars}).
	
\begin{figure}[hp]
\centering
\includegraphics[scale=0.55]{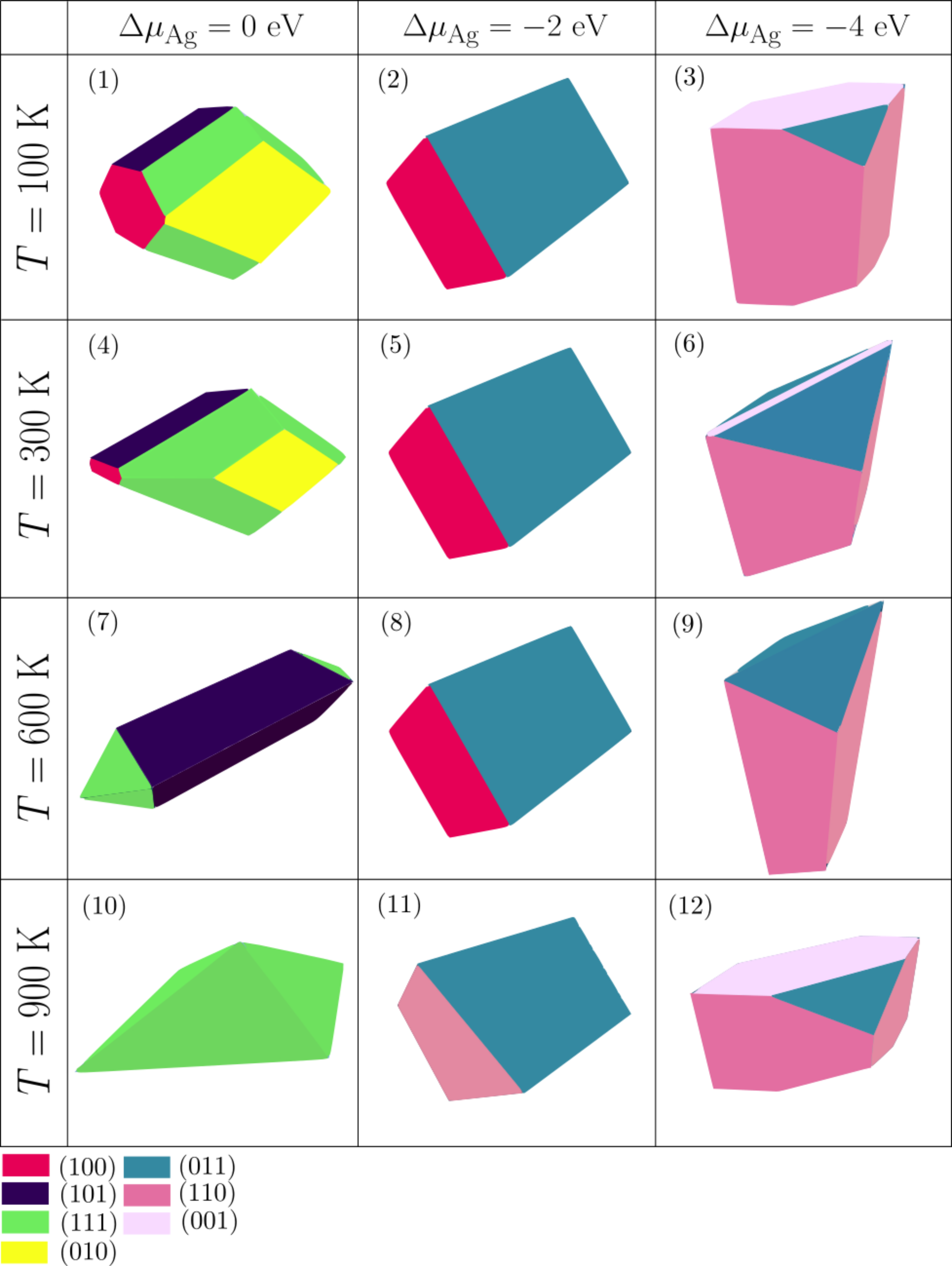}
\caption{Ideal Wulff morphologies for \ce{Ag2CrO4} as a function of temperature and silver chemical potential.}
\label{fig:wulff}
\end{figure}

\begin{figure}[h]
\centering
\includegraphics[scale=0.5]{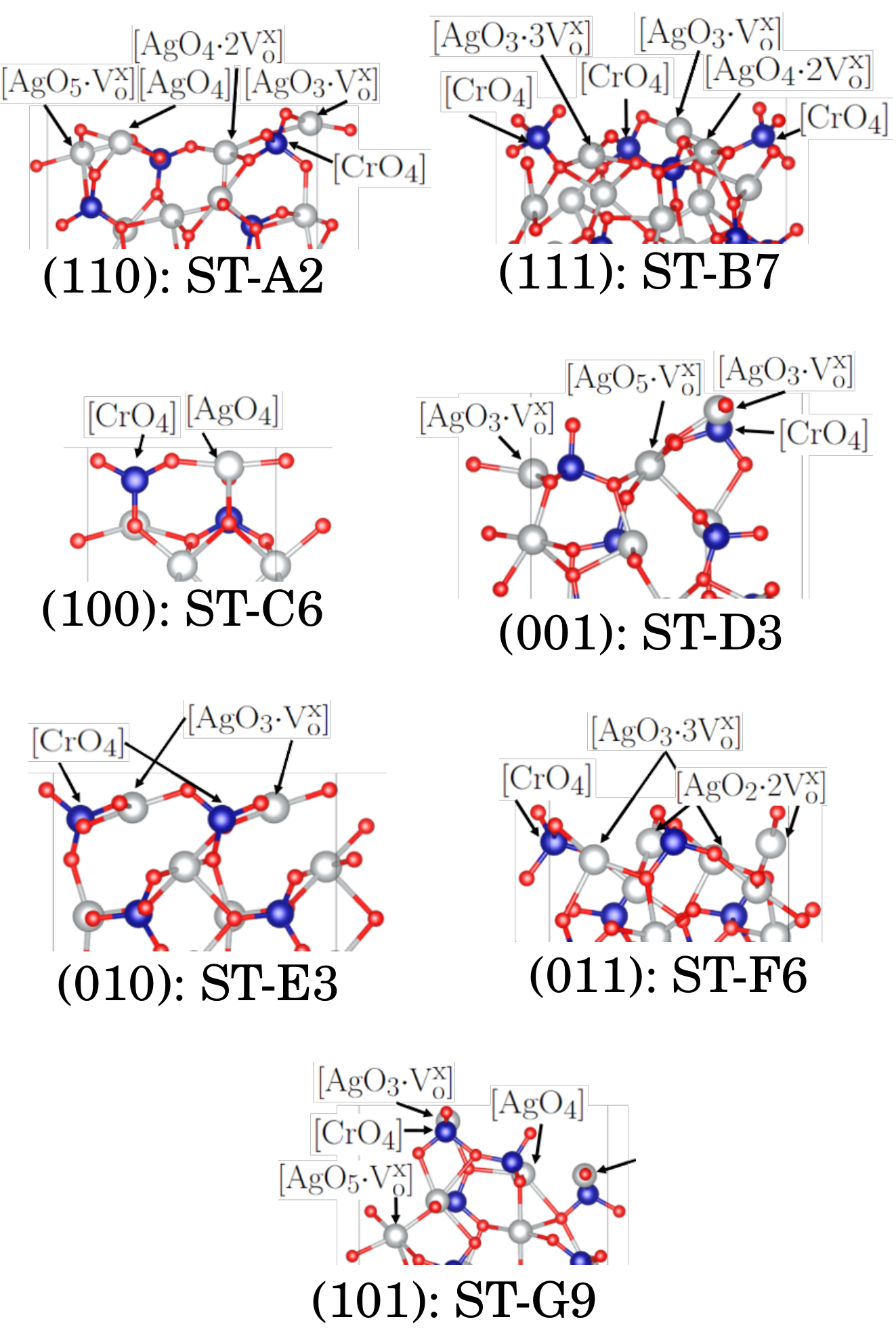}
\caption{Relaxed slab models for the most stable terminations present in the Wulff ideal morphologies shown in \Cref{fig:wulff}.}
\label{fig:wulff-contcars}
\end{figure}

		To explore this, we computed the Wulff morphologies as a function of temperature and silver chemical potential, as shown in \Cref{fig:wulff}.
At 100 K under silver-rich conditions (morphology 1), the exposed facets correspond to the (100), (101), (111), and (010) orientations, represented by terminations ST-C6, ST-G9, ST-B7, and ST-E3, respectively. When the difference in silver chemical potential is decreased to $-2$ eV (morphology 2), the exposure of the (100) facet increases, while the others are replaced by (011), forming a tetragonal crystal dominated by terminations ST-C6 ((100)) and ST-F6 ((011)). In silver-poor conditions (morphology 3), the (011) facet becomes significantly less exposed, with the (110) facet emerging as the most dominant, followed by (001). This corresponds to the presence of terminations ST-F6 ((011)), ST-A2 ((110)), and ST-D3 ((001)), yielding a crystal with plane faces at the extremities and elongated lateral surfaces.
At 300 K, changes under silver-rich conditions (morphology 4) are subtle, with a slight reduction in the exposure of (100) and (010) and an increase in (101) and (111), while maintaining the same terminations as at 100 K, given their high stability regardless of thermodynamic conditions. For intermediate silver conditions at the same temperature (morphology 5), no significant changes are observed compared to 100 K, and the same is true for morphology 8 at 600 K. Morphology 6 is characterized by a reduction in (001) surface, limited to narrow regions on the top and bottom crystal faces. However, under these conditions, the predicted surface free energies ($\gamma$) fall outside the stability region of silver chromate (beyond the colored region in \Cref{fig:general-diagram}); thus, this morphology is not predicted as stable by our model. Nevertheless, the (011) facet becomes more prominent, while the overall geometry remains similar to that at 100 K.
At 600 K under silver-rich conditions (morphology 7), only (111) and (101) facets remain exposed, with the latter dominating the crystal surface. This morphology is markedly different, displaying a trigonal geometry at the extremities (corresponding to (111)) connected by three extended planar faces (corresponding to (101)). For the opposite extreme condition (morphology 9), the system is again predicted not to form, as with morphologies 11 and 12. Nonetheless, these morphologies exhibit the complete disappearance of the (001) plane, replaced by broader exposures of the (110) and (011) planes.
At the highest temperature considered, morphology 10 consists exclusively of (111) surfaces, eliminating the planar facets present in previous conditions and producing a more compact crystal. Increasing the silver chemical potential difference (morphology 11) slightly distorts the morphology relative to the equivalent condition at 100 K, due to the replacement of the (100) plane by the (110) plane. Finally, morphology 12 exhibits a significant narrowing of the (110) facet, a reduction in the (011) facet, and the reappearance of the (001) facet on the top and bottom faces, producing a shape similar to that of morphology 3.

Analyzing the most probable morphologies (\Cref{fig:wulff}) in terms of their relaxed atomic surface structures (\Cref{fig:wulff-contcars}) provides additional insights into stability trends. The structurally simplest slabs (ST-C6 and ST-E3) are favored only at low temperatures ($T < 300$ K) and under high silver chemical potential (minimum of $-2$ eV), corresponding to morphologies 1, 2, 4, and 5, with the sole exception of morphology 8. This behavior likely reflects the greater availability of oxygen and silver atoms from the external environment under these conditions, which facilitates the completion of surface metal clusters, producing silver- and oxygen-rich terminations.
At higher temperatures (lower oxygen chemical potential), these terminations disappear, replaced by more complex and distorted slabs. This suggests that some systems remain stable despite reduced oxygen availability, particularly those with less coordinated silver clusters, as seen in morphologies 7 and 10 (terminations ST-B7 and ST-G9).

In silver-poor conditions, decreasing the oxygen chemical potential favors the exposure of the ST-A2 termination of the (110) facet (morphologies $3 \rightarrow 6 \rightarrow 9 \rightarrow 12$), which is characterized by fewer surface atoms and larger vacancies. Simultaneously, the ST-D3 termination of (001) (which includes silver clusters with one oxygen vacancy each) is progressively disfavored compared to ST-F6 of (011), where silver atoms exhibit slightly lower coordination. This trend likely reflects the decreasing availability of oxygen in the external environment (morphologies $3 \rightarrow 6 \rightarrow 9$).

Finally, at extremely high temperatures (corresponding to highly oxygen-poor conditions), the predicted morphologies exhibit either pronounced structural modifications (morphologies 10 and 11) or virtually none, as in morphology 12.
Overall, changes in silver chemical potential induce more pronounced morphological variations at constant temperature than changes in temperature. This indicates that, among the most stable slabs, silver coordination (and, to some extent, the degree of structural distortion) plays the dominant role in stability trends, while modifications to chromium clusters remain relatively subtle.

Although the present work does not explicitly consider solvent effects or surface reactions with water, it is well established for many metal oxide materials that exposure to aqueous environments can significantly modify the surface chemistry through adsorption and dissociation of water molecules. Surface hydroxylation, i.e., the formation of surface OH groups by dissociative chemisorption of \ce{H2O} at undercoordinated metal cations, is a common process on oxide surfaces and plays a central role in defining the solid/liquid interface structure and reactivity (e.g., dissociative adsorption at Lewis acid sites) \cite{tamura2001mechanism}. Furthermore, experimental and theoretical studies on reducible oxides show that the presence of oxygen vacancies and surface defects can enhance water dissociation and hydroxylation at lower relative humidities, with the distribution of hydroxyl groups strongly dependent on facet and defect density \cite{lahiri2023interplay}. Surface hydroxyls not only influence wetting and charge transfer characteristics but also affect catalytic and photocatalytic processes at the solid–liquid interface \cite{groh2021hydroxylation}. In the specific case of \ce{Ag2CrO4}, experimental observations of morphology changes in different aqueous synthesis conditions imply that the arrangement of surface terminations and undercoordinated sites (which would be potential adsorption and hydroxylation centers in water) vary with solvent and temperature \cite{assis_surface-dependent_2021, xu2014effects}. Therefore, the variety of stable surface terminations predicted here suggests that their interaction with water and the consequent formation of surface hydroxyl species (with potential implications for photocatalytic and biological activity in aqueous environments) would be strongly termination-dependent, motivating future theoretical and experimental investigation.

	\section{\label{sec:conclusions}Conclusion}
		In this work, we applied and extended an established first-principles thermodynamic framework to quantify the surface energetics and structural stability of \ce{Ag2CrO4} for all seven low-index orientations: (001), (010), (100), (011), (101), (110), and (111). The surface Gibbs free energy, expressed as a function of the oxygen and silver chemical potentials, provided a rigorous means to map the relative stability domains of distinct terminations and to identify the thermodynamic transitions among them.

		The computed trends reveal that surface stability is governed primarily by the local coordination environment and the electronic nature of the bonding network. \ce{Ag}–\ce{O} bonds, characterized by a more ionic and delocalized charge distribution, undergo larger relaxations upon cleavage, whereas the more covalent \ce{Cr}–\ce{O} bonding imposes a stronger energetic penalty for bond disruption. Consequently, surfaces exposing compact \ce{Cr}–\ce{O} clusters exhibit lower surface Gibbs free energies, particularly when relaxation preserves the underlying lattice symmetry. This structural rigidity accounts for the dominance of \ce{Cr}-terminated facets under most equilibrium conditions.

		The Wulff construction based on the calculated surface energies predicts that the equilibrium morphology evolves continuously with the chemical potentials: high $\mu_{\ce{Ag}}$ and $\mu_{\ce{O}}$ stabilize low-index, highly coordinated terminations, while reduced chemical potentials favor more open and distorted facets.
		Additionally, by applying the Wulff construction method to the computed surface energies, we predicted ideal crystal morphologies under varying thermodynamic conditions. These predictions revealed that silver chemical potential exerts a stronger influence on morphological evolution than temperature (which reflects oxygen chemical potential).

		Overall, this analysis delineates the complex stability landscape of \ce{Ag2CrO4} surfaces within the two-dimensional ($\mu_{\ce{Ag}},\mu_{\ce{O}}$) chemical potential space, allowing direct identification of the most stable configurations under arbitrary external constraints. The resulting topology of the surface Gibbs free energy surface establishes a quantitative connection between local bonding anisotropy, surface relaxation energetics, and macroscopic morphology. This framework, coupling first principles energetics with a grand-canonical thermodynamic formalism, provides a generalizable route for predicting equilibrium morphologies and stability crossovers in multicomponent oxide systems.
	
	\section*{Acknowledgments}

		All authors are grateful for the financial support of Brazilian Funding Agencies. 
		A.F. acknowledges funding from Conselho Nacional de Desenvolvimento Científico e 
		Tecnológico, CNPq (grant 131853/2022-8). T.T.D. acknowledges funding from 
		Fundação de Amparo à Pesquisa do Estado de São Paulo, FAPESP (2023/03447-8). 
		M.A.S. acknowledges support from Aperfeiçoamento de Pessoal de 	Nível Superior - 
		Brasil, CAPES (Finance Code 001), Fundação de Amparo à Pesquisa do Estado de São Paulo, FAPESP 
		(2016/23891-6, 2017/26105-4), and Centro de 
		Desenvolvimento de Materiais Funcionais (CDMF/CEPID/FAPESP 2013/07296-2). 
		This work utilized computational resources provided by the Centro Nacional 
		de Processamento de Alto Desempenho em São Paulo (CENAPAD) and the 
		Centro de Computação John David Rogers (CCJDR-UNICAMP).

	\section*{Author Contributions}
		A.F. and T.T.D. performed the data curation, formal analysis, investigation, methodology, 
		software, validation, visualization, writing of the original draft, and revision and editing 
		the manuscript. T.W.Z. contributed to the formal analysis, investigation, writing 
		of the original draft and revision and editing of the manuscript. M.A.S. performed 
		the conceptualization, funding acquisition, project administration, resources, 
		supervision, and editing of the manuscript.

	\bibliographystyle{apsrev4-2}
	\bibliography{bibliography}
\end{document}


\title{\textbf{Supplementary Materials: Understanding oxide surface stability: Theoretical insights from silver chromate} 
}%

\author{A. Facundes}
\author{T. T. Dorini}%
\author{T. W. von Zuben}
\author{M. A. San-Miguel} 
\affiliation{Department of Physical Chemistry, Institute of Chemistry, Universidade Estadual de Campinas, Campinas, São Paulo, Brazil}
\email{smiguel@unicamp.br}

\date{\today}
\maketitle


    

    \newpage
    \section{\label{sec:surface_cluster_models}Surface Cluster Models}
        \begin{figure}[H]
            \centering
            \includegraphics[scale=0.35]{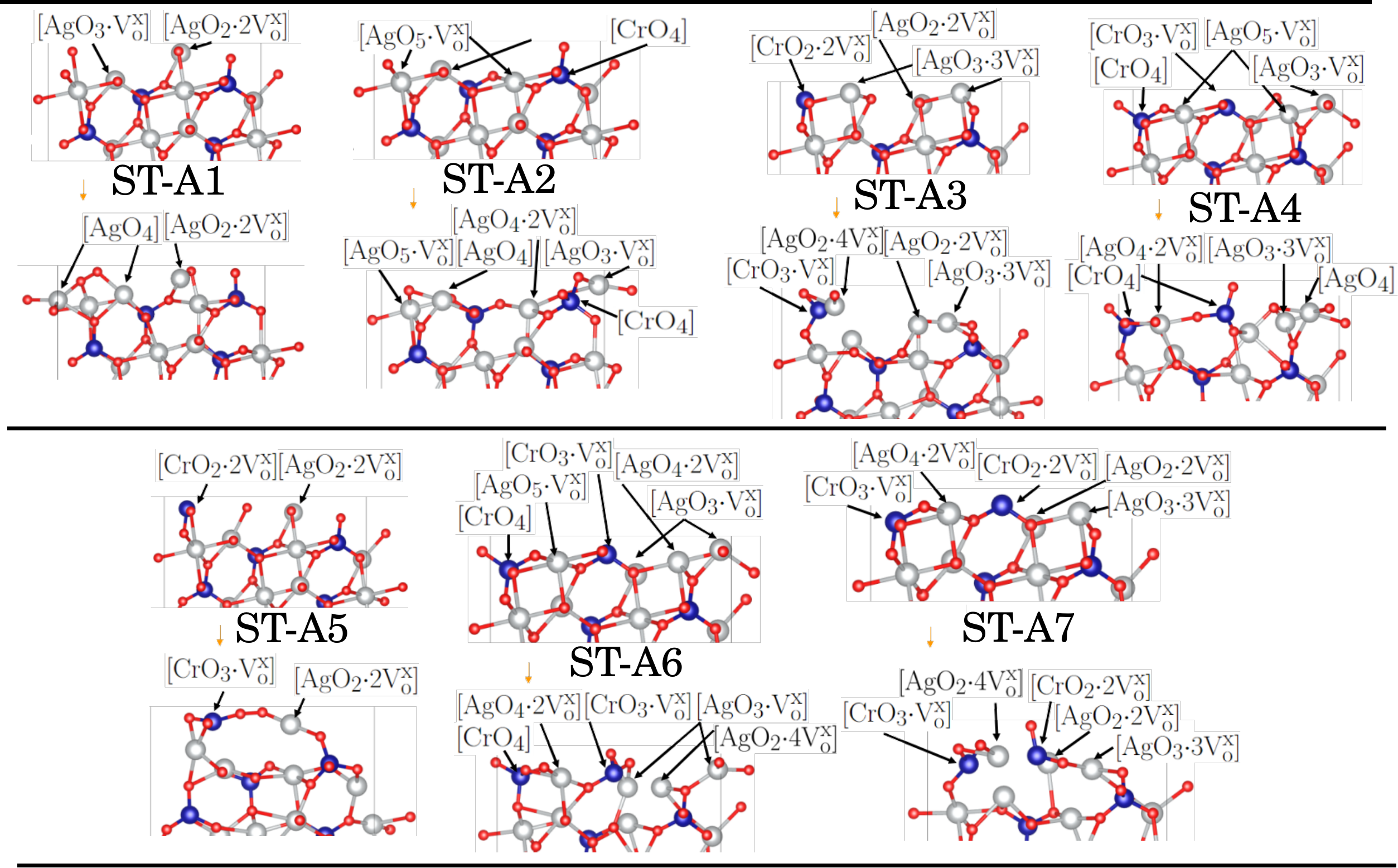}
            \caption{A representation of the surface models for the orientation 
			(110), before and after relaxation. The orange arrow represents the relaxation step.
            The surface clusters are presented using the 
			Kröger-Vink notation.}
            \label{fig:110-rel-slabs}
        \end{figure}
    
        \begin{figure}[H]
            \centering
            \includegraphics[scale=0.35]{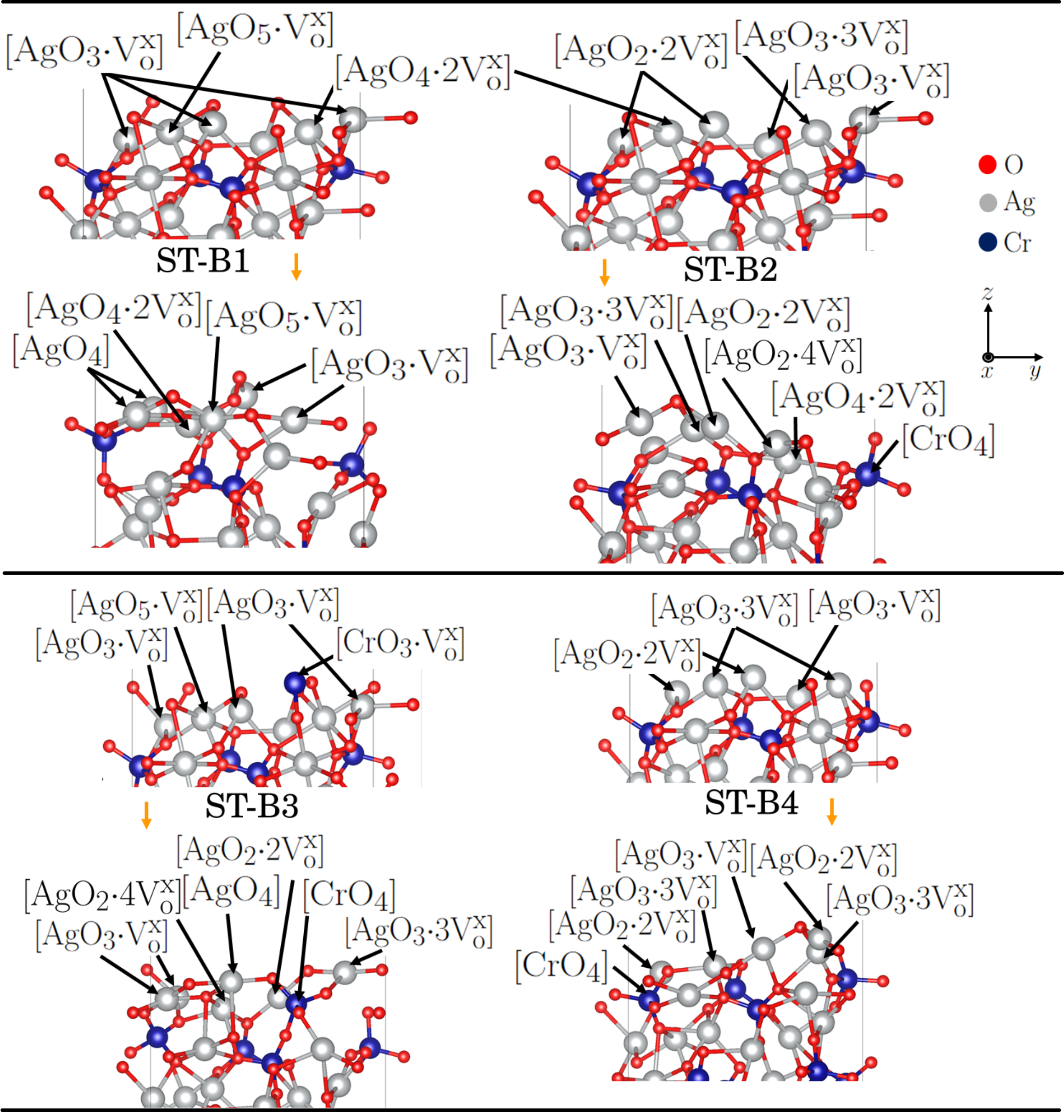}
            \caption{A representation of the surface models for the orientation 
			(111), before and after relaxation. The orange arrow represents the relaxation step.
            The surface clusters are presented using the 
			Kröger-Vink notation.}
            \label{fig:111-rel-slabs1}
        \end{figure}
        \begin{figure}[H]
            \centering
            \includegraphics[scale=0.35]{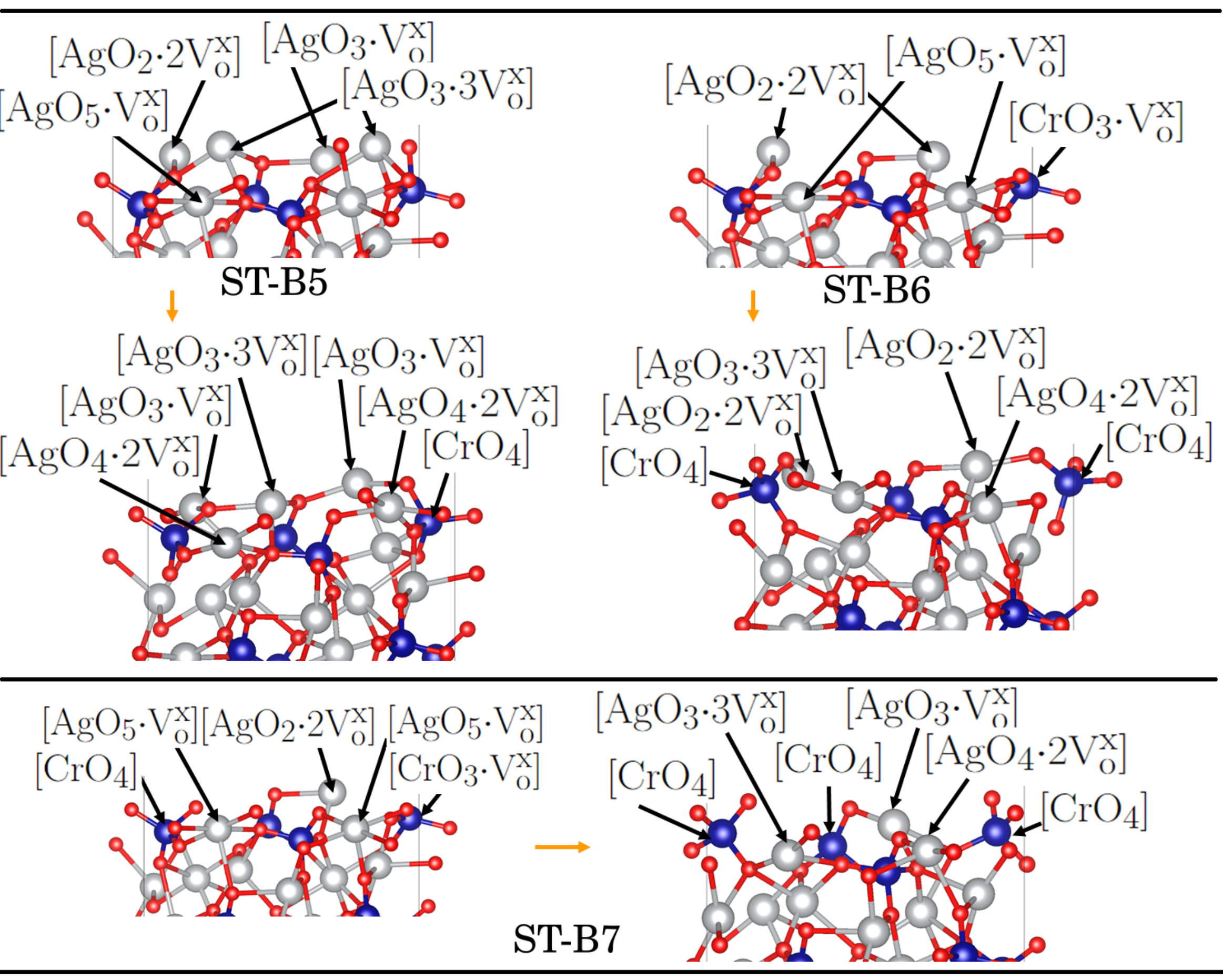}
            \caption{A representation of the surface models for the orientation 
			(111), before and after relaxation. The orange arrow represents the relaxation step. 
            The surface clusters are presented using the 
			Kröger-Vink notation.}
            \label{fig:111-rel-slabs2}
        \end{figure}

        \begin{figure}[H]
            \centering
            \includegraphics[scale=0.3]{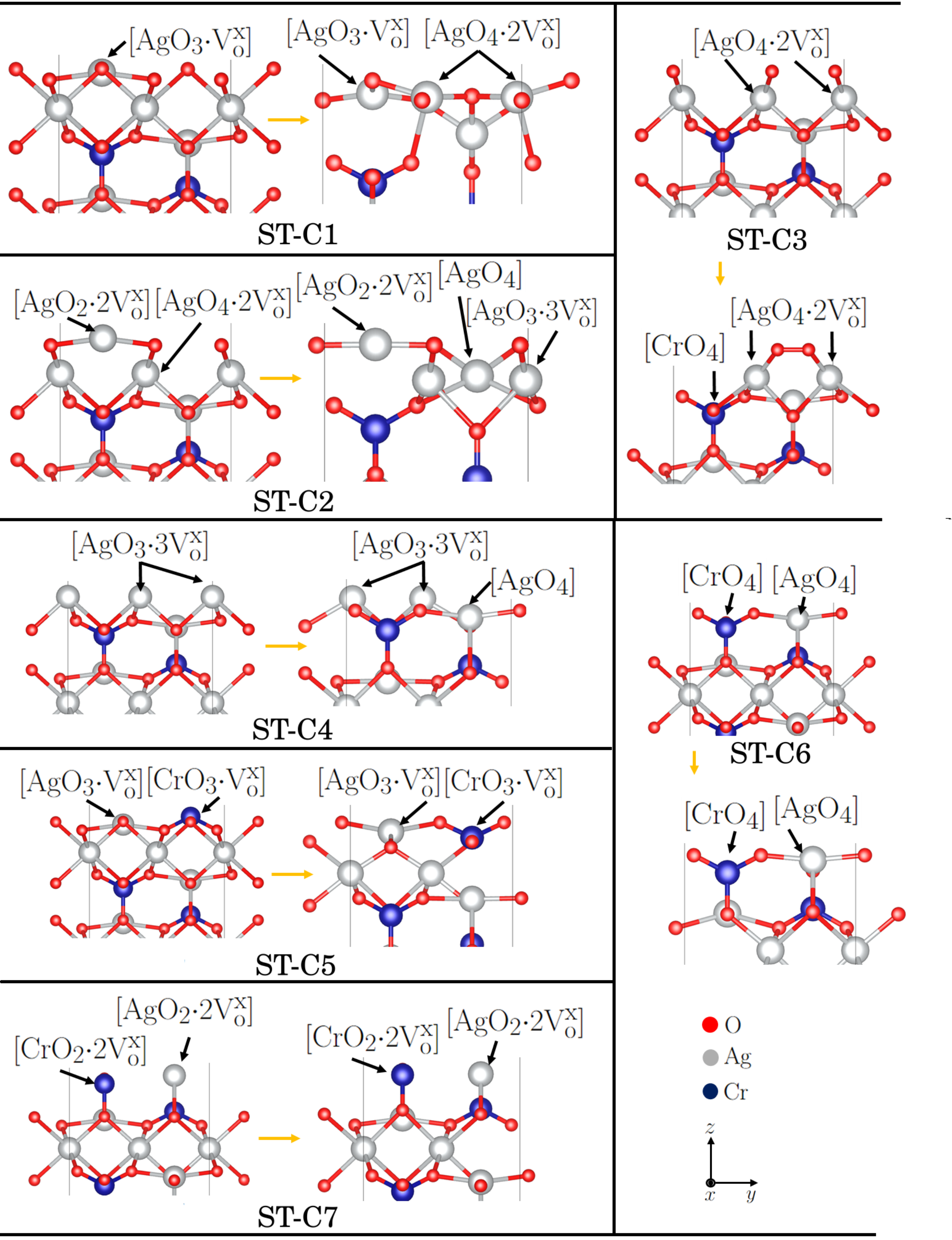}
            \caption{A representation of the surface models for the orientation 
			(100), before and after relaxation. The orange arrow represents the relaxation step. 
            The surface clusters are presented using the 
			Kröger-Vink notation.}
            \label{fig:100-rel-slabs}
        \end{figure}

        \begin{figure}[H]
            \centering
            \includegraphics[scale=0.35]{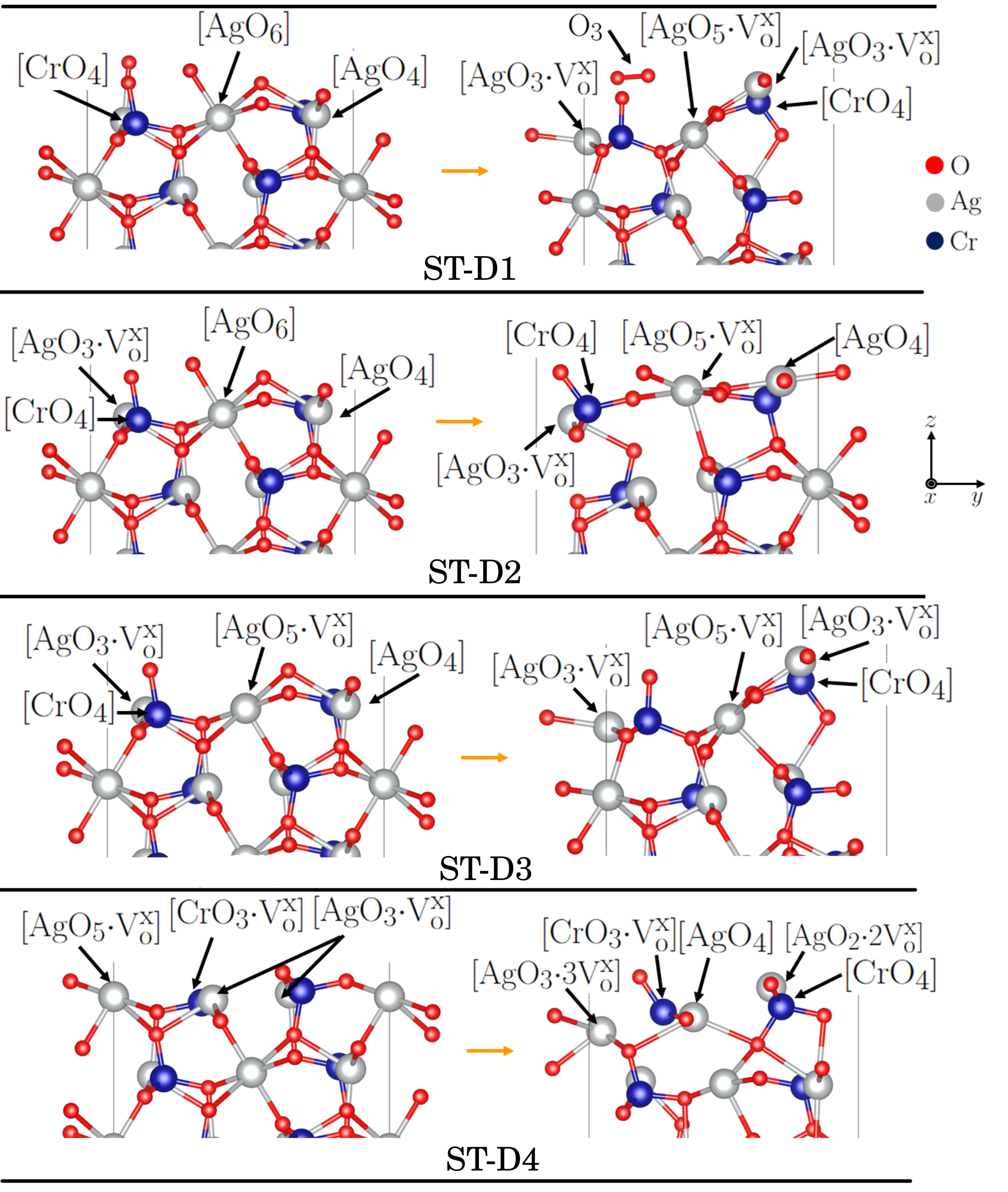}
            \caption{A representation of the surface models for the orientation 
			(001), before and after relaxation. The orange arrow represents the relaxation step. 
            The surface clusters are presented using the 
			Kröger-Vink notation.}
            \label{fig:001-rel-slabs}
        \end{figure}

        \begin{figure}[H]
            \centering
            \includegraphics[scale=0.35]{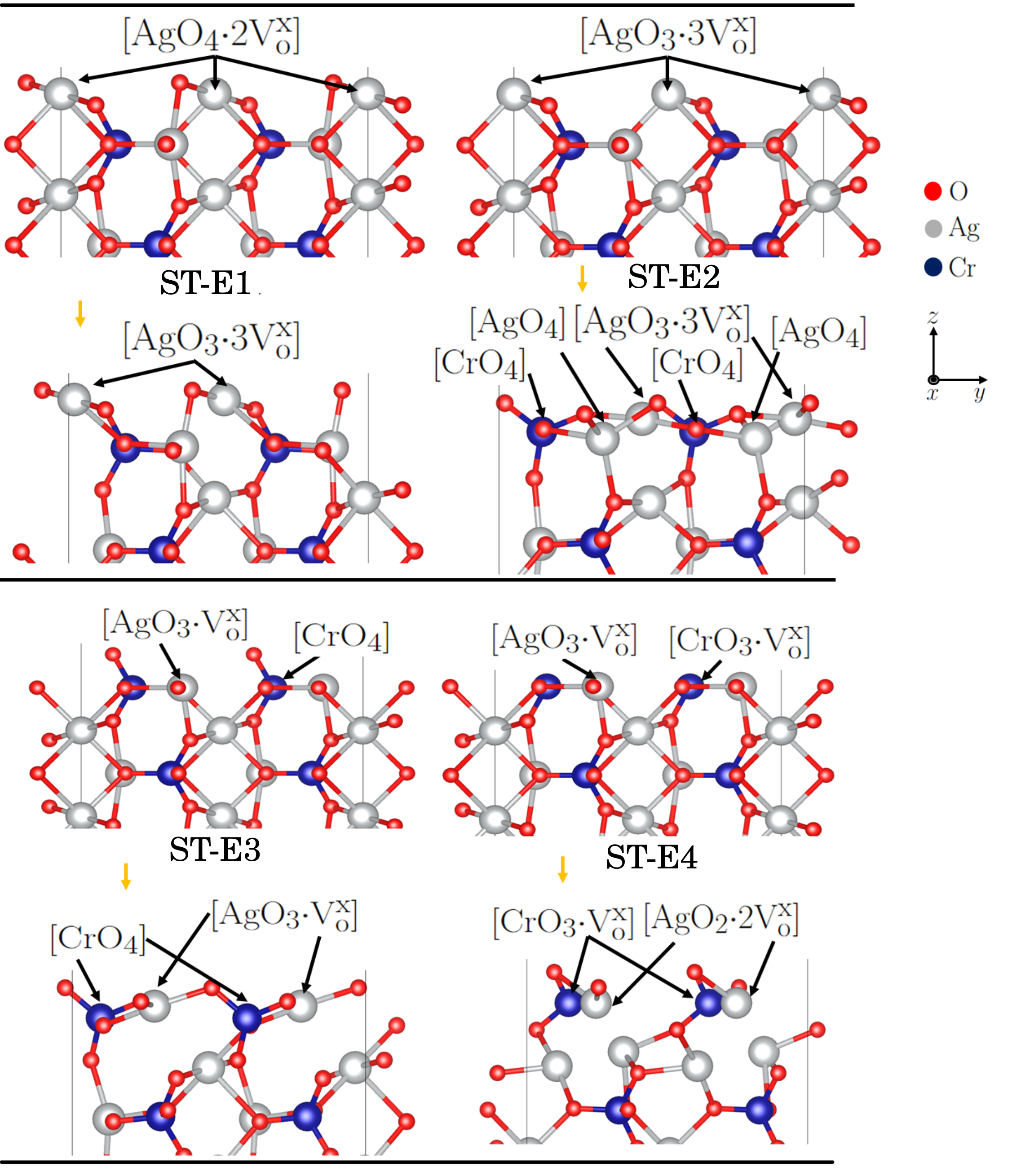}
            \caption{A representation of the surface models for the orientation 
			(010), before and after relaxation. The orange arrow represents the relaxation step. 
            The surface clusters are presented using the 
			Kröger-Vink notation.}
            \label{fig:010-rel-slabs}
        \end{figure}

        \begin{figure}[H]
            \centering
            \includegraphics[scale=0.3]{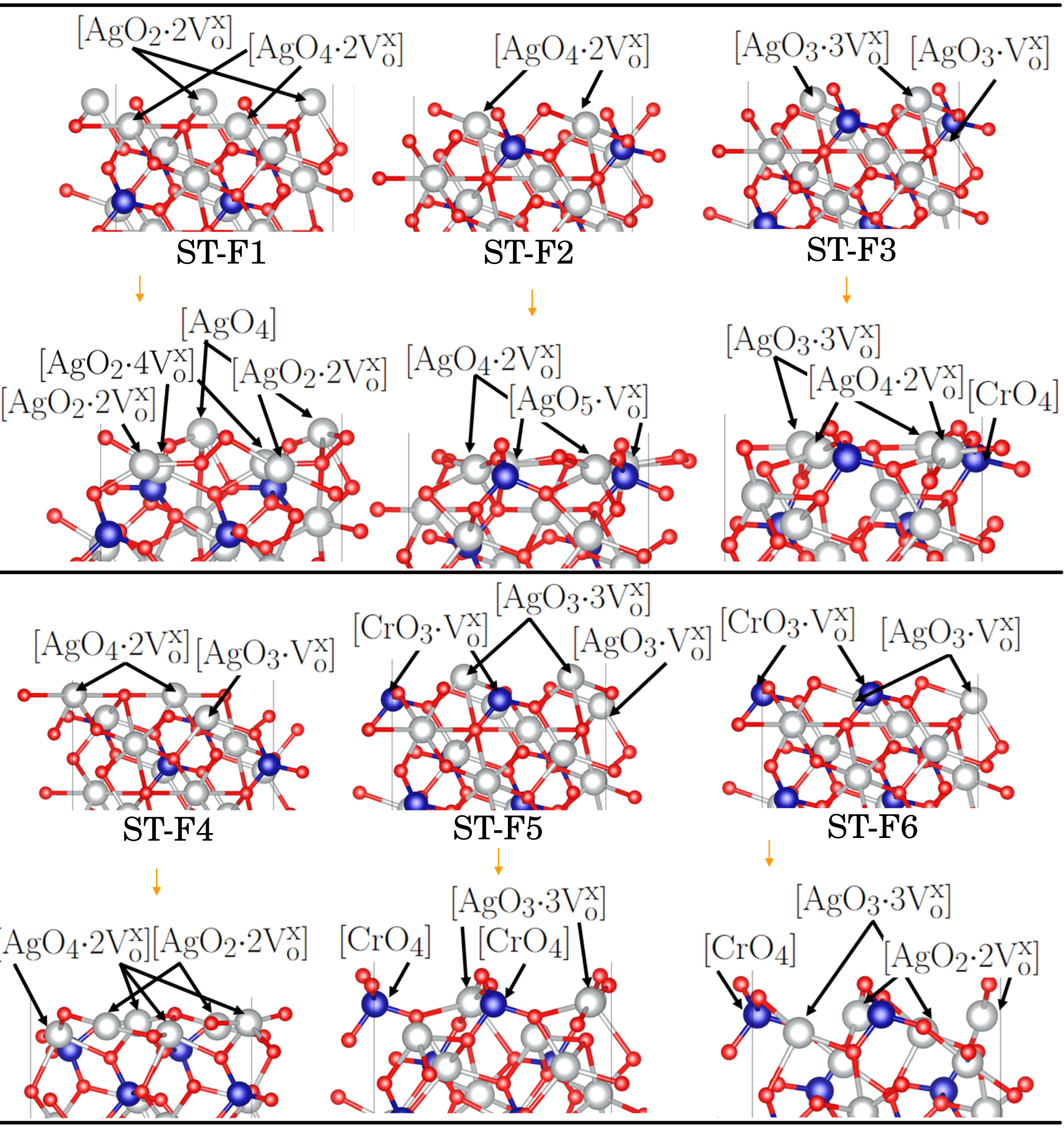}
            \caption{A representation of the surface models for the orientation (011), before and after relaxation. The orange arrow represents the relaxation step. 
            The surface clusters are presented using the 
			Kröger-Vink notation.}
            \label{fig:011-rel-slabs1}
        \end{figure}
        \begin{figure}[H]
            \centering
            \includegraphics[scale=0.35]{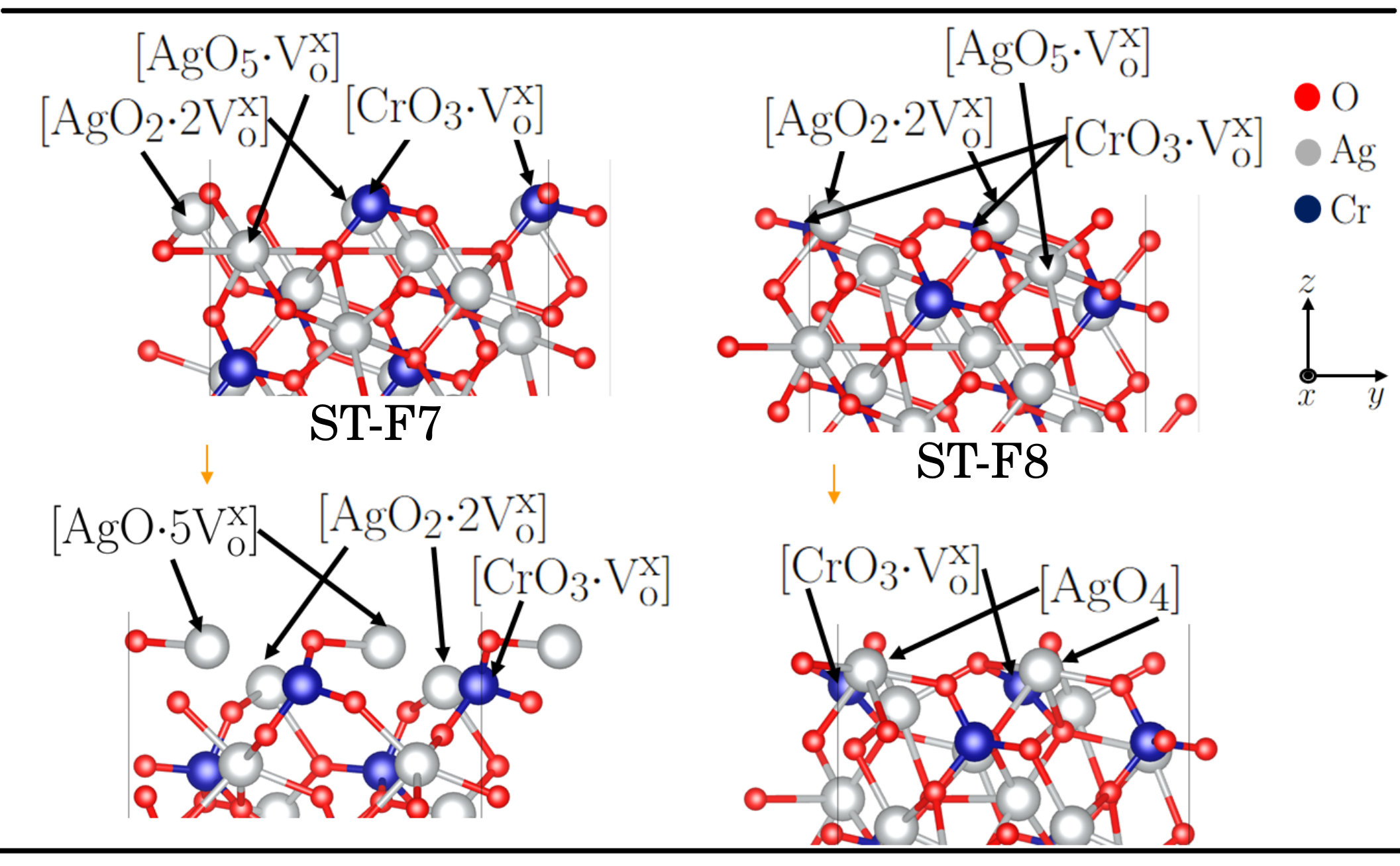}
            \caption{A representation of the surface models for the orientation (011), before and after relaxation. The orange arrow represents the relaxation step. 
            The surface clusters are presented using the 
			Kröger-Vink notation.}
            \label{fig:011-rel-slabs2}
        \end{figure}

        \begin{figure}[H]
            \centering
            \includegraphics[scale=0.3]{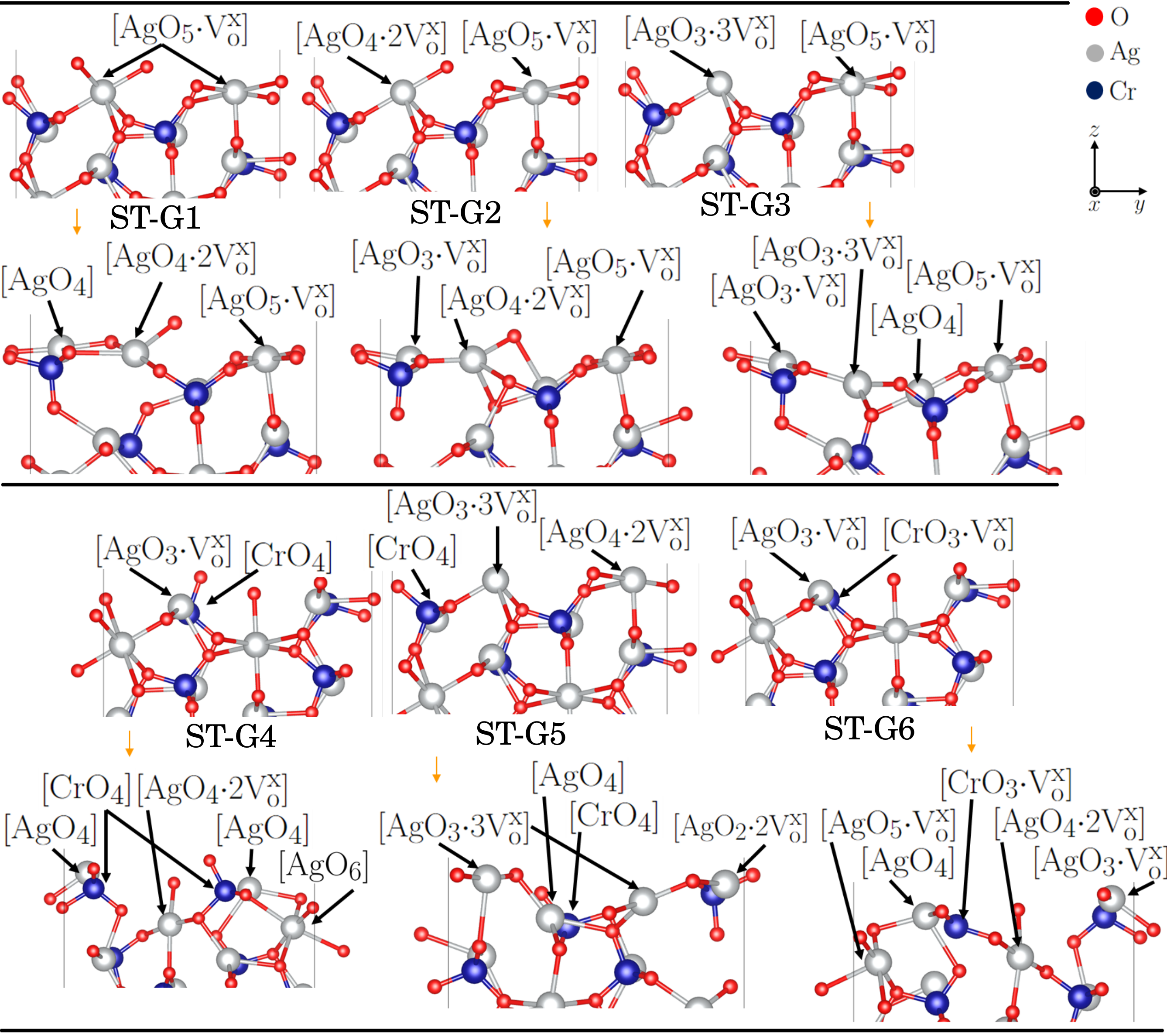}
            \caption{A representation of the surface models for the orientation (011), before and after relaxation. The orange arrow represents the relaxation step.
             The surface clusters are presented using the 
			Kröger-Vink notation.}
            \label{fig:101-rel-slabs1}
        \end{figure}
        \begin{figure}[H]
            \centering
            \includegraphics[scale=0.27]{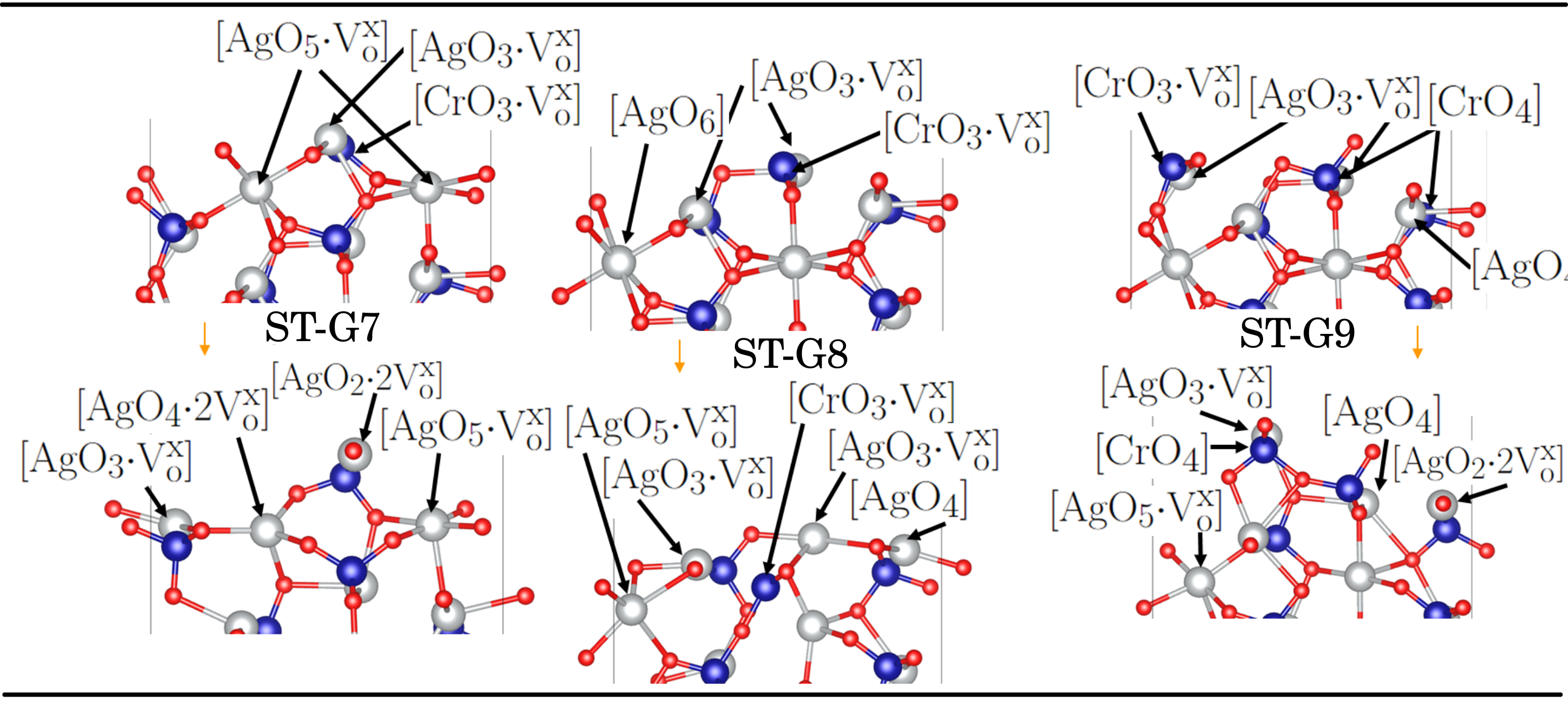}
            \caption{A representation of the surface models for the orientation (101), before and after relaxation. The orange arrow represents the relaxation step.
             The surface clusters are presented using the 
			Kröger-Vink notation.}
            \label{fig:101-rel-slabs2}
        \end{figure}

    \newpage
    \section{\label{sec:thermodynamic-model-derivation}Thermodynamic Model Derivation}
    The model derived here was based on References \cite{reuter_composition_2001, rogal2007ab, bottin_stability_2003, mcquarrie}.
    \subsection{\label{subsec:surface-gibbs-free-energy}Surface Gibbs free energy}
	The Gibbs free energy ($G$) of a solid system interacting with a gas phase is given by the sum of 
	the contributions of each phase, including the interacting region, characterizing the surface 
	\begin{equation}
		G = G_{\text{solid}} + G_{\text{gas}} + G_{\text{surface}}.
		\label{eq:gibbs-total}
	\end{equation}
	
	Since we are interested in the chemistry and physics of the surface,
	\begin{equation}
		G_{\text{surface}} = G - G_{\text{solid}} - G_{\text{gas}},
	\end{equation}
	which, for the case of \ce{Ag2CrO4}, normalized by the surface 
    area ($A$) of a symmetric slab, is
    \begin{multline}
        \gamma(T, P) = \frac{1}{2A}\big[G^{\text{slab}}(T, P, N_{\ce{Ag}}, N_{\ce{Cr}}, N_{\ce{O}}) - 
        N_{\ce{Ag}}\mu_{\ce{Ag}}(T, P) - N_{\ce{Cr}}\mu_{\ce{Cr}}(T, P) - \\
        N_{\ce{O}}\mu_{\ce{O}}(T, P)\big],
        \label{eq:gamma_normalized}
    \end{multline}
    where the Gibbs free energies of the solid and gas phases were described by the respective 
    number of atoms of the species $i$ ($N_i,\, i =$ \ce{Ag}, \ce{Cr}, \ce{O}), 
    and its chemical potential ($\mu_i$). Assuming a mass thermodynamic equilibrium, 
    the chemical potentials are no longer independent, satisfying
    \begin{equation}
        \mu_{\ce{Ag2CrO4}}^{\text{bulk}}(T, P) = 2\mu_{\ce{Ag}}(T, P) + 
        \mu_{\ce{Cr}}(T, P) + 4\mu_{\ce{O}}(T, P) =
        g_{\ce{Ag2CrO4}}^{\text{bulk}}(T, P),
        \label{eq:mass_equilibrium}
    \end{equation}
    where $g_{\ce{Ag2CrO4}}^{\text{bulk}}$ is the Gibbs free energy of the 
    \ce{Ag2CrO4} bulk. Isolating $\mu_{\ce{Cr}}(T, P)$ from 
    equation \Cref{eq:mass_equilibrium} and substituting in equation 
    \Cref{eq:gamma_normalized}, 
    \begin{multline}
        \gamma(T, P) = \frac{1}{2A}\big[G^{\text{slab}}(T, P, N_{\ce{Ag}}, N_{\ce{Cr}}, N_{\ce{O}}) - 
        N_{\ce{Cr}}g_{\ce{Ag2CrO4}}^{\text{bulk}}(T, P) - \mu_{\ce{Cr}}(T, P)N_{\ce{Ag,Cr}} - \\
        \mu_{\ce{O}}(T, P)N_{\ce{O,Cr}}\big],
        \label{eq:gamma_almostfinal}
    \end{multline}
    where
    \begin{gather}
        N_{\ce{Ag,Cr}} = N_{\ce{Ag}} - 2N_{\ce{Cr}}\\
        N_{\ce{O,Cr}} = N_{\ce{O}} - 4N_{\ce{Cr}}.
    \end{gather}

    Defining differences of chemical potentials as 
    \begin{gather}
        \Delta\mu_{\ce{Ag}}(T, P) = \mu_{\ce{Ag}}(T, P) - \mu_{\ce{Ag}}^{\text{bulk}}(T, P)
        \label{eq:delta_ag}\\
        \Delta\mu_{\ce{Cr}}(T, P) = \mu_{\ce{Cr}}(T, P) - \mu_{\ce{Cr}}^{\text{bulk}}(T, P)
        \label{eq:delta_cr}\\
        \Delta\mu_{\ce{O}}(T, P) = \mu_{\ce{O}}(T, P) - \frac{1}{2}\mu_{\ce{O2}}^{\text{free}}(T, P),
        \label{eq:delta_o}
    \end{gather}
    where the superscript ``free" stands for ``free molecule". Using the mass equilibrium 
    assumption and substituting equations \Cref{eq:delta_ag} and \Cref{eq:delta_o} 
    in equation \Cref{eq:gamma_almostfinal}, 
    \begin{equation}
        \gamma(T, P) = \frac{1}{2A}\big[\Gamma(T, P) - \Delta\mu_{\ce{Ag}}(T, P)N_{\ce{Ag,Cr}} - 
        \Delta\mu_{\ce{O}}(T, P)N_{\ce{O,Cr}}\big],
        \label{eq:gamma}
    \end{equation}
    where
    \begin{multline}
        \Gamma(T, P) = G^{\text{slab}}(T, P, N_i) - 
        N_{\ce{Cr}}g_{\ce{Ag2CrO4}}^{\text{bulk}}(T, P) - 
        g_{\ce{Ag}}^{\text{bulk}}(T, P)N_{\ce{Ag,Cr}} - \\
        g_{\ce{O2}}^{\text{free}}(T, P)N_{\ce{O,Cr}}.
    \end{multline}

    To use the equations above to study the stability of silver chromate 
    slabs, it is important to work with energy values that correspond to thermodynamic 
    conditions in which the silver chromate crystal is stable, i.e., no oxygen 
    molecules are released from the bulk to the external environment, and no 
    silver and chromium metal clusters are precipitated on the slab surface. This 
    means that the chemical potential of each component of the silver chromate 
    phase is lower than the respective pure phase, which corresponds to oxygen-poor 
    condition (vacuum). Therefore, from equations 
    \Cref{eq:delta_ag} $-$ \Cref{eq:delta_o},
    \begin{gather}
        \max\left[\mu_{\ce{Ag}}(T, P)\right] = g_{\ce{Ag}}^{\text{bulk}}(0,0) \implies 
        \Delta\mu_{\ce{Ag}}(T, P) < 0,\\
        \max\left[\mu_{\ce{Cr}}(T, P)\right] = g_{\ce{Cr}}^{\text{bulk}}(0,0) \implies 
        \Delta\mu_{\ce{Cr}}(T, P) < 0,\\
        \max\left[\mu_{\ce{O}}(T, P)\right] = \frac{1}{2}g_{\ce{O2}}^{\text{free}}(0,0) \implies 
        \Delta\mu_{\ce{O}}(T, P) < 0.
    \end{gather} 
    
    The minimum values of these functions can be defined by equation 
    \eqref{eq:mass_equilibrium} and the definitions of $\Delta\mu_i(T, P)$ as
    \begin{gather*}
        g^{\text{bulk}}_{\ce{Ag2CrO4}}(T, P) = 
        2\left(\Delta\mu_{\ce{Ag}}(T, P) + g^{\text{bulk}}_{\ce{Ag}}\right) + 
        \Delta\mu_{\ce{Cr}}(T, P) + g^{\text{bulk}}_{\ce{Cr}} + 
        4\left(\Delta\mu_{\ce{O}}(T, P) + \frac{1}{2}g^{\text{free}}_{\ce{O2}}\right)\\
        2\Delta\mu_{\ce{Ag}}(T, P) + 4\Delta\mu_{\ce{O}}(T, P) = 
        g^{\text{bulk}}_{\ce{Ag2CrO4}}(T, P) - \left(2g^{\text{bulk}}_{\ce{Ag}} + 
        g^{\text{bulk}}_{\ce{Cr}} + \frac{1}{2}g^{\text{free}}_{\ce{O2}}\right) - 
        \Delta\mu_{\ce{Cr}}(T, P).
    \end{gather*}

    Since $\Delta\mu_{\ce{Cr}}(T, P) < 0$,
    \begin{equation}
        2\Delta\mu_{\ce{Ag}}(T, P) + 4\Delta\mu_{\ce{O}}(T, P) > 
        g^{\text{bulk}}_{\ce{Ag2CrO4}}(T, P) - \left(2g^{\text{bulk}}_{\ce{Ag}} + 
        g^{\text{bulk}}_{\ce{Cr}} + \frac{1}{2}g^{\text{free}}_{\ce{O2}}\right).
        \label{eq:potential-minimum-initial}
    \end{equation}

    Similarly, 
    the silver chromate crystal must not do 
    desintegrate in simpler oxides, such as \ce{Ag2O} and \ce{Cr2O3}. Therefore,
    \begin{gather}
        2\Delta\mu_{\ce{Ag}}(T, P) + \Delta\mu_{\ce{O}}(T, P) > 
        g^{\text{bulk}}_{\ce{Ag2O}}(T, P) - \left(2g^{\text{bulk}}_{\ce{Ag}} + 
        \frac{1}{2}g^{\text{free}}_{\ce{O2}}\right)
        \label{eq:ag2o-minium-initial}\\
        2\Delta\mu_{\ce{Cr}}(T, P) + 3\Delta\mu_{\ce{O}}(T, P) > 
        g^{\text{bulk}}_{\ce{Cr2O3}}(T, P) - \left(2g^{\text{bulk}}_{\ce{Cr}} + 
        \frac{3}{2}g^{\text{free}}_{\ce{O2}}\right).
        \label{eq:cr2o3-minium-initial}
    \end{gather}

    \subsection{\label{subsec:gas-phase}Gas phase}
        Assuming a thermodynamic equilibrium of the gas phase, which is pure in 
        oxygen molecule, the oxygen chemical potential is given by
        \begin{equation}
            \mu_{\ce{O}}(T, P) = -\frac{1}{2}\frac{k_B T \ln Z^{T}_{\ce{O2}} + PV}{N},
            \label{eq:oxygen-chemical-potential}
        \end{equation}
        where $Z^T_{\ce{O2}}$ is the partition function of $N$ oxygen molecules as 
        an ideal gas, which, using the Born-Oppenheimer approximation, has the form
        \begin{equation}
            Z^T_{\ce{O2}} = \frac{1}{N!}(q^{\text{trans}}q^{\text{vib}}
            q^{\text{elect}}q^{\text{nucl}})^N
            \label{eq:partition-function}
        \end{equation}

        From the energy eigenvalues of a particle in a box with dimensions approaching 
        to infinity,
        \begin{equation}
            q^{\text{trans}} = V\left(\frac{2\pi mk_BT}{h^2}\right)^{\frac{3}{2}},
        \end{equation}
        where $m$ is the mass of the particle. From the rigid rotor aproximation, 
        \begin{equation}
            q^{\text{rot}} = \sum_{J = 0}^\infty(2J + 1)\exp\left[-\frac{J(J + 1)B_0}{k_BT}\right],
        \end{equation}
        with $B_0 = \hbar^2/2I$ being the rotation constant, depending on the momentum of 
        inertia $I$. For the case of diatomic molecules, under appropriate temperature 
        values, the rotational term can be simplified using a symmetry number $\sigma^s$, 
        representing the number of distinguishable orientations that the molecule can have. 
        If the rotational levels are sufficiently small, 
        \begin{equation}
            \mu^{\text{rot}} \approx -k_BT\ln\left(\frac{k_BT}{\sigma^sB_0}\right).
        \end{equation}

        The vibrational contribution can be derived using the harmonic approximation 
        to write the partition function as a sum of the harmonic oscillators 
        ($\omega_i$) of all fundamental modes $M$ of the particle. Therefore, 
        \begin{equation}
            q^{\text{vib}} = \sum_{i = 1}^M\sum_{n = 0}^\infty\exp\left[-\left(n + \frac{1}{2}\right)\frac{\hbar\omega_i}{k_BT}\right].
        \end{equation}

        The resolution of the geometric series results in
        \begin{equation}
            \mu^{\text{vib}} = E^{\text{ZPE}} + \Delta\mu^{\text{vib}} = 
            \sum_{i = 1}^M\bigg\{\frac{\hbar\omega_i}{2} + k_BT\ln\left[1 - \exp\left(\frac{\hbar\omega_i}{k_BT}\right)\right]\bigg\},
        \end{equation}
        where $E^{\text{ZPE}}$ is the zero point vibration energy. In relation to the electronic 
        contributions, since the internal excitation energies of the molecule are much 
        lower than $k_BT$, the electronic contribution is mainly due to its fundamental 
        state. Considering a possible degeneracy,
        \begin{equation}
            \mu^{\text{elect}} \approx E^T_{\ce{O2}} - k_BT\ln\left(I^{\text{spin}}\right).
        \end{equation}

        The same form is obtained for the nuclear degrees of freedom, but it is neglected 
        here, since nuclear changes in chemical processes are not important, in the 
        majority of the cases. Hence, with all $\mu^i$, using equation 
        \Cref{eq:partition-function} into \Cref{eq:oxygen-chemical-potential},
        \begin{equation}
            \mu_{\ce{O}}(T, P) = -\frac{1}{2N}\bigg\{k_BT\ln\left[\frac{1}{N!}\left(q^{\text{trans}}\right)^N\right] - PV\bigg\} 
            + \frac{1}{2}\mu^{\text{rot}} + \frac{1}{2}\mu^{\text{vib}} + 
            \frac{1}{2}\mu^{\text{elect}},
        \end{equation}
        which can be rewritten as 
        \begin{equation}
            \mu_{\ce{O}}(T, P) = \frac{1}{2}E^T_{\ce{O2}} + \frac{1}{2}E^{\text{ZPE}} + 
            \Delta_{\ce{O}}(T, P),
        \end{equation}
        where
        \begin{multline}
            \Delta\mu_{\ce{O}}(T, P) = -\frac{1}{2}k_BT\bigg\{
                \ln\left[\left(\frac{2\pi m}{h^2}\right)^{3/2}\frac{\left(k_BT\right)^{5/2}}{P}\right] +
                \ln\left(\frac{k_BT}{\sigma^sB_0}\right) - \\
                \ln\left[1 - \exp\left(\frac{\hbar\omega_0}{k_BT}\right)\right] + 
                \ln\left(I^{\text{spin}}\right)\bigg\},
        \end{multline}
        with $\sigma^s = 2$, $B_0 = 0.18$ meV and $I^{\text{spin}} = 3$. 
%

    \subsection{\label{subsec:solid-phase}Solid phase}
        The Gibbs free energy $G$ is defined as the sum of the total ($E^T$), 
        vibrational ($F^{\text{vib}}$), and conformational ($F^{\text{conf}}$)
        energies, the pressure and the volume of the system, 
        \begin{multline}
            G(T, P, N_{\ce{Ag}}, N_{\ce{Cr}}, N_{\ce{O}}) = 
            E^T(V, N_{\ce{Ag}}, N_{\ce{Cr}}, N_{\ce{O}}) + 
            F^{\text{vib}}(T, V, N_{\ce{Ag}}, N_{\ce{Cr}}, N_{\ce{O}}) +\\
            F^{\text{conf}}(T, V, N_{\ce{Ag}}, N_{\ce{Cr}}, N_{\ce{O}}) + 
            PV(T, P, N_{\ce{Ag}}, N_{\ce{Cr}}, N_{\ce{O}}),
        \end{multline}
        where $E^T + F^{\text{vib}} + F^{\text{conf}} = F$ is the Helmholtz 
        free energy. 
        
        The vibrational contribution can be defined from the phonon density of 
        states of the system ($\sigma(\omega)$), formulating $F^{\text{vib}}$ as an integral 
        over the significant vibrational modes ($\omega$) of the system,
        \begin{equation}
            F^{\text{vib}}(T, V, N_{\ce{Ag}}, N_{\ce{Cr}}, N_{\ce{O}}) = 
            \int d\omega f^{\text{vib}}(T, \omega)\sigma(\omega),
        \end{equation}
        where
        \begin{equation}
            f^{\text{vib}}(T, \omega) = \hbar\omega\left(\frac{1}{2} + 
            \frac{1}{e^{\beta\hbar\omega} - 1}\right) - 
            k_BT\left[\frac{\beta\hbar\omega}{e^{\beta\hbar\omega} - 1} - 
            \ln\left(1 - e^{-\beta\hbar\omega}\right)\right].
        \end{equation}

        For simplification, we considered the Einstein model to define $\sigma(\omega) \propto \delta(\omega - \omega_0)$. 
        Therefore,
        \begin{multline}
            F^{\text{vib}}(T, V, N_{\ce{Ag}}, N_{\ce{Cr}}, N_{\ce{O}}) = 
            \int d\omega\;\delta(\omega - \omega_0)\bigg\{\hbar\omega\left(
                \frac{1}{2} + \frac{1}{e^{\beta\hbar\omega} - 1}\right) - \\
                k_BT\left[\frac{\beta\hbar\omega}{e^{\beta\hbar\omega} - 1} - 
                \ln\left(1 - e^{-\beta\hbar\omega}\right)\right]\bigg\},
        \end{multline}
        which results in
        \begin{equation}
            F^{\text{vib}}(T, V, N_{\ce{Ag}}, N_{\ce{Cr}}, N_{\ce{O}}) = 
            \frac{1}{2}\hbar\omega_0 + 
            k_B T\ln\left(1 - e^{-\hbar\omega_0/k_BT}\right).
        \end{equation}
        
        The conformational energy, at low temperatures, depends on the conformational 
        entropy ($S^{\text{conf}}$) related to a limited number of 
        defects in the structure ($n_d$) among a significantly higher quantity of 
        surface sites $N_s$. The normalized form reads 
        \begin{equation}
            \frac{TS^{\text{conf}}}{N_sA} = \frac{k_BT}{N_sA}
            \ln\left[\frac{\left(N_s + n_d\right)!}{N_s!n_d!}\right],
        \end{equation}
        which, if $n_d \ll N_s$, simplifies by the Stirling formula to
        \begin{equation}
            \frac{TS^{\text{conf}}}{N_sA} = \frac{k_BT}{A}
            \left[\ln\left(1 + \frac{n_d}{N_s}\right) + 
            \frac{n_d}{N_s}\ln\left(1 + \frac{N_s}{n_d}\right)\right].
        \end{equation}

        Therefore, for the whole solid system,
        \begin{equation}
            G = \frac{1}{2}\hbar\omega_0 + 
            k_B T\ln\left(1 - e^{-\hbar\omega_0/k_BT}\right) + k_BT
            \left[\ln\left(1 + \frac{n_d}{N_s}\right) + 
            \frac{n_d}{N_s}\ln\left(1 + \frac{N_s}{n_d}\right)\right] + PV.
        \end{equation}

        Defining the adimensional quantities
        \begin{equation}
            \xi \equiv \frac{k_BT}{E^T} \eqcomma \lambda \equiv \frac{\hbar\omega_0}{E^T} \eqcomma \chi \equiv \frac{n_d}{N_s} \eqcomma \pi \equiv \frac{PV}{E^T},
        \end{equation}
        one gets
        \begin{equation}
            \frac{G}{E^T} = 1 + \frac{\lambda}{2} + \xi\ln\left(1 - e^{-\lambda/\xi}\right) + \xi\left[(1 + \chi)\ln(1 + \chi) - \chi\ln\chi\right] + \pi.
        \end{equation}

        As long as the surface defects are diluted, e.g. $\xi \ll 1$, if
        \begin{equation}
            \xi \ll 1 \eqcomma \lambda \ll 1 \eqcomma  \xi\chi|\ln\chi| \ll 1 \eqcomma \pi \ll 1,
        \end{equation}
        then
        \begin{equation}
            G(T, P, N_{\ce{Ag}}, N_{\ce{Cr}}, N_{\ce{O}}) \approx 
            E^T(V, N_{\ce{Ag}}, N_{\ce{Cr}}, N_{\ce{O}}),
            \label{eq:approx}
        \end{equation}
        where the total energy term can be intepreted as the DFT-total energy. 
        The condition for the point-zero energy term $\xi/2$ is always satisfied, 
        since $\mathcal{O}(k_BT) \ll E^T$, which is of the order of hundreds of eV. 
        The condition for $\lambda$ is also satisfied for the same reason, since 
        $\left|\ln\left(\frac{\lambda}{\xi}\right)\right|$ is small for any temperature 
        value considered in this study. The conformational condition is satisfied under 
        the constraint of diluted surface defects, e.g., $\chi \leq 0.1$. At last 
        the condition for $\pi$ is also satisfied, which can be viewed from a simple 
        dimensional analysis. The $PV$ term has the dimension 
        $\left[\text{atm}\times\text{Å}^3/\text{Å}^2\right]$, which is about 
        $10^{-3}$ meV/Å$^2$. Therefore, even under high pressures of the order of 100 atm, 
        such a quantity would be around 0.1 meV/Å$^2$, which is negligible 
        when compared to the other energy terms.
        
\begin{figure}[h]
            \centering
            \includegraphics[scale=0.7]{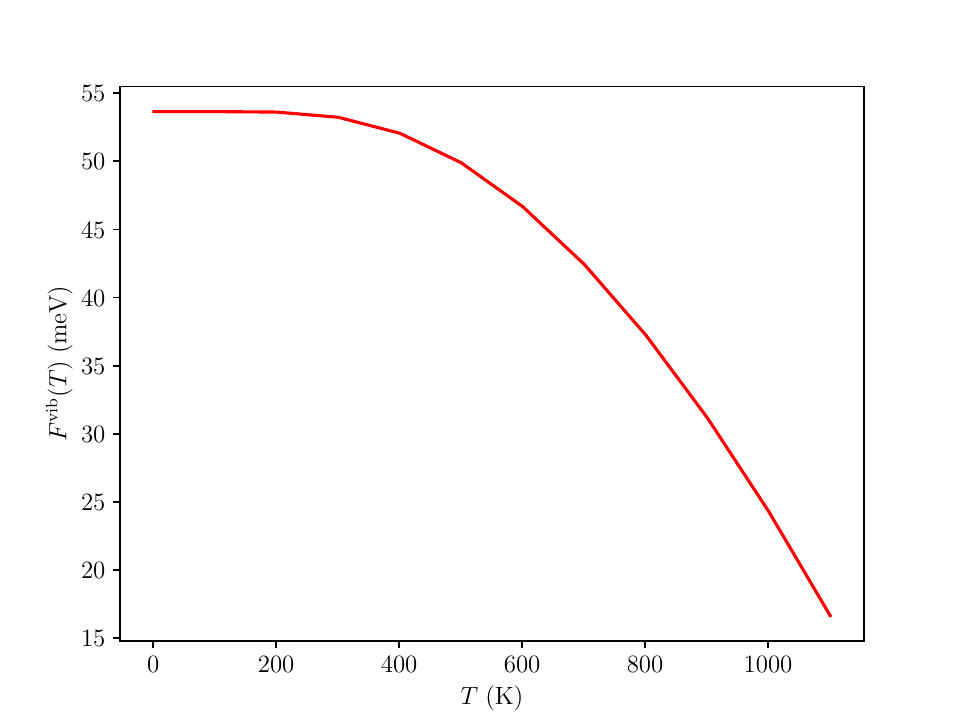}
            \caption{Vibrational Helmholtz free energy as a function of temperature. 
            The highest phonon vibrational mode was calculated as $\omega_0 = 26$ THz.}
            \label{fig:helmholtz}
        \end{figure}

        Therefore, \Cref{eq:approx} holds for our system (and, in a more general way, 
        for solid systems under the influence of temperatures up to 1000 K), which 
        we also show numerically below. 
        The Helmholtz vibrational contribution for the \ce{Ag2CrO4} bulk system is 
        of the order of 10 meV for the whole temperature range considered in this 
        study (\Cref{fig:helmholtz}), which is much lower than the total energy (shown in \Cref{tab:thermo-data1}), 
        close to $-$42 eV.
        
\begin{figure}[h]
            \centering
            \includegraphics[scale=0.7]{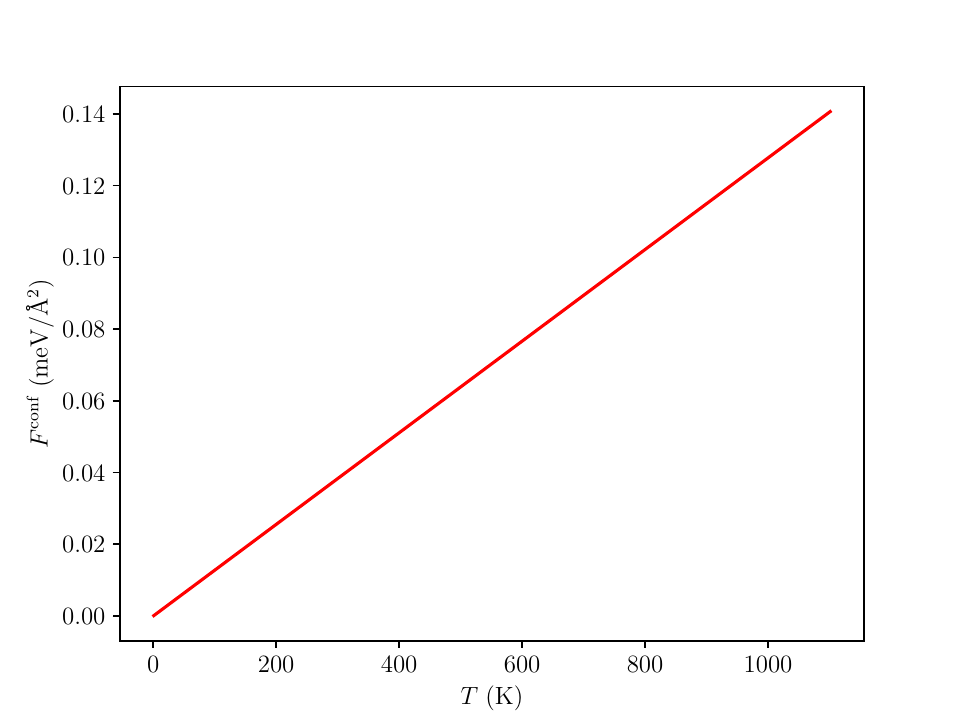}
            \caption{Conformational Helmholtz energy as a function of temperature 
            assuming 10 \% of surface defects.}
            \label{fig:entropy}
        \end{figure}

        The conformational contribution, assuming $n_s/N_s \approx 10\%$, will be 
        of the order of 0.1 meV/Å$^2$ (\Cref{fig:entropy}), which is much lower than the total 
        energy per surface area, close to $-$500 meV/Å$^2$.

        Therefore, the $\Gamma$ term in equation \Cref{eq:gamma} is rewritten as
        \begin{equation}
            \Gamma(T, P) = \Gamma \approx E^{\text{slab}} - 
            N_{\ce{Cr}}E^{\text{bulk}}_{\ce{Ag2CrO4}} - E^{\text{bulk}}_{\ce{Ag}}N_{\ce{Ag,Cr}} - 
            \frac{1}{2}E^T_{\ce{O2}}N_{\ce{O,Cr}},
        \end{equation}
        while the limit defined in \Cref{eq:potential-minimum-initial} is
        \begin{equation}
            2\Delta\mu_{\ce{Ag}}(T, P) + 4\Delta\mu_{\ce{O}}(T, P) > 
        E^{\text{bulk}}_{\ce{Ag2CrO4}}(T, P) - \left(2E^{\text{bulk}}_{\ce{Ag}} + 
        E^{\text{bulk}}_{\ce{Cr}} + \frac{1}{2}E^T_{\ce{O2}}\right),
        \end{equation}
        which is the same as
        \begin{equation}
            2\Delta\mu_{\ce{Ag}}(T, P) + 4\Delta\mu_{\ce{O}}(T, P) > E^f_{\ce{Ag2CrO4}},
            \label{eq:minium-final}
        \end{equation}
        where $E^f_{\ce{Ag2CrO4}}$ is the formation energy of the \ce{Ag2CrO4} crystla. 
        The same rationale applies for the simpler oxides, that is
        \begin{gather}
            2\Delta\mu_{\ce{Ag}}(T, P) + \Delta\mu_{\ce{O}}(T, P) < E^f_{\ce{Ag2O}}
        \label{eq:ag2o-minium-final}\\
        2\Delta\mu_{\ce{Cr}}(T, P) + 3\Delta\mu_{\ce{O}}(T, P) < E^f_{\ce{Cr2O3}}.
        \label{eq:cr2o3-minium-final}
        \end{gather}

        The \Cref{eq:cr2o3-minium-final} is dependent on $\Delta\mu_{\ce{Cr}}(T, P)$, 
        which is used as the dependent variable of the model (i. e. $\gamma$ depends 
        explicitly on $\Delta\mu_{\ce{Ag}}(T, P)$ and $\Delta\mu_{\ce{O}}(T, P)$). 
        Hence, it is important to adapt the expression as a dependence of 
        $\Delta\mu_{\ce{Ag}}(T, P)$ and $\Delta\mu_{\ce{O}}(T, P)$ only, so the 
        corresponding precipitation line can be indicated directly on the surface 
        diagrams. From \Cref{eq:mass_equilibrium},
        \begin{gather*}
            2\Delta\mu_{\ce{Cr}}(T, P) + 3\Delta\mu_{\ce{O}}(T, P) < E^f_{\ce{Cr2O3}}\\
            2\left(\mu_{\ce{Cr}} - E_{\ce{Cr}}^{\text{bulk}}\right) + 3\Delta\mu_{\ce{O}}(T, P) < E^f_{\ce{Cr2O3}}.
        \end{gather*}

        From the definition of $\Delta\mu_i(T, P)$,
        \begin{gather*}
            2\left[E_{\ce{Ag2CrO4}}^{\text{bulk}} - 2\left(\Delta\mu_{\ce{Ag}} + E_{\ce{Ag}}^{\text{bulk}}\right) - 
            4\left(\Delta\mu_{\ce{o}} + \frac{1}{2}E_{\ce{O2}}^{\text{T}}\right) - E_{\ce{Cr}}^{\text{bulk}}\right] + 
            3\Delta\mu_{\ce{O}} > E_{\ce{Cr2O3}}^f\\
            2E_{\ce{Ag2CrO4}}^{\text{bulk}} - 4\Delta\mu_{\ce{Ag}} - 
            4E_{\ce{Ag}}^{\text{bulk}} - 5\Delta\mu_{\ce{O}} - 4E_{\ce{O2}}^{\text{T}} - 
            2E_{\ce{Cr}}^{\text{bulk}} > E_{\ce{Cr2O3}}^f\\
            -4\Delta\mu_{\ce{Ag}} - 5\Delta\mu_{\ce{O}} > 2E_{\ce{Ag2CrO4}}^{\text{bulk}} - 
            4E_{\ce{Ag}}^{\text{bulk}} - 2E_{\ce{Cr}}^{\text{bulk}} - 4E_{\ce{O2}}^{\text{T}} - 
            E_{\ce{Cr2O3}}^f,
        \end{gather*}
        which results in
        \begin{equation}
            4\Delta\mu_{\ce{Ag}} + 5\Delta\mu_{\ce{O}} < 2E_{\ce{Ag2CrO4}}^f - E_{\ce{Cr2O3}}^f.
        \end{equation}

    \section{\label{sec:thermodynamic-data}Thermodynamic Data}
        \begin{table}[htbp]
            \centering
            \caption{Total energy per unit cell calculated for 
            \ce{Ag2CrO4} (Pnma), \ce{Ag} (Fm-3m), Cr (Im$\bar{3}$m), 
            O2, Ag2O (Pn$\bar{3}$m1), and Cr2O3 (R$\bar{3}$c).}
            \begin{tabular}{cc}
                \hline
                \textbf{Compound} & \textbf{Total energy (eV)}\\
                \hline
                \ce{Ag2CrO4} & -42.67\\
                \ce{Ag} & -2.72\\
                \ce{Cr} & -9.48\\
                \ce{O2} & -9.87\\
                \ce{Ag2O} & -10.73\\
                \ce{Cr2O3} & -43.39\\
                \hline
                \label{tab:thermo-data1}
            \end{tabular}
        \end{table}

        \begin{table}[htbp]
            \centering
            \caption{Total energy calculated for each termination of \ce{Ag2CrO4} in this study.}
            \resizebox{\textwidth}{!}{%
            \begin{tabular}{cccccc}
                \hline
                \textbf{Termination} & \textbf{Total energy (eV)} & \textbf{Termination} & \textbf{Total energy (eV)} & \textbf{Termination} & \textbf{Total energy (eV)}\\
                \hline
                ST-A1 & -622.14 & ST-C3 & -276.72 & ST-F4 & -698.26\\
                ST-A2 & -599.91 & ST-C4 & -255.73 & ST-F5 & -678.10\\
                ST-A3 & -590.14 & ST-C5 & -331.82 & ST-F6 & -837.92\\
                ST-A4 & -670.71 & ST-C6 & -331.72 & ST-F7 & -808.45\\
                ST-A5 & -662.37 & ST-C7 & -298.71 & ST-F8 & -808.83\\
                ST-A6 & -661.59 & ST-D1 & -624.41 & ST-G1 & -564.11\\
                ST-A7 & -626.50 & ST-D2 & -618.87 & ST-G2 & -552.97\\
                ST-B1 & -738.36 & ST-D3 & -600.17 & ST-G3 & -534.14\\
                ST-B2 & -707.76 & ST-D4 & -661.57 & ST-G4 & -533.45\\
                ST-B3 & -787.81 & ST-E1 & -441.98 & ST-G5 & -525.43\\
                ST-B4 & -695.00 & ST-E2 & -425.72 & ST-G6 & -518.93\\
                ST-B5 & -687.48 & ST-E3 & -501.18 & ST-G7 & -604.89\\
                ST-B6 & -826.61 & ST-E4 & -472.63 & ST-G8 & -566.44\\
                ST-B7 & -822.53 & ST-F1 & -725.43 & ST-G9 & -653.93\\
                ST-C1 & -298.32 & ST-F2 & -718.23 & - & -\\
                ST-C2 & -281.60 & ST-F3 & -695.92 & - & -\\
                \hline
                \label{tab:thermo-data2}
            \end{tabular}}
        \end{table}

    \section{\label{sec:energy-diagrams}Energy Profiles and Surface Diagrams}
        \begin{figure}[H]
            \centering
            \includegraphics[scale=0.6]{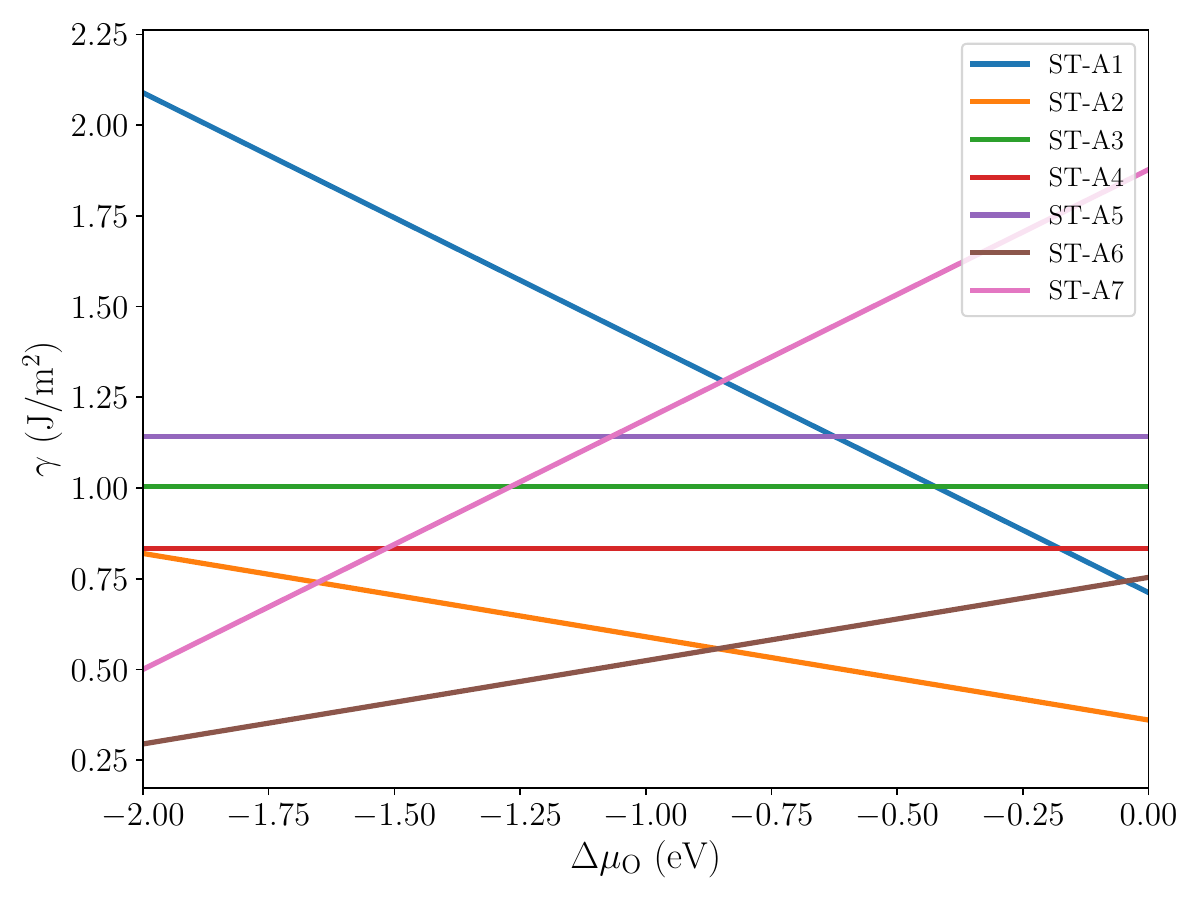}
            \caption{Gibbs surface free energy as a function of 
            $\Delta\mu_{\ce{O}}(T, P)$ for the terminations of the (110) orientation. 
            $\Delta\mu_{\ce{Ag}}(T, P) = 0$ eV.}
        \end{figure}

        \begin{figure}[H]
            \centering
            \includegraphics[scale=0.6]{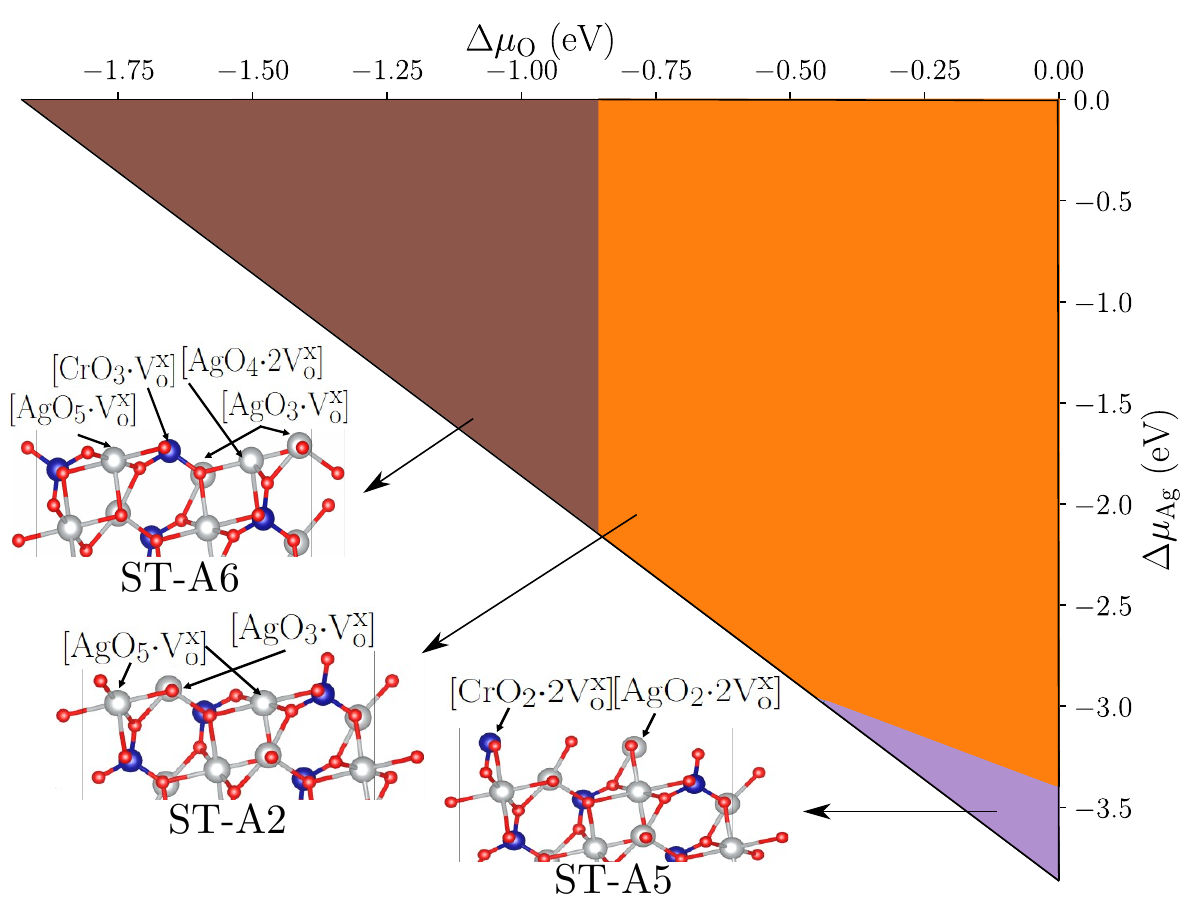}
            \caption{Surface diagram for the terminations of the (110) orientation. 
            Arrows indicate the corresponding non-relaxed slabs.}
        \end{figure}

        \begin{figure}[H]
            \centering
            \includegraphics[scale=0.6]{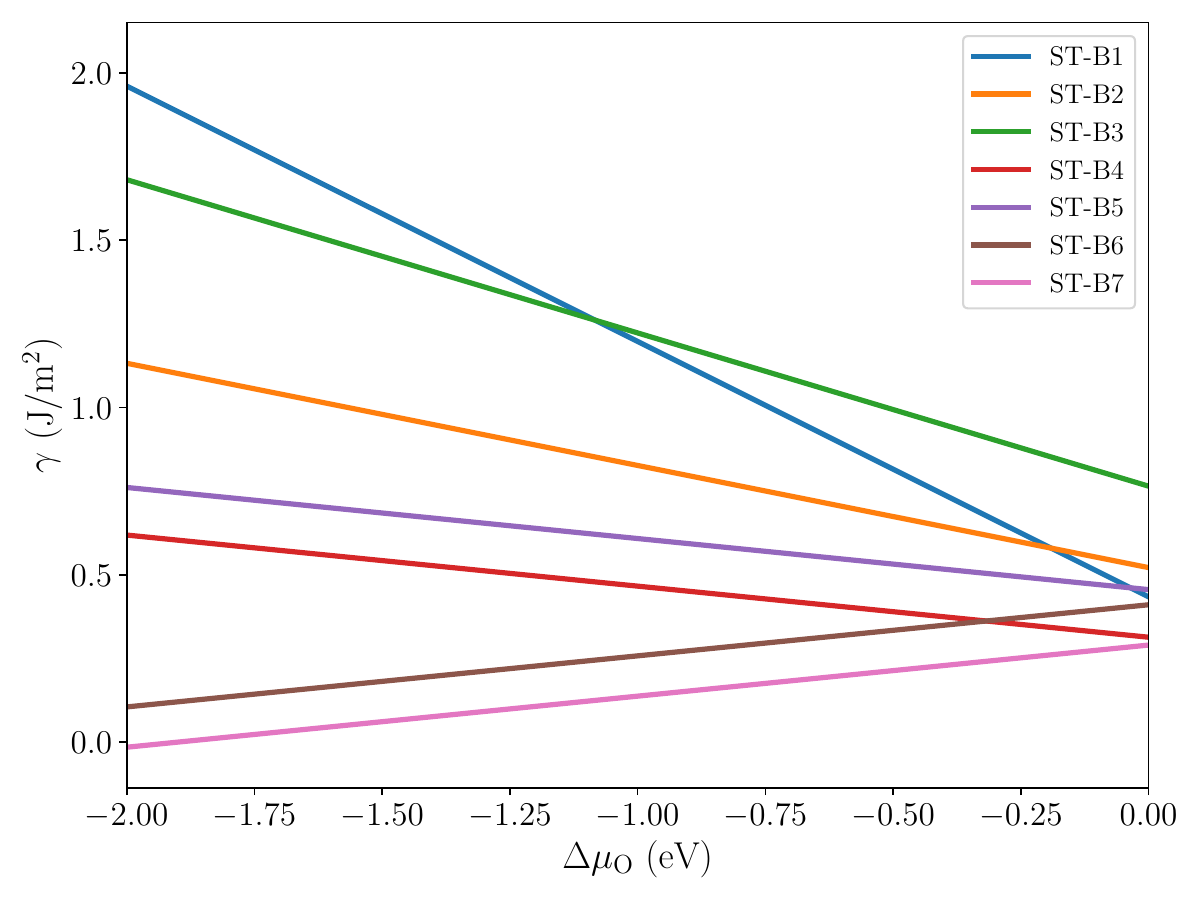}
            \caption{Gibbs surface free energy as a function of 
            $\Delta\mu_{\ce{O}}(T, P)$ for the terminations of the (111) orientation. 
            $\Delta\mu_{\ce{Ag}}(T, P) = 0$ eV.}
        \end{figure}

        \begin{figure}[H]
            \centering
            \includegraphics[scale=0.6]{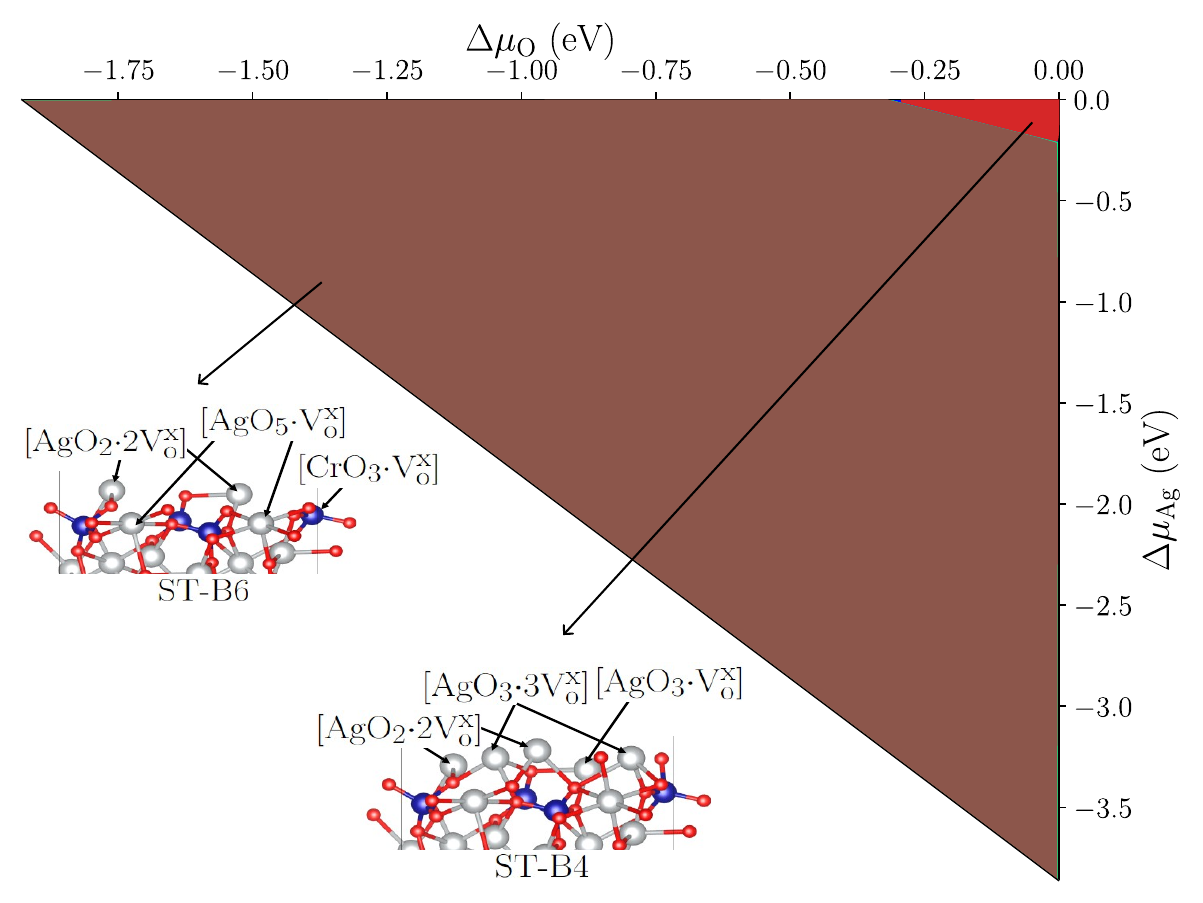}
            \caption{Surface diagram for the terminations of the (111) orientation. 
            Here, the second most stable structures are decpicted, as ST-B7 remains
            the most stable regardless of the thermodynamic conditions. 
            Arrows indicate the corresponding non-relaxed slabs.}
        \end{figure}

        \begin{figure}[H]
            \centering
            \includegraphics[scale=0.6]{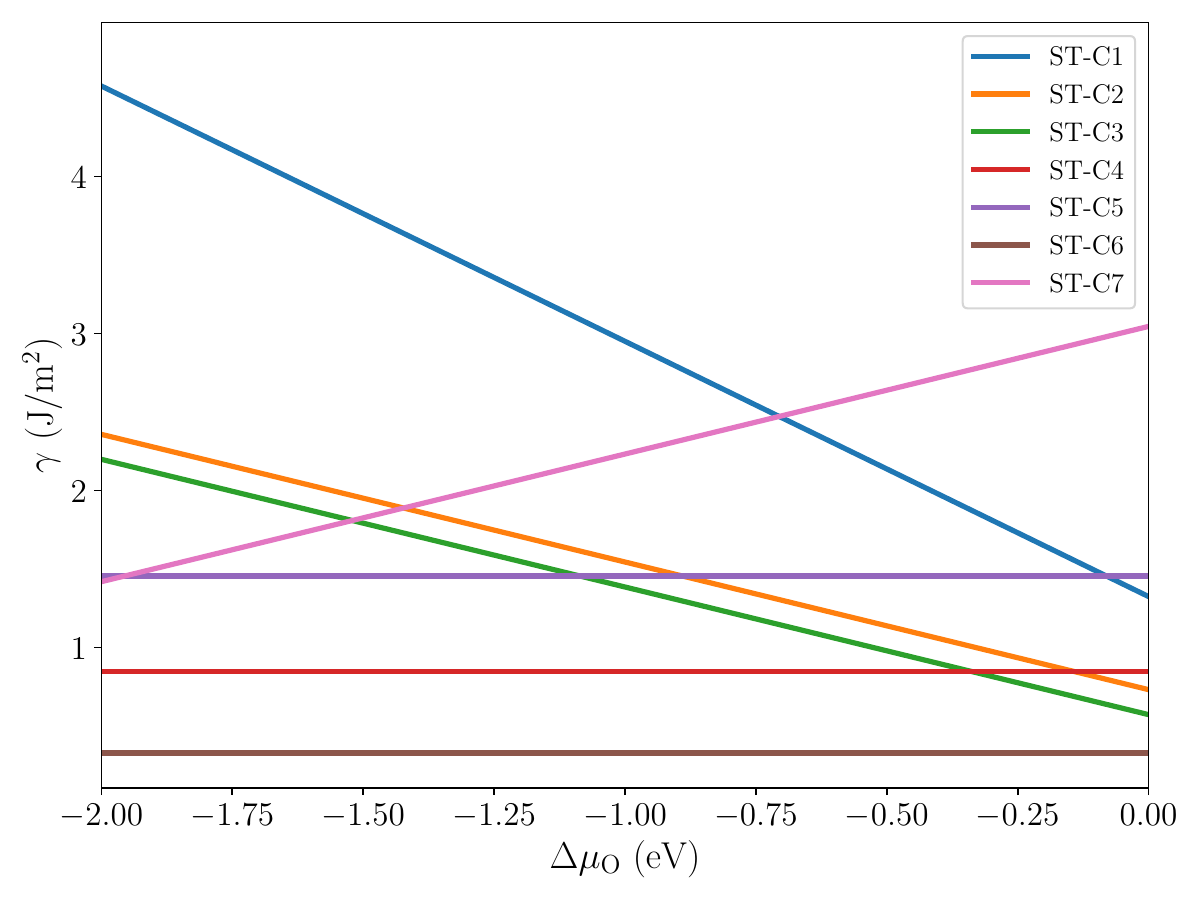}
            \caption{Gibbs surface free energy as a function of 
            $\Delta\mu_{\ce{O}}(T, P)$ for the terminations of the (100) orientation. 
            $\Delta\mu_{\ce{Ag}}(T, P) = 0$ eV.}
        \end{figure}

        \begin{figure}[H]
            \centering
            \includegraphics[scale=0.6]{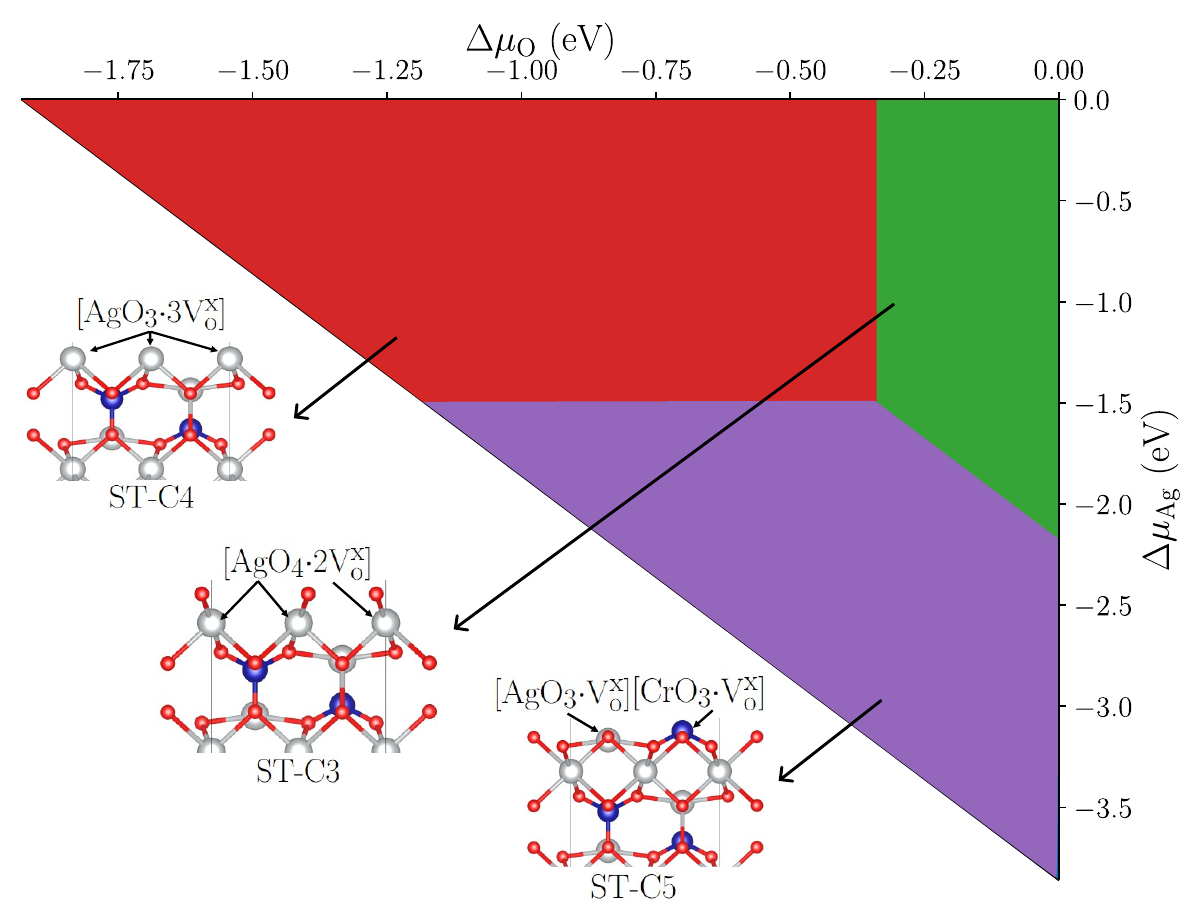}
            \caption{Surface diagram for the terminations of the (100) orientation. 
            Here, the second most stable structures are depicted, as ST-C7 remains
            the most stable regardless of the thermodynamic conditions. 
            Arrows indicate the corresponding non-relaxed slabs.}
        \end{figure}

        \begin{figure}[H]
            \centering
            \includegraphics[scale=0.6]{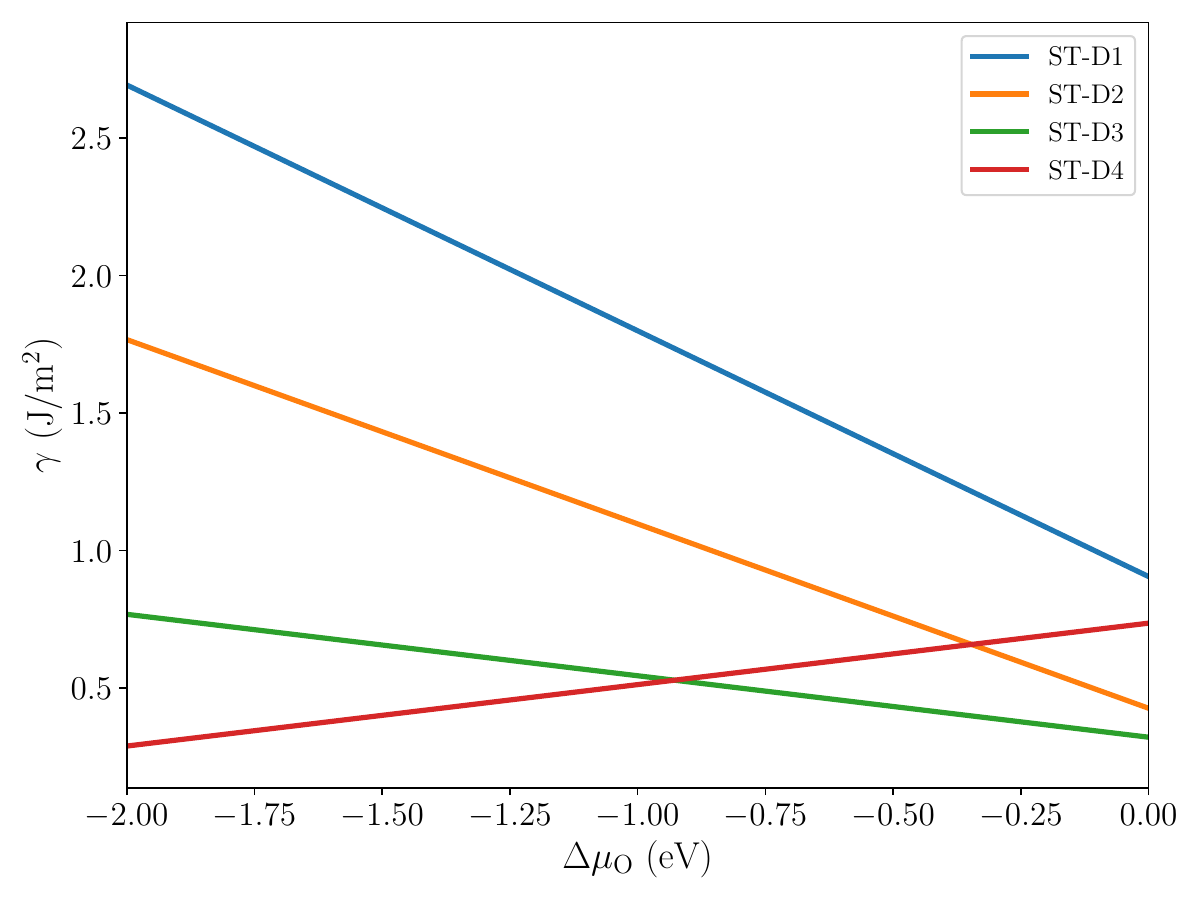}
            \caption{Gibbs surface free energy as a function of 
            $\Delta\mu_{\ce{O}}(T, P)$ for the terminations of the (001) orientation. 
            $\Delta\mu_{\ce{Ag}}(T, P) = 0$ eV.}
        \end{figure}

        \begin{figure}[H]
            \centering
            \includegraphics[scale=0.6]{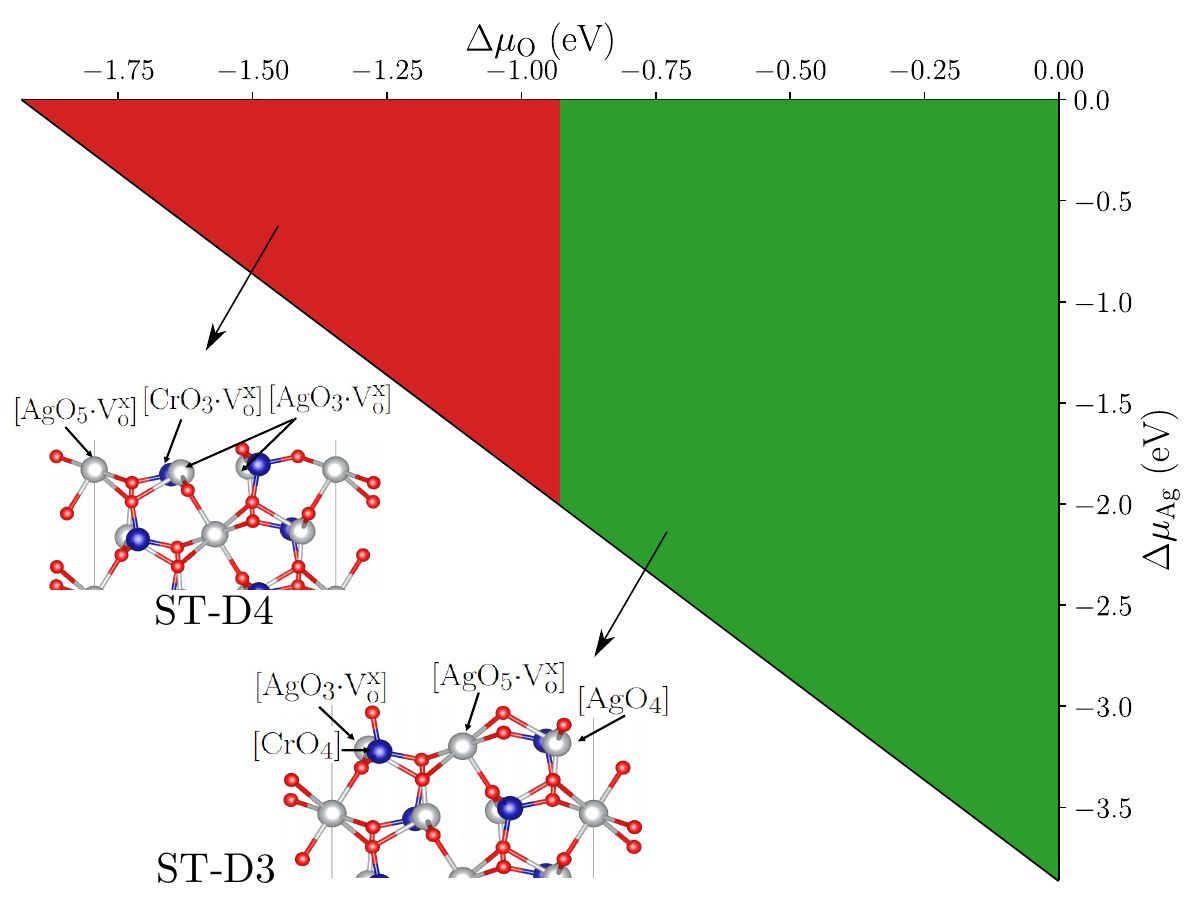}
            \caption{Surface diagram for the terminations of the (001) orientation. 
            Arrows indicate the corresponding non-relaxed slabs are indicated by arrows.}
        \end{figure}

        \begin{figure}[H]
            \centering
            \includegraphics[scale=0.6]{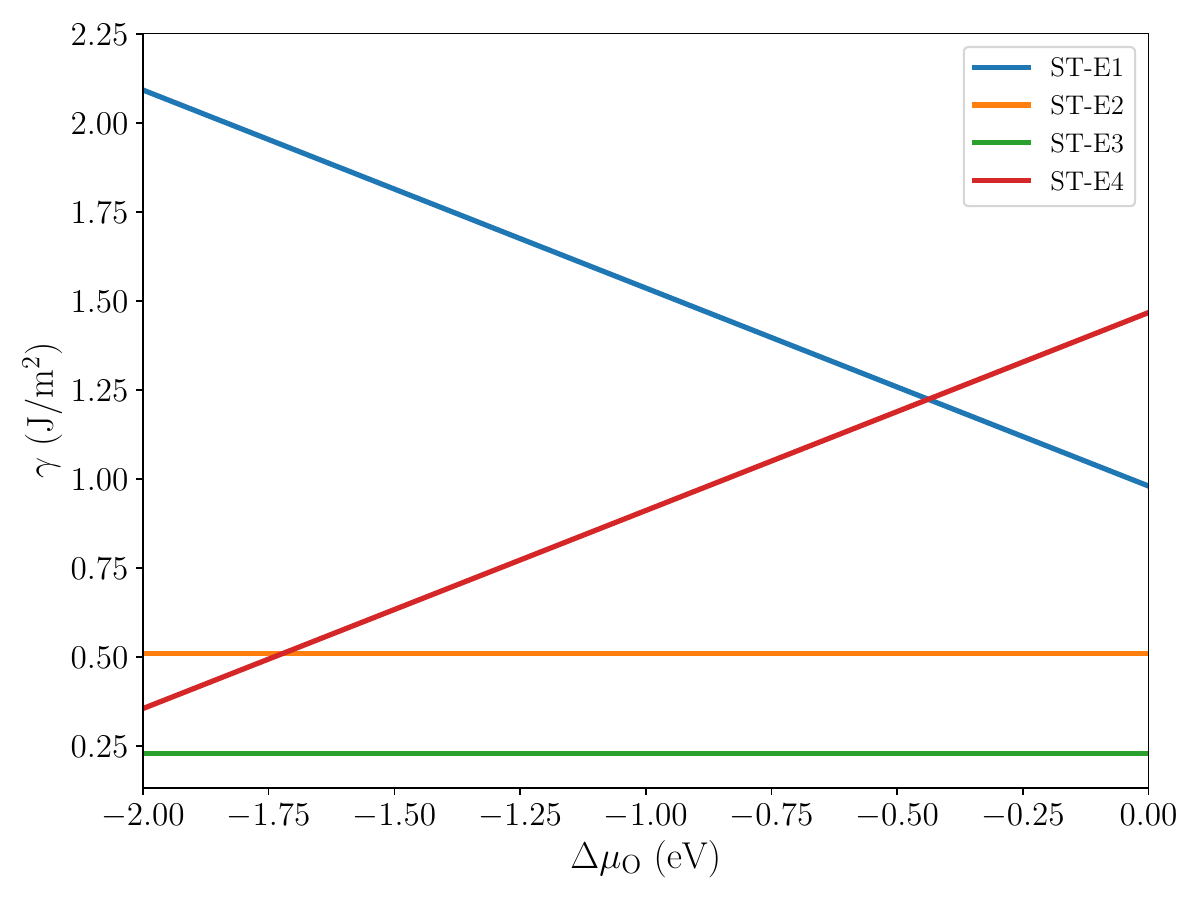}
            \caption{Gibbs surface free energy as a function of 
            $\Delta\mu_{\ce{O}}(T, P)$ for the terminations of the (010) orientation. 
            $\Delta\mu_{\ce{Ag}}(T, P) = 0$ eV.}
        \end{figure}

        \begin{figure}[H]
            \centering
            \includegraphics[scale=0.6]{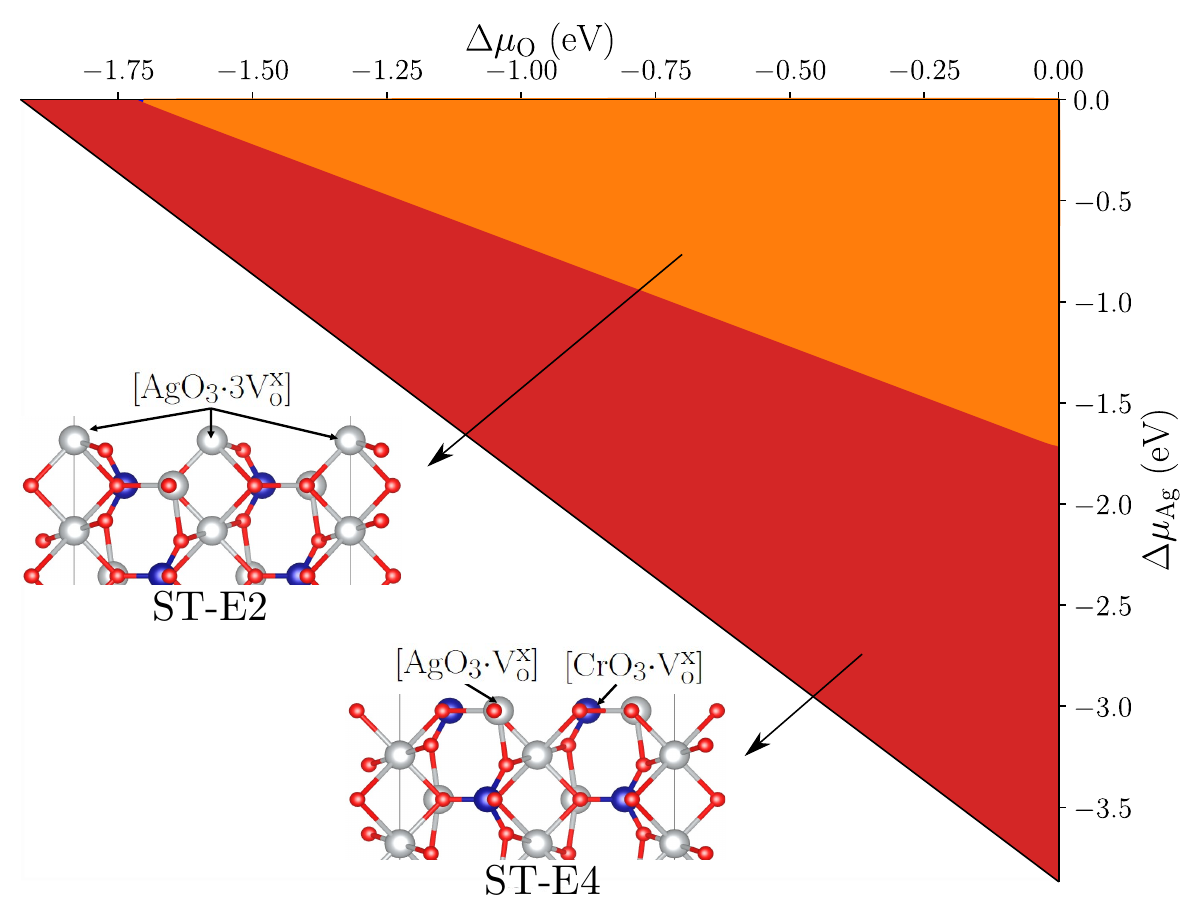}
            \caption{Surface diagram for the terminations of the (010) orientation. 
            Here, the second most stable structures are depicted, as ST-E3 remains
            the most stable regardless of the thermodynamic conditions. 
            Arrows indicate the corresponding non-relaxed slabs.}
        \end{figure}

    \begin{figure}[H]
        \centering
        \includegraphics[scale=0.6]{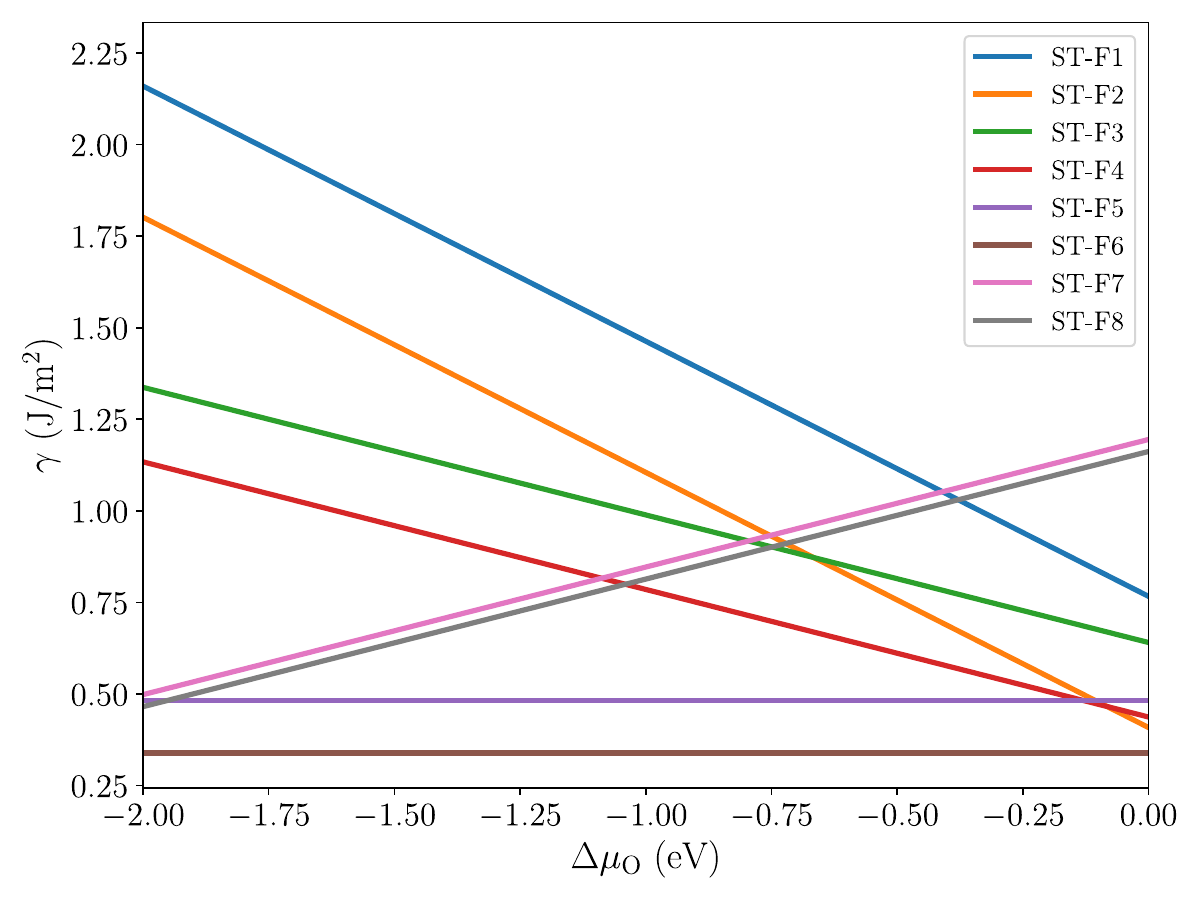}
        \caption{Gibbs surface free energy as a function of 
        $\Delta\mu_{\ce{O}}(T, P)$ for the terminations of the (011) orientation. 
        $\Delta\mu_{\ce{Ag}}(T, P) = 0$ eV.}
    \end{figure}

    \begin{figure}[H]
        \centering
        \includegraphics[scale=0.6]{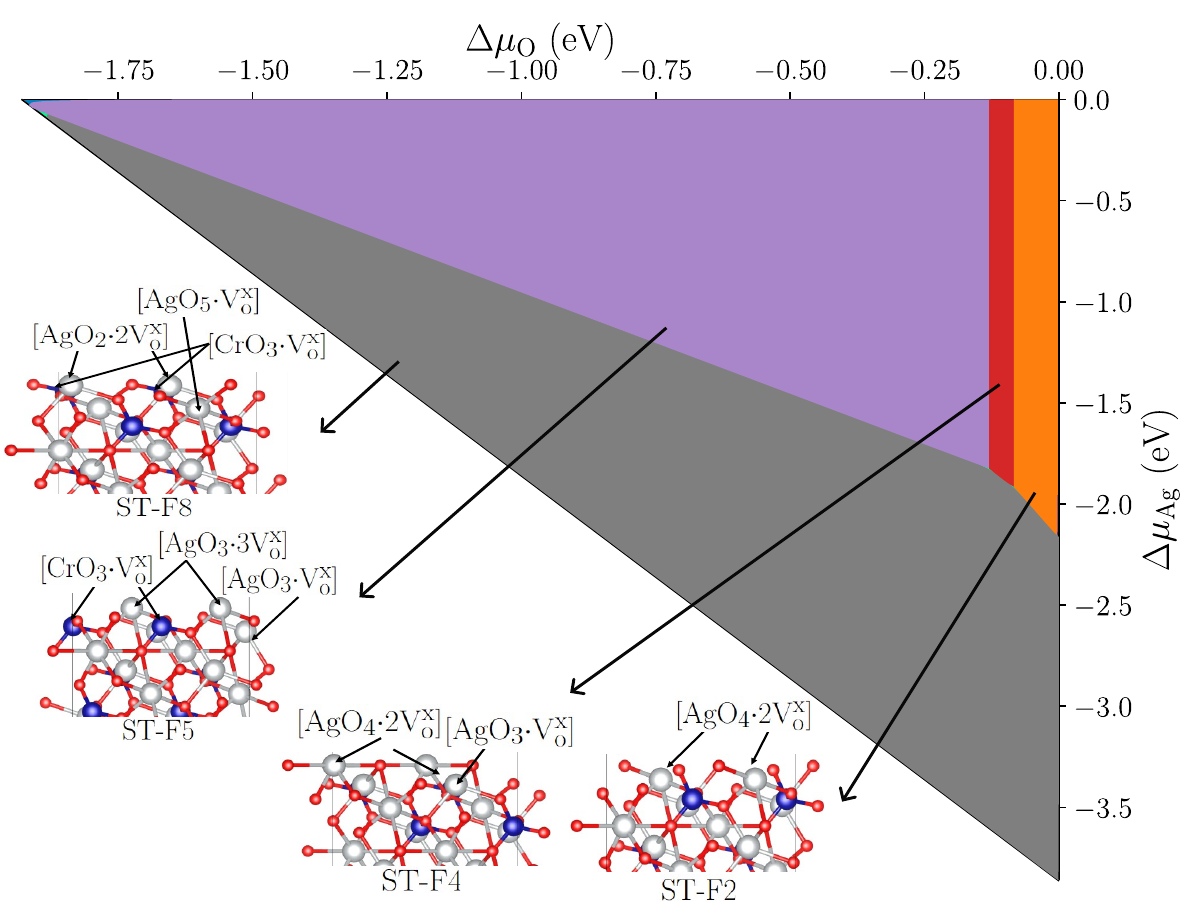}
        \caption{Surface diagram for the terminations of the (011) orientation. 
        Here, the second most stable structures are depicted, as ST-F6 remains
        the most stable regardless of the thermodynamic conditions. 
        Arrows indicate the corresponding non-relaxed slabs.}
    \end{figure}

    \begin{figure}[H]
        \centering
        \includegraphics[scale=0.6]{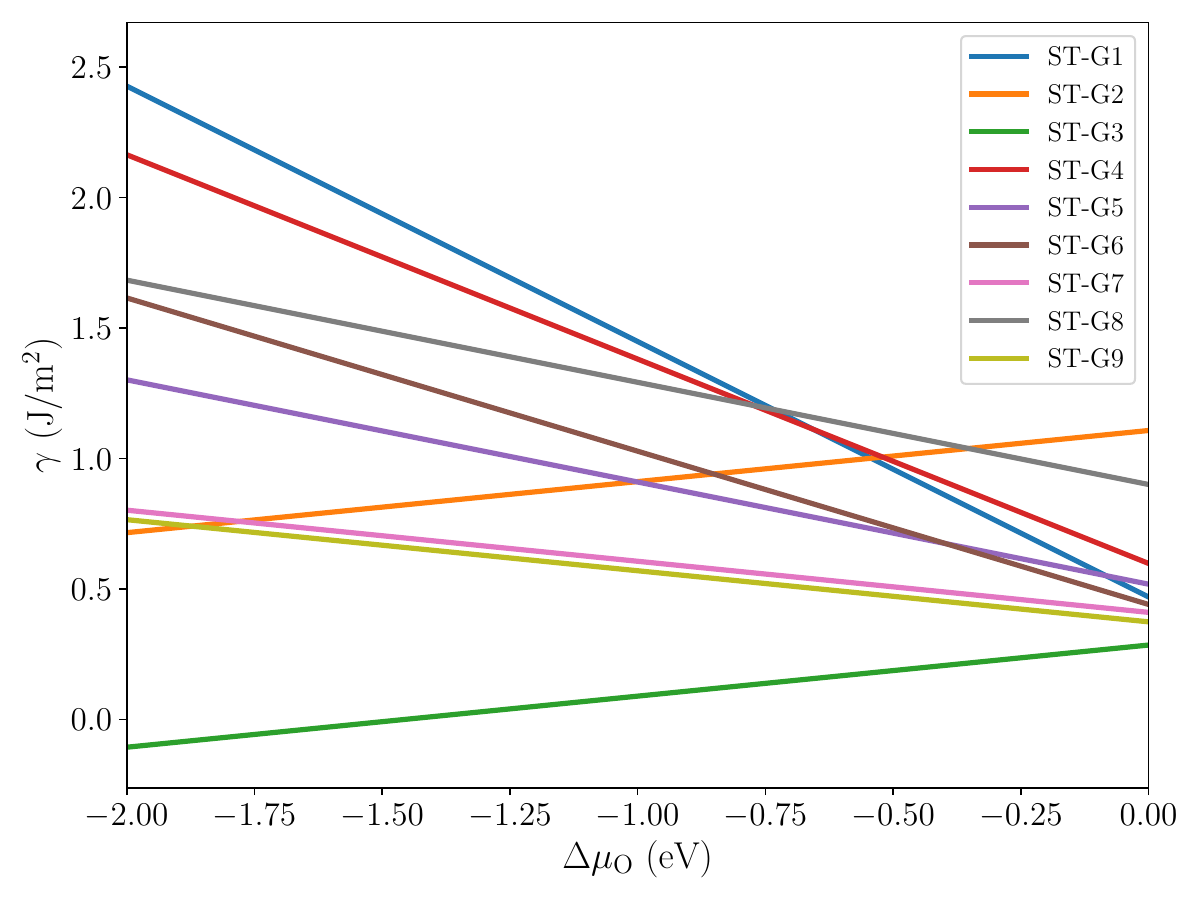}
        \caption{Gibbs surface free energy as a function of 
            $\Delta\mu_{\ce{O}}(T, P)$ for the terminations of the (101) orientation. 
            $\Delta\mu_{\ce{Ag}}(T, P) = 0$ eV.}
    \end{figure}
    
    \begin{figure}[H]
        \centering
        \includegraphics[scale=0.6]{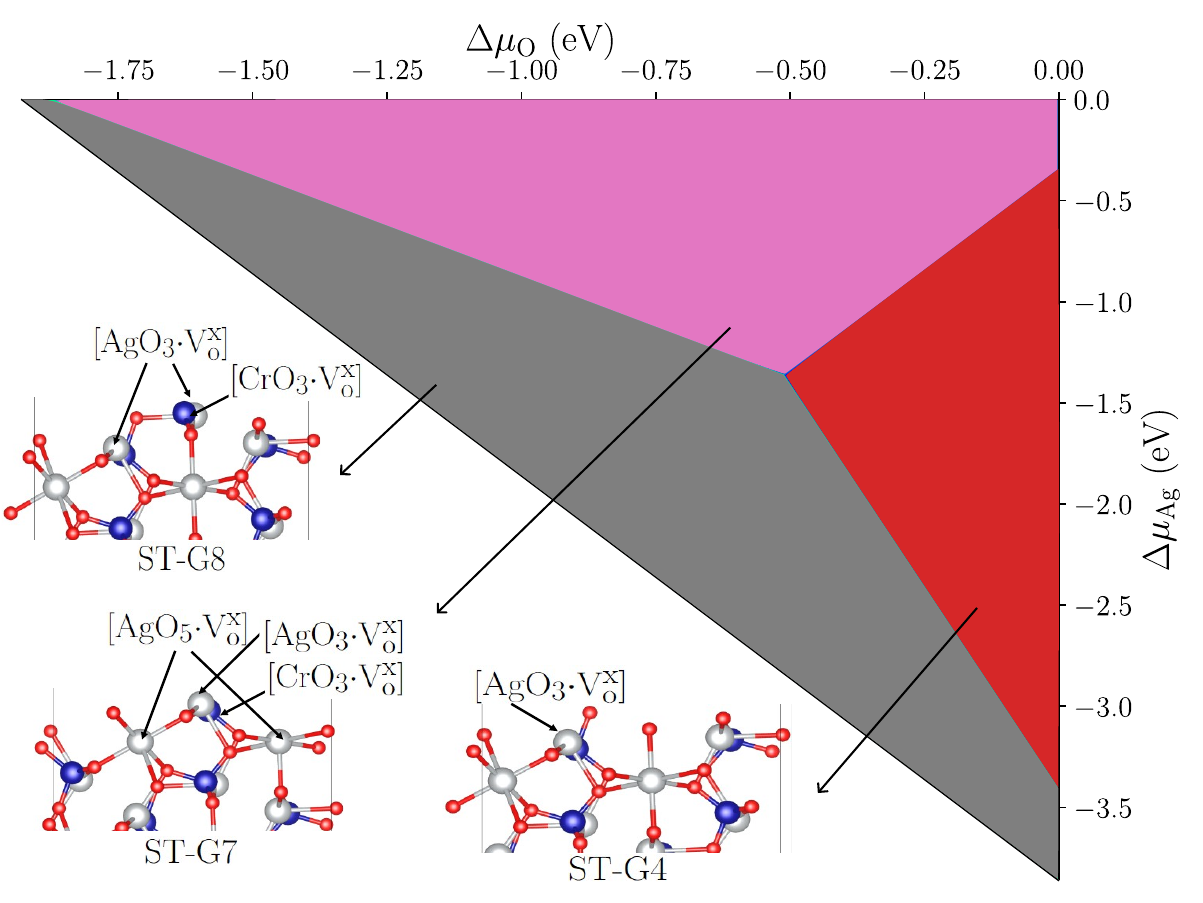}
        \caption{Surface diagram for the terminations of the (101) orientation. 
        Here, the second most stable structures are decpicted, as ST-G9 remains
        the most stable regardless of the thermodynamic conditions. 
        Arrows indicate the corresponding non-relaxed slabs.}
    \end{figure}

    \section{\label{sec:stabiliy-tendencies}Stability Trends}
    \subsection{\label{subsec:110}(110) orientation}
    The most stable terminations of the (110) orientation are ST-A2, 
    ST-A5, and ST-A6. After the relaxation step, ST-A2 undergoes minimal structural 
    modifications, including the addition of one oxygen in one oxygen atom in both a [\ce{AgO3}$\cdot\mathrm{V_o^x}$] 
    and a [\ce{AgO4}$\cdot2\mathrm{V_o^x}$] cluster, therby completing the [\ce{AgO4}] coordination 
    and nearly completing the [\ce{AgO6}] one. ST-A5 also shows minimal structural 
    modifications, with the coordination of the [\ce{CrO2}$\cdot2\mathrm{V_o^x}$] 
    cluster increasing to [\ce{CrO3}$\cdot\mathrm{V_o^x}$]. ST-A6 presents more noticeable structural changes, generaly 
    deacreasing the coordination of silver clusters, while exposing chromium clusters at the surface 
    (though without altering their coordination). 

    The leaste stable terminations are ST-A1 and ST-A7. In ST-A1, only minor structural modifications occur, namely the
    completion of the [\ce{AgO4}] coordination. However, this termination contains a high concentration of 
    silver atoms, while the chromium atoms are located only in the second atomic layer. In ST-A7, the degree of cluster 
    coordination remains unchanged, but the atomic displacements 
    are so large that the final structure appears poorly sustained: the first atomic layer is connected to the second by only 3 chemical bonds, which 
    can directly compromise the system's stability.

\subsection{\label{sec:111}(111) orientation}
    For (111) orientation, the ST-B7 termination is the most 
    stable, regardless of the thermodynamic conditions. Relaxation 
    reduces the coordination of its [\ce{AgO5}$\cdot\mathrm{V_o^x}$] 
    clusters, producing [\ce{AgO3}$\cdot3\mathrm{V_o^x}$] and 
    [\ce{AgO4}$\cdot2\mathrm{V_o^x}$]. At the same time, one oxygen atom is added to a [\ce{AgO2}$\cdot2\mathrm{V_o^x}$] cluster, 
    yielding [\ce{AgO3}$\cdot\mathrm{V_o^x}$]. 

    The least stable terminations are ST-B1 and ST-B3. ST-B1 exhibits significant structural distortion, with part of 
    the second atomic layer in the unrelaxed structure shifting towards the first layer, therby increasing the exposure of 
    silver atoms while leaving chromium absent. 
    ST-B3 also undergoes strong distortion, reducing the coordination of 
    some [\ce{AgO6}] clusters, fully exposing one [\ce{AgO4}], and completing the 
    coordination of [\ce{CrO3}$\cdot\mathrm{V_o^x}$]. Moreover, highly undercoordinated [\ce{AgO2}$\cdot4\mathrm{V_o^x}$] 
    clusters are present.
    
\subsection{\label{sec:100}(100) orientation}
    For the (100) orientation, ST-C6 termination is the most stable termination under all thermodynamic 
    conditions, followed by ST-C4 and ST-C3. ST-C6 exhibits minimal distorion  with almost no atomic displacement, 
    and preserves fully coordinated surface silver and chromium tetraheda. ST-C4 shows greater structural distortion: 
    the second atomic layer moves toward the first one, increasing the exposure of chromium clusters (not accessible 
    in the unrelaxed structure), similarly to ST-C3.

    In contrast, ST-C1 and ST-C7 are the least stable. In ST-C1, 
    atomic displacements increase the number of silver atoms 
    near the first layer, forming [\ce{AgO4}$\cdot2\mathrm{V_o^x}$] clusters. 
    ST-C7, however, shows neither significant atomic displacements nor coordination 
    changes. Yet, similiar ST-A7 in the (110) 
    orientation, it features a [\ce{CrO2}$\cdot2\mathrm{V_o^x}$] 
    cluster at the surface. This suggests that chromium clusters with two 
    oxygen vacancies tend to destabilize the system, even if the structure 
    is close to an optimal configuration.

\subsection{\label{sec:001}(001) orientation}
    All (001) terminations are all structurally similar, being composed of the same three types of clusters before and after relaxation, with 
    differences only in silver coordination (all showing reduced oxygen coordination). Therefore, their stability 
    variations are primarily associated with the extent of structural distortion.
    
    ST-D3 and ST-D4 are the most stable terminations, showing 
    lower structural distortions. The least stable one, 
    regardless of thermodynamic conditions, is ST-C1, 
    which undergoes such severe structural distortion that some oxygen atoms are 
    expelled from the lattice, forming \ce{O3} molecules. 
    This occurs only in this case and may indicate that the loss of 
    species from the surface leads to pronounced instability.

\subsection{\label{sec:010}(010) orientation}
    For the (010) orientation, ST-E3 is the most stable termination under all thermodynamic 
    conditions. Despite significant distortions, it remains the least distorted of this set 
    and preserves its surface [\ce{CrO4}] clusters. The second most stable case, ST-E2, increases the 
    number of chromium atoms in the first atomic layer.
    
    Conversely, ST-E1 is the least stable termination, being 
    the only one with a first atomic layer composed exclusively of 
    silver clusters.

\subsection{\label{sec:011}(011) orientation}
    The most stable (011) termination is ST-F6, which undergoes structural distortion that 
    increases the exposure of its chromium clusters, while completing their coordination 
    (previoulsy [\ce{CrO3}$\cdot\mathrm{V_o^x}$]). 

    Among the least stable cases, ST-F1 exhibits significant distortion, but does 
    not expose chromium clusters. It also contains [\ce{AgO2}$\cdot4\mathrm{V_o^x}$] clusters, 
    as observed in ST-B3 of the (111) orientation. ST-F7, on the other hand, 
    exposes chromium clusters with one oxygen vacancy, but its severe distortion 
    degree generates highly undercoordinated [\ce{AgO}$\cdot5\mathrm{V_o^x}$] 
    clusters. This behavior resembles that of [\ce{AgO2}$\cdot4\mathrm{V_o^x}$] clusters, 
    suggesting that surface [\ce{AgO6}] clusters with coordination numbers below three strongly 
    destabilize the system.

\subsection{\label{sec:101}(101) orientation}
    In the (101) orientation, the stability depends on the chemical potential of silver. ST-G3 is the most stable termination at 
    $\Delta\mu_{\ce{Ag}} = 0$ eV, while ST-G9 becomes more stable when 
    this condition is not satisfied. Thus, when the silver atoms in the \ce{Ag2CrO4} 
    phase are in equilibrium with a pure silver phase, the system that 
    brings the atomic layers closer together and exposes fully coordinated chromium clusters 
    (ST-G3) is the most stable. Upon relaxation, ST-G9 
    completes the coordination of its [\ce{CrO3}$\cdot\mathrm{V_o^x}$] 
    clusters, therby exposing a [\ce{AgO5}$\cdot\mathrm{V_o^x}$] cluster 
    in the second layer. These modifications likely enhance its stability under 
    $\Delta\mu_{\ce{Ag}} \neq 0$ eV, which corresponds to a wider range of thermodynamic conditions.

    The least stable termination is ST-G1. Although its atomic positions undergo only minor displacements, 
    the silver clusters lose coordination while the chromium clusters ramain unchanged. Furthermore, 
    the non-planar surface reduces chromium exposure, and the second atomic layer 
    appears poorly connected to the first one, leading to a less stable crystal structure.

    \bibliographystyle{apsrev4-2}
	\bibliography{bibliography}